\documentclass[modern]{aastex7}

\usepackage{amsmath, amssymb, bm}
\usepackage{graphicx}

\DeclareMathOperator*{\argmin}{arg\,min}

\begin{document}

\title{Observing Rayleigh-Taylor stable and unstable accretion through a Kalman filter analysis of X-ray pulsars in the Small Magellanic Cloud}

\author[orcid=0000-0002-6547-2039]{Joseph O'Leary}
\affiliation{School of Physics, University of Melbourne, Parkville, VIC 3010, Australia.}
\affiliation{Australian Research Council Centre of Excellence for Gravitational Wave Discovery (OzGrav), Parkville, VIC 3010, Australia.}
\email[show]{joe.oleary@unimelb.edu.au}

\author[orcid=0000-0003-4642-141X
]{Andrew Melatos}
\affiliation{School of Physics, University of Melbourne, Parkville, VIC 3010, Australia.}
\affiliation{Australian Research Council Centre of Excellence for Gravitational Wave Discovery (OzGrav), Parkville, VIC 3010, Australia.}
\email{}

\author[orcid=0000-0002-6542-6032]{Tom Kimpson}
\affiliation{School of Physics, University of Melbourne, Parkville, VIC 3010, Australia.}
\affiliation{Australian Research Council Centre of Excellence for Gravitational Wave Discovery (OzGrav), Parkville, VIC 3010, Australia.}
\email{}

\author[orcid=0000-0002-7652-2206]{Dimitris M. Christodoulou}
\affiliation{University of Massachusetts Lowell, Kennedy College of Sciences, Lowell, MA, 01854, USA.}
\affiliation{Lowell Centre for Space Science and Technology, Lowell, MA 01854, USA.}
\email{}

\author[orcid=0009-0007-3868-5227]{Nicholas J. O'Neill}
\affiliation{School of Physics, University of Melbourne, Parkville, VIC 3010, Australia.}
\affiliation{Australian Research Council Centre of Excellence for Gravitational Wave Discovery (OzGrav), Parkville, VIC 3010, Australia.}
\email{}

\author[orcid=0000-0002-2689-0190]{Patrick M. Meyers}
\affiliation{Theoretical Astrophysics Group, California Institute of Technology, Pasadena, CA 91125, USA.}
\email{}

\author[orcid=0000-0001-8572-8241]{Sayantan Bhattacharya}
\affiliation{University of Massachusetts Lowell, Kennedy College of Sciences, Lowell, MA, 01854, USA.}
\affiliation{Lowell Centre for Space Science and Technology, Lowell, MA 01854, USA.}
\email{}

\author[orcid=0000-0002-8427-0766]{Silas G.T. Laycock}
\affiliation{University of Massachusetts Lowell, Kennedy College of Sciences, Lowell, MA, 01854, USA.}
\affiliation{Lowell Centre for Space Science and Technology, Lowell, MA 01854, USA.}
\email{}

\begin{abstract}
Global, three-dimensional, magnetohydrodynamic simulations of Rayleigh-Taylor instabilities at the disk-magnetosphere boundary of rotating, magnetized, compact stellar objects reveal that accretion occurs in three regimes: the stable regime, the chaotic unstable regime, and the ordered unstable regime. Here we track stochastic fluctuations in the pulse period $P(t)$ and aperiodic X-ray luminosity $L(t)$ time series of 24 accretion-powered pulsars in the Small Magellanic Cloud using an unscented Kalman filter to analyze Rossi X-ray Timing Explorer data. We measure time-resolved histories of the magnetocentrifugal fastness parameter $\omega(t)$ and we connect $\omega(t)$ with the three Rayleigh-Taylor accretion regimes. The 24 objects separate into two distinct groups, with 10 accreting in the stable regime, and 14 accreting in the ordered unstable regime. None of the 24 objects except SXP 293 visit the chaotic unstable regime for sustained intervals, although several objects visit it sporadically. The Kalman filter output also reveals a positive temporal cross-correlation between $\omega(t)$ and the independently measured pulse amplitude $A(t)$, which agrees with simulation predictions regarding the pulse-forming behavior of magnetospheric funnel flows in the three accretion regimes.
\end{abstract}

\section{Introduction} \label{Sec:Intro}
Disk accretion onto magnetized compact objects \citep{Frank_2002,Longair_2010,Lyne_2012} such as pre-main sequence stars, e.g.\ classical T Tauri stars \citep{Koenigl_1991,Bouvier_2006,Hartmann_2016}, white dwarfs, e.g.\ cataclysmic binaries \citep{Smak_1984,Paczynski_1985}, and neutron stars, e.g.\ accretion-powered pulsars \citep{Ghosh_1978,White_1983,Lamb_1989}, is a nonlinear, time-dependent, and three-dimensional process. The accretion disk in such systems is truncated at the time-dependent Alfv\'en radius $R_{\rm{m}}(t)$, where magnetospheric Maxwell stresses are balanced by the disk ram pressure \citep{Ghosh_1979}. Accretion onto the compact object occurs, when the Kepler corotation radius $R_{\rm{c}}(t)$ is comparable to or exceeds $R_{\rm{m}}(t)$. The disk-magnetosphere interaction is complicated geometrically and is driven by complex, stochastic, hydromagnetic processes. These include the emergence of twisted magnetic field lines \citep{Amari_1996,Lai_2014}, magnetic reconnection \citep{Aly_1990,Parfrey_2012}, and disk warping and precession due to the misalignment of the magnetic and rotation axes \citep{Foucart_2011}, as well as self-healing Rayleigh-Taylor \citep{Arons_1976,Anzer_1980}, Kelvin-Helmholtz \citep{Burnard_1983}, and magnetorotational \citep{Balbus_1991} instabilities. Modeling this complex microphysics necessitates state-of-the-art numerical simulations \citep{Romanova_2003,Stone_2007,Stone2007a,Bachetti_2010,Romanova_2012,Dyda_2013,Romanova_2015}.

Three-dimensional magnetohydrodynamic simulations studying Rayleigh-Taylor instabilities at the disk-magnetosphere boundary \citep{Romanova_2008,Kulkarni_2008,Blinova_2016} reveal that accretion occurs in three regimes: (i) the stable regime; (ii) the Rayleigh-Taylor chaotic unstable regime; and (iii) the Rayleigh-Taylor ordered unstable regime. The time-dependent fastness parameter $\omega(t) = [R_{\rm{m}}(t)/R_{\rm{c}}(t)]^{3/2}$ is the main factor governing what regime applies at time $t$ \citep{Romanova_2015,Blinova_2016}.  In the stable regime, for $\omega(t) \gtrsim 0.6$, disk material is transported along magnetic field lines and deposited at two hotspots near the two magnetic poles; see Figure 1 in \cite{Kulkarni_2008} and Figures 1 and 2 in \cite{Romanova_2009} for examples of traditional, stable funnel flows. In the chaotic unstable regime, for $0.45 \lesssim \omega(t) \lesssim 0.6$, matter penetrates the magnetosphere, forming several accretion tongues terminating in hotspots moving randomly on the stellar surface between the equator and magnetic poles; see Figures 1--3 in \cite{Romanova_2008} and Figures 2 and 3 in \cite{Blinova_2016} for examples of accretion in the chaotic unstable regime, as well as Figure 6 in \cite{Kulkarni_2008} for an example of hotspot migration caused by Rayleigh-Taylor instabilities. In the ordered unstable regime, for $\omega(t) \lesssim 0.45$, the accretion tongues merge, forming one or two ordered equatorial accretion streams close to the star, which move randomly over the surface; see Figures 4 and 5 in \cite{Blinova_2016} for examples of ordered unstable accretion as well as an example of a simulated light curve produced by ordered unstable hotspots. Due to computational cost, the foregoing references focus on compact magnetospheres with $R_{\rm{m}}(t)/R \lesssim 7$, where $R$ is the stellar radius. Generalizing the analysis to (for example) accretion-powered pulsars with $2 \lesssim  \log_{10} [R_{\rm{m}}(t)/R] \lesssim  3$ is computationally intractable at present and must be explored by other means.

Recently it has been shown that one can track $R_{\rm m}(t)$, $R_{\rm c}(t)$, and hence $\omega(t)$ simultaneously as functions of $t$ in accretion-powered pulsars by analyzing measured time series of the aperiodic X-ray flux $L(t)$ and pulse period $P(t)$ with a Kalman filter \citep{Melatos_2022}. As a bonus, one can also estimate the static parameters of the classic magnetocentrifugal model of disk accretion \citep{Ghosh_1979}, e.g.\ the radiative efficiency of accretion and the star's magnetic moment \citep{OLeary_2024b,OLeary_2024a}, by combining the Kalman filter with a nested sampler \citep{Skilling_2004,Speagle_2020,Ashton_2022}. Extracting information of this sort is impossible in many traditional analyses of astronomical data, where one assumes magnetocentrifugal equilibrium, i.e.\ $\omega(t)=1$ for all $t$. The Kalman filter framework builds on earlier work employing autoregressive moving average models to study torque-luminosity correlations \citep{Baykal_1993} and testing for consistency with random walk and shot-noise processes \citep{deKool_1993,Baykal_1997,Lazzati_1997}. It complements related but different X-ray pulsar parameter estimation studies employing nonoptimal $\chi^2$ estimators \citep{Takagi_2016,Yatabe_2018} as well as Bayesian analyses of accretion torque models \citep{karaferias_2023}. The Kalman filter framework has been verified and calibrated against synthetic data \citep{Melatos_2022}. It has been applied to Rossi X-ray Timing Explorer (RXTE) Proportional Counter Array [PCA; \cite{Jahoda_2006}] measurements of the Small Magellanic Cloud (SMC) X-ray transient SXP 18.3 \citep{Corbet_2003}, yielding the first independent measurement of the magnetic dipole moment of SXP 18.3 without any attendant assumptions about the radiative efficiency of accretion \citep{OLeary_2024a}. The analysis was extended by \cite{OLeary_2024b} to the nonlinear regime (rotational disequilibrium) using an unscented Kalman filter \citep{Wan_2000,Challa_2011}, yielding new data products, e.g.\ auto- and cross-correlation coefficients involving hidden magnetospheric variables, for a subpopulation of 24 X-ray transients in the SMC; see Sections 5--8 in \cite{OLeary_2024b} for details.

In this paper we employ the Kalman filter framework developed by \cite{Melatos_2022} to track the evolution of $\omega(t)$ using RXTE PCA $P(t)$ and $L(t)$ time series from 24 SMC X-ray pulsars \citep{OLeary_2024b}. The goal is to connect the results with the three accretion regimes observed in global, three-dimensional, magnetohydrodynamic numerical simulations of rotating, magnetized stars, as itemized above \citep{Romanova_2015,Blinova_2016}. The paper is structured as follows. In Section \ref{Sec:Obs} we introduce the 24 RXTE PCA $P(t)$ and $L(t)$ time series analyzed in this paper. In Section \ref{Sec:KFAnalysis} we summarize (i) the magnetospheric state variables associated with the canonical picture of magnetocentrifugal accretion (the angular velocity of the star, the mass accretion rate, the Maxwell stress at the disk-magnetosphere boundary, and the radiative efficiency of accretion) as well as the measurement equations which map the state variables to $P(t)$ and $L(t)$; (ii) the nonlinear, stochastic differential equations which govern how the state variables evolve; and (iii) the unscented Kalman filter \citep{Wan_2000,Wan_2001,Challa_2011} and nested sampling algorithms \citep{Speagle_2020,Ashton_2022} employed here for nonlinear state and parameter estimation. In Section \ref{Sec:RTRegimes} we apply the Kalman filter and nested sampler to measure the time-dependent fastness parameters $\omega(t)$ of the 24 SMC X-ray pulsars in Section \ref{Sec:Obs}. We present and discuss the main result of the paper: 10 of the 24 objects have $\omega(t) \gtrsim 0.60$ and spend most of their time accreting in the stable regime, while the other 14 objects have $\omega(t) \lesssim 0.45$ and spend most or all of their time accreting in the ordered unstable regime. In Section \ref{Sec:MagFunnels} we cross-correlate the measured $\omega(t)$ time series against the independently measured pulse amplitude $A(t)$ of each object as a function of time, noting that the pulse amplitude is a signature of the number and geometry of the accretion funnels in the magnetosphere \citep{Kulkarni_2008}. The results are interpreted in terms of the previously published results from global, three-dimensional, magnetohydrodynamic numerical simulations of Rayleigh-Taylor accretion regimes in rotating, magnetized stars \citep{Kulkarni_2008,Romanova_2015,Blinova_2016}. The astrophysical implications are canvassed briefly in Section \ref{Sec:Concl}.

\section{RXTE Observations of SMC X-ray pulsars}\label{Sec:Obs}
The RXTE was launched on 1995 December 30 from the National Aeronautics and Space Administration's (NASA) Kennedy Space Centre, carrying onboard three payloads: the PCA \citep{Jahoda_2006}, the High Energy X-ray Timing Experiment \citep{Gruber_1996,Rothschild_1998}, and the All-Sky Monitor \citep{Levine_1996}. Their combined goal is to probe the timing properties of astronomical X-ray sources on time scales of microseconds to months in the 2--250 keV energy range. For $\sim$ 16 years, the spacecraft collected X-ray timing observations of the SMC, which are publicly available via the High Energy Astrophysics Science Archive Research Centre,\footnote{\url{https://heasarc.gsfc.nasa.gov/}} together with relevant RXTE data reduction tools, e.g.\ the \texttt{FTOOLS} and \texttt{PCABASKET} software packages. 

In this paper we analyze the time-domain RXTE data products published by \cite{Yang_2017} and recently analyzed by \cite{OLeary_2024b} for 24 accretion-powered pulsars in the SMC. The pulsars' names and timing properties are listed in Table \ref{Table:SourceProperties} for the convenience of the reader. The data products include time series sampled at times $t_n$ ($1 \leq n \leq N$) of the post-processed pulse period $P(t)$, aperiodic X-ray luminosity $L(t)$, and pulse amplitude $A(t)$. \cite{Yang_2017} generated light curves in the 3--10 keV energy range using a total of 36316 PCA X-ray timing measurements. They connected the timing properties of previously discovered X-ray transients with the X-ray sources detected in the PCA field of view using their known fundamental and harmonic frequencies and a Lomb-Scargle analysis, generating a $P(t)$ time series for each source. For every $P(t_n)$ sample, they folded the RXTE PCA light curves at the estimated frequency, yielding one $A(t_n)$ sample per $P(t_n)$ sample. The reader is referred to Table 2 in \cite{Yang_2017} and Figure 1 in \cite{OLeary_2024b} for per-object summaries of sample sizes, as well as \cite{VanderPlas_2018} for a practical guide on analyzing irregularly sampled time series with Lomb-Scargle periodograms. The number of $P(t)$, $L(t)$, and $A(t)$ samples $N$ is different for each source and is listed in the fifth column of Table \ref{Table:SourceProperties}. 

The time-domain data products employed in the present paper [as well as in \cite{OLeary_2024a} and \cite{OLeary_2024b}], i.e.\ the pulse period $P(t)$ time series in \cite{Yang_2017}, are not corrected for orbital motion. The latter authors analyzed the long-term, secular pulse period variations of 52 SMC HMXBs, while restricting attention to X-ray timing points with Lomb-Scargle statistical significance $\geq 99\%$ termed ``significant detections'' \citep{Yang_2017,VanderPlas_2018}. In general, significant detections are separated by $t_{n+1} - t_n \gg P_{\rm b}$, where $10 \lesssim P_{\rm b}/(1 \, \rm{day}) \lesssim 100$ denotes the typical binary orbital periods of SMC HMXBs \citep{Coe_2015,Kennea_2018}. Without finer sampling, it is challenging to correct the foregoing time-domain data products for orbital motion by tracking the harmonic Doppler modulation of $P(t)$. One can correct for the orbital motion independently using multi-wavelength spectroscopy of the companion star \citep{Killestein_2023}, but such an analysis lies outside the scope of this paper. Looking forward to the Kalman filter framework in Section \ref{Sec:KFAnalysis}, pulse period variations due to binary motion are comparable to or smaller than the pulse period measurement uncertainties $\sigma_P$ (see Table \ref{Table:SourceProperties}) and hence are absorbed by the Gaussian measurement noise process of Equation (\ref{Eq:PulsePeriodNoise}). To check the matter further, we present two preliminary tests. First, in Appendix \ref{App:B}, we assess how sensitive the Kalman filter in Section \ref{Sec:KFAnalysis} is to changes in $\sigma_P$, e.g. due to binary motion. Second, in Appendix \ref{App:PPVariations}, we present an order-of-magnitude analysis of the typical pulse period variations due to binary motion observed in some SMC HMXBs, e.g. SXP 2.37 \citep{LaPalombara_2016}. The provisional results of Appendix \ref{App:B} as well as the order-of-magnitude estimates of Appendix \ref{App:PPVariations} suggest that the sensitivity is low. A full battery of tests is postponed, until the Kalman filter analysis is repeated on more objects with higher $N$, to increase the statistical significance of the exercise.   

The spin distribution as well as the rotational state (equilibrium or otherwise) of the 24 objects drawn from \cite{Yang_2017} are visualized in Figure 2 in \cite{OLeary_2024b} using a traditional $P$--$\dot{P}$ diagram, where $\dot{{P}}= d P/dt$ is calculated by \cite{Yang_2017} from a linear fit of $P(t_n)$ versus $t_n$. The rotational state of each star is classified using $\dot{P}$ divided by its measurement error, denoted as $\epsilon$ in the third column of Table \ref{Table:SourceProperties}. Stars classified in Table 3 of \cite{Yang_2017}  as spinning up ($ \epsilon \leq -1.5$), rotational equilibrium ($-1.5  < \epsilon < 1.5$), and spinning down ($\epsilon \geq 1.5$), are reported in the top, middle, and bottom sections of Table \ref{Table:SourceProperties} respectively. The reader is referred to Sections 2 and 3 in \cite{Yang_2017} for details about the RXTE data reduction procedure and the overall properties of the population of accretion-powered pulsars in the SMC. 

\begin{table}
\centering
\hspace{-1.5cm}
\begin{tabular}{c c c c c}
\tableline
\tableline
Name (SXP) & $\dot{P}$ $(\rm s \; day^{-1})$ & $\epsilon$ & $\sigma_P$ $(\rm s)$ & $N$ \\
\tableline
    4.78 & $-2.0\times 10^{-6}$ & $-2.00$ & $0.004$ & $804$ \\
    11.9 & $-1.2\times 10^{-5}$ & $-3.00$ & $0.025$ & $394$ \\
    59.0 & $-4.4\times 10^{-5}$ & $-4.40$ & $0.110$ & $902$ \\
    172 & $-3.0\times 10^{-4}$ & $-4.33$ & $0.642$ & $863$ \\
    323 & $-1.5\times 10^{-3}$ & $-5.44$ & $1.829$ & $649$ \\
    756 & $-4.1\times 10^{-3}$ & $-2.48$ & $10.226$ & $391$ \\
\tableline
    6.85 & $-1.0\times 10^{-6}$ & $-0.50$ & $0.011$ & $616$ \\
    11.5 & $1.4\times 10^{-5}$ & $0.875$ & $0.040$ & $599$ \\
    18.3 & $3.0\times 10^{-6}$ & $0.75$ & $0.025$ & $854$ \\
    82.4 & $5.3\times 10^{-5}$ & $0.65$ & $0.283$ & $863$ \\
    101 & $-1.6\times 10^{-4}$ & $-0.48$ & $0.86$ & $411$ \\
    152 & $-1.1\times 10^{-4}$ & $-1.47$ & $0.338$ & $561$ \\
    202A & $-2.9\times 10^{-4}$ & $-1.06$ & $1.236$ & $567$ \\
    214 & $-8.9\times 10^{-5}$ & $-0.45$ & $1.135$ & $854$ \\
    264 & $1.4\times 10^{-4}$ & $0.20$ & $1.564$ & $650$ \\
    292 & $2.7\times 10^{-4}$ & $0.81$ & $0.868$ & $644$ \\
    293 & $-5.0\times 10^{-5}$ & $-0.29$ & $0.774$ & $944$ \\
    523 & $-1.8\times 10^{-3}$ & $-0.87$ & $5.313$ & $552$ \\
    565 & $2.3\times 10^{-4}$ & $0.20$ & $4.497$ & $872$ \\
    893 & $-6.8\times 10^{-4}$ & $-0.59$ & $9.591$ & $642$ \\
\tableline
    8.88 & $3.0\times 10^{-6}$ & $3.0$ & $0.011$ & $861$ \\
    51.0 & $5.3\times 10^{-5}$ & $3.12$ & $0.155$ & $653$ \\
    95.2 & $2.2\times 10^{-4}$ & $1.56$ & $0.234$ & $867$ \\
    138 & $4.6\times 10^{-4}$ & $1.75$ & $0.577$ & $898$ \\
\tableline
\end{tabular}
\caption{Names and timing properties of the 24 SMC accretion-powered pulsars analyzed in this paper. The pulse period derivative (column 2), proximity to rotational equilibrium (column 3), pulse period standard deviation (column 4), and total number of RXTE PCA observations (column 5) are denoted by $\dot{P}$, $\epsilon$, $\sigma_P$, and $N$, respectively. The top, middle, and bottom sections contain X-ray pulsars classified in Table 3 in \cite{Yang_2017} as spinning up ($\epsilon \leq -1.5$), near rotational equilibrium ($-1.5  < \epsilon < 1.5$), and spinning down ($\epsilon \geq 1.5$), respectively.}
	\label{Table:SourceProperties}
\end{table}

\section{Kalman filter parameter estimation}\label{Sec:KFAnalysis}
The unscented Kalman filter \citep{Julier_1997,Wan_2000,Wan_2001,Zarchan_2005} is a recursive Bayesian estimator, whose primary goal is to compute the posterior density of the hidden state variables $\bm{X}(t_n)$, given a stream of time-ordered, noisy measurements $\bm{Y}(t_n)$, and a set of fixed model parameters $\bm{\Theta}$.  Here we vary $\bm{\Theta}$, subject to suitable astrophysical priors, and employ a nested sampler \citep{Skilling_2004,Skilling_2006,Ashton_2019,Speagle_2020} to determine the value of $\bm{\Theta}$ that maximize the Bayesian likelihood, conditional on $\bm{Y}(t_1),\hdots,\bm{Y}(t_N)$. In this section, we explain how to adapt the Kalman filter components to magnetocentrifugal parameter estimation for accretion-powered pulsars. In Section \ref{SubSec:Obs_and_States}, we define the stochastic magnetospheric variables $\bm{X}$,  the noisy observables $\bm{Y}$, as well as the nonlinear measurement equations, which map $\bm{Y}$ to $\bm{X}$. The equations of motion obeyed by $\bm{X}$ are written down in Section \ref{SubSec:AccretionTorque}, viz. the canonical,  magnetocentrifugal accretion torque law \citep{Ghosh_1977,Ghosh_1979}, as well as a phenomenological, idealized model of
the stochastic driving forces associated with hydromagnetic instabilities at the disk-magnetosphere boundary \citep{Melatos_2022}. The nested sampling algorithm is summarized in the context of Kalman parameter estimation in Section \ref{SubSec:KF_NS}. The technical information in Sections \ref{SubSec:Obs_and_States}--\ref{SubSec:KF_NS} is presented in abridged form for the purpose of reproducibility. We refer the reader to Appendix C of \cite{Melatos_2022} for fuller details about the complex physics at the disk-magnetosphere boundary, and to Section 4.4 of \cite{OLeary_2024a} and Section 2.3 of \cite{OLeary_2024b} for information about the systematic and model uncertainties, as well as some of the simplifications associated with Equations (\ref{Eq:PulsePeriodNoise})--(\ref{Eq:EtaConstant}) below.

\subsection{Observables and state variables}\label{SubSec:Obs_and_States}

The post-processed $P(t)$ and $L(t)$ time series published by \cite{Yang_2017} are ingested by the Kalman filter as observables. Specifically one has $\bm{Y}_{n} = [P(t_n), L(t_n)]$ at time $t_n$ from 24 accretion-powered pulsars in the SMC (see Table \ref{Table:SourceProperties}).

We express the standard model of magnetocentrifugal disk accretion \citep{Ghosh_1977,Ghosh_1979} in terms of four time-dependent magnetospheric variables, three of which are hidden: (i) the angular velocity of the star $\Omega(t)$ (units: $\rm{rad \, s^{-1}}$), which is inversely proportional to the pulse period $P(t)$ and is essentially measured directly, as discussed below; (ii) the mass accretion rate $Q(t)$ (units: $\rm{g \, s^{-1}}$); (iii) the Maxwell stress at the disk-magnetosphere boundary $S(t)$ (units: $\rm{g \, cm^{-1} \, s^{-2}}$); and (iv) the dimensionless radiative efficiency of accretion $\eta(t)$. 

The foregoing magnetospheric variables are related indirectly to the RXTE observables $P(t)$ and $L(t)$ by
\begin{equation}\label{Eq:PulsePeriodNoise}
    P(t) = 2 \pi/\Omega(t) + N_{P}(t),
\end{equation}
and
\begin{equation}\label{Eq:LuminosityNoise}
    L(t) = GM Q(t) \eta(t)/R + N_{L}(t),
\end{equation}
where Newton's gravitational constant and the mass of the star are denoted by $G$ and $M$ respectively. In Equations (\ref{Eq:PulsePeriodNoise}) and (\ref{Eq:LuminosityNoise}), $N_{P}(t)$ and $N_{L}(t)$ denote Gaussian noise processes, satisfying the following ensemble statistics: $\langle N_{P}(t_n) \rangle$ = 0, $\langle N_{L}(t_n) \rangle$ = 0, $\langle N_{P}(t_n) N_{P}(t_{n'}) \rangle = \Sigma_{PP}^2 \delta_{nn'}$, $\langle N_{L}(t_n) N_{L}(t_{n'}) \rangle = \Sigma_{LL}^2 \delta_{n n'}$, and $\langle N_{L}(t_n) N_{P}(t_{n'}) \rangle = 0$, where $\delta_{nn'}$ denotes the Kronecker delta. 
\subsection{Magnetocentrifugal accretion dynamics}\label{SubSec:AccretionTorque}
The two time-dependent characteristic radii in the canonical picture of magnetocentrifugal accretion, i.e.\ the Alfv\'en radius $R_{\rm{m}}(t)$ and corotation radius $R_{\rm{c}}(t)$, are expressed in terms of the magnetospheric variables in Section \ref{SubSec:Obs_and_States} by \citep{Melatos_2022}
\begin{equation}\label{Eq:AlfRad}
R_{\rm m} (t) = (2 \pi^{2/5})^{-1} (GM)^{1/5} Q(t)^{2/5} S(t)^{-2/5},
\end{equation}
and 
\begin{equation}\label{Eq:CoRot}
R_{\rm c}(t) = (GM)^{1/3}\Omega(t)^{-2/3}.
\end{equation}
It may seem that $R_{\rm{m}}(t)$ in Equation (\ref{Eq:AlfRad}) increases with $Q(t)$, whereas the opposite is expected physically, e.g.\ as inferred from Equation (7) in \cite{Klus_2014}. However, $S(t) = (2 \pi)^{-1} \mu^2 R_{\rm{m}}(t)^{-6}$ contains $R_{\rm{m}}(t)$ and the magnetic dipole moment $\mu$ implicitly. Upon substituting the latter definition for $S(t)$ into Equation (\ref{Eq:AlfRad}), we recover the standard expression for $R_{\rm{m}}(t)$ in terms of $Q(t)$ and $\mu$, and it becomes apparent that $R_{\rm m}(t)$ decreases, as $Q(t)$ increases, as expected.

In the standard model of magnetocentrifugal accretion, a rotating, magnetized compact object experiences a combination of material and hydromagnetic torques, described by the phenomenological torque law \citep{Ghosh_1977,Ghosh_1979}
\begin{equation}\label{Eq:AccretionTorque}
    I \frac{d \Omega}{dt} = [GM R_{\rm m}(t)]^{1/2} Q(t) [1 - \omega(t)],
\end{equation}
where the star's moment of inertia is denoted by $I$. We remind the reader that the fastness parameter is defined in terms of the time-dependent characteristic radii according to $\omega(t) = [R_{\rm{m}}(t)/R_{\rm{c}}(t)]^{3/2}$. For $R_{\rm{m}}(t) < R_{\rm{c}}(t)$, material strikes the stellar surface, and the star spins up, as angular momentum is exchanged between the disk material and the star, with $d \Omega/dt > 0$. For $R_{\rm{m}}(t) > R_{\rm{c}}(t)$, material in the disk-magnetosphere boundary orbits slower than the corotating magnetosphere, and matter is ejected centrifugally, spinning the star down, with $d \Omega/dt < 0$. For $R_{\rm{m}}(t) \approx R_{\rm{c}}(t)$, the star is said to be near rotational equilibrium, with $d \Omega/dt \approx 0$. All three scenarios are captured approximately by Equation (\ref{Eq:AccretionTorque}) and the three parts of Table \ref{Table:SourceProperties}.

Motivated by the promising results reported previously \citep{Melatos_2022,OLeary_2024a,OLeary_2024b}, we adopt a phenomenological, idealized, Ornstein-Uhlenbeck model \citep{Gardiner_1985} of the stochastic driving forces associated with hydromagnetic instabilities at the disk-magnetosphere boundary. Specifically, we assume that $Q(t)$ and $S(t)$ execute mean-reverting random walks driven by white-noise stochastic fluctuations and satisfy the Langevin equations
\begin{equation}\label{Eq:LangevinQ}
    \frac{d Q}{dt} = -\gamma_Q [Q(t) - \bar{Q}] + \xi_Q(t),
\end{equation}
\begin{equation}\label{Eq:LangevinS}
    \frac{d S}{dt} = -\gamma_S [S(t) - \bar{S}] + \xi_S(t),
\end{equation}
where $\gamma_Q$ and $\gamma_S$ denote damping constants, and the white-noise driving terms, $\xi_Q(t)$ and $\xi_S(t)$, satisfy the following ensemble statistics: $\langle \xi_{Q}(t)\rangle = 0$, $\langle \xi_S(t) \rangle = 0$, $\langle \xi_Q(t) \, \xi_Q(t') \rangle = \sigma_{QQ}^2 \, \delta (t - t')$, $\langle \xi_S(t) \, \xi_S(t')  \rangle = \sigma_{SS}^2 \, \delta(t - t')$, and $\langle \xi_Q(t) \, \xi_S (t') \rangle = 0$. Equations (\ref{Eq:LangevinQ}) and (\ref{Eq:LangevinS}) ensure that $Q(t)$ and $S(t)$ do not undergo long-term secular drifts, and fluctuate stochastically about their asymptotic, ensemble-averaged values denoted by $\langle Q(t) \rangle \approx \bar{Q}$ and $\langle S(t) \rangle \approx \bar{S}$ with characteristic time scales of mean-reversion $\gamma_Q^{-1}$ and $\gamma_S^{-1}$, and rms fluctuations $\sim \gamma_{Q}^{-1/2}\sigma_{QQ}$ and $\sim \gamma_{S}^{-1/2}\sigma_{SS}$, respectively.  Equations (\ref{Eq:AccretionTorque})--(\ref{Eq:LangevinS}) are supplemented with a deterministic equation for the radiative efficiency, 
\begin{equation}\label{Eq:EtaConstant}
    \eta(t) = \bar{\eta}.
\end{equation}
That is, we assume $\eta(t)=\bar{\eta}=\rm{constant}$ but leave its value free to be estimated by the Kalman filter. 

Equations (\ref{Eq:AccretionTorque})--(\ref{Eq:EtaConstant}) represent a compromise dictated by the modest explanatory power of the available volume of RXTE PCA post-processed data $(N \leq 10^{3})$. They oversimplify many important aspects of the complex accretion physics associated with X-ray pulsars. As just one example, we draw the reader's attention to the canonical magnetocentrifugal torque law, Equation (\ref{Eq:AccretionTorque}). Within the traditional magnetocentrifugal paradigm, it is implicitly assumed that the rotation and magnetic axes are aligned, i.e. $\Theta = 0^\circ$, and hence, we approximate $\Omega(t)$ as a single, time-dependent, scalar variable. In practice, however, the star's angular velocity is a vector $\bm{\Omega(t)} = [\Omega_1(t), \Omega_2(t), \Omega_3(t)]$, with $\Omega(t) \approx \Omega_3(t)$ assumed in this paper. It is possible in principle to modify Equation (\ref{Eq:AccretionTorque}) for $\Theta \neq 0^\circ$ using (for example) Equation 5.5 of \cite{Lai_1999}. However, such modifications give rise to complicated accretion torques as well as disk warping and precession due to the misalignment of the magnetic and rotation axes \citep{Lai_1999,Foucart_2011,Lai_2014, Romanova_2021}. Accordingly we persevere with Equation (\ref{Eq:AccretionTorque}) as a first pass at the problem until larger volumes of data become publicly available. We refer the reader to (i) Appendix C of \cite{Melatos_2022} for a detailed discussion about the disk-magnetosphere boundary and how one might modify Equation (\ref{Eq:AccretionTorque}) to account for several important time-dependent phenomena not captured fully by Equation (\ref{Eq:AccretionTorque}), e.g.\ episodic accretion in the weak propeller regime  \citep{Dangelo_2010,Dangelo_2012,Dangelo_2015,Dangelo_2017}; (ii) Section 2.4 of \cite{Melatos_2022} for a summary of the simplifications inherent in Equations (\ref{Eq:LangevinQ}) and (\ref{Eq:LangevinS}); and (iii) Section 2.3 of \cite{OLeary_2024b} for situations where the assumption $\eta(t)=\bar{\eta}=\rm{constant}$ may not hold.

\subsection{Kalman filter and nested sampler}\label{SubSec:KF_NS}
Given a RXTE PCA time series $\bm{Y}_{n} = [P(t_n), L(t_n)]$ with $1 \leq n \leq N$, a Kalman filter sequentially estimates three  quantities: the hidden state $\bm{X}_{n} = [\Omega(t_n),Q(t_n), S(t_n), \eta(t_n)]$, the hidden state error covariance $\bm{P}_{n}$, and a Gaussian likelihood $\mathcal{L} = p(\{\bm{Y}\}_{n=1}^{N}| \bm{\Theta})$. The recursive updates of $\bm{X}_n$ and $\bm{P}_{n}$ depend on the respective dynamical and measurement uncertainties, quantified by the process and measurement noise covariances, denoted by $\bm{Q}_n$ and $\bm{\Sigma}_n$ at time $t_n$, respectively. With respect to $\mathcal{L}$, the Kalman filter computes the expected value of the RXTE PCA observables $\bm{Y}^{-}_{n}$, from the noiseless terms on the right-hand side of Equations (\ref{Eq:PulsePeriodNoise}) and $(\ref{Eq:LuminosityNoise})$. The measurement residual $\bm{e}_n = \bm{Y}_n - \bm{Y}^{-}_{n}$ and associated covariance $\langle \bm{e}_n \, \bm{e}_n^{\rm{T}} \rangle$ are then used to calculate $\mathcal{L}$, where the superscript `T' denotes the matrix transpose. For brevity, we do not write out the Kalman recursion equations explicitly and their numerous auxiliary quantities, e.g.\ $\mathcal{L}$, $\bm{Q}_n$, $\bm{\Sigma}_n$, and the Kalman gain $\bm{k}_n$, as this is done elsewhere. Specifically, we refer the reader to Appendices A and B of \cite{OLeary_2024b} for more details about the unscented Kalman filter algorithm \citep{Julier_1997,Wan_2000,Wan_2001} and a summary of its output, respectively.

Nested sampling \citep{Skilling_2004,Skilling_2006} is an iterative, computational tool for evaluating multidimensional integrals. In the context of Bayesian inference, it approximates the marginal likelihood, i.e.\ Bayesian evidence, as well as the posterior distribution of $\bm{\Theta}$ via weighted histograms or other density estimation techniques. The \texttt{dynesty} sampler \citep{Speagle_2020} used in this paper has two main inputs:  $\mathcal{L}$, and the prior distribution $p(\bm{\Theta})$ of $\bm{\Theta}$. It also has two tunable controls: the number of ``live points'' $N_{\rm{live}}$, and a stopping condition or tolerance $\Delta$. 

Kalman filter parameter estimation with nested sampling proceeds as follows. The sampler is initialized with an ensemble of $N_{\rm{live}}$ live points, denoted by $\bm{\Theta}^{(1)}_{1}$,$\hdots$,$\bm{\Theta}^{(1)}_{N_{\rm{live}}}$, drawn randomly from the prior $p(\bm{\Theta})$.  At step $k$ in the iterative process, the nested sampler calculates the Kalman filter likelihood $\mathcal{L}^{(k)}_{m} = p[\{\bm{Y}_n\}_{n=1}^{N}|\bm{\Theta}^{(k)}_{m}]$ for $1 \leq m \leq N_{\rm{live}}$, and replaces $\bm{\Theta}_{m'}^{(k)}$ with a new live point $\bm{\Theta}_{m'}^{(k+1)}$ drawn from $p(\bm{\Theta})$, subject to the condition $\mathcal{L}^{(k+1)}_{m'} > \mathcal{L}^{(k)}_{m'}$ with $m' = \argmin_{m} \mathcal{L}^{(k)}_{m}$. At step $k+1$, the sampler keeps track of $N_{\rm{live}}$ sequentially updated live points, as well as $k$ discarded ``dead'' points. When $\Delta$ is satisfied, the sampler uses the live and discarded points to estimate the Bayesian evidence and the posterior distribution of $\bm{\Theta}$ via (for example) Equations (16) and (18) in \cite{Speagle_2020}, respectively. The reader is referred to Appendices A.4 and A.5 of \cite{OLeary_2024b} for details about the astrophysical priors and nested sampler settings employed here, as well as to \cite{Meyers_2021} and \cite{ONeill_2024} for overviews of a related but different parameter estimation problem in pulsar astrophysics, namely analyzing radio pulsar timing noise with a Kalman filter. 

\section{Rayleigh-Taylor Accretion Regimes}\label{Sec:RTRegimes}
Hydromagnetic accretion onto a rotating, magnetized, compact object mediated by instabilities at the disk-magnetosphere boundary involves a complex interplay between the plasma and magnetic field in the magnetosphere and disk. In this section, we present new observational evidence for the Rayleigh-Taylor ordered unstable and stable accretion regimes predicted by three-dimensional numerical simulations \citep{Kulkarni_2008,Blinova_2016,Burdonov_2022}. In Section \ref{SubSec:TimeDepOmega} and Appendix \ref{App:A}, we measure the fastness history $\omega(t_1), \dots, \omega(t_N)$ of every one of the 24 objects in Table \ref{Table:SourceProperties} using the inference scheme in Section \ref{Sec:KFAnalysis} and show that, remarkably, they separate into two distinct classes. We then discuss in detail the qualitative features of $\omega(t)$ for one representative object from each class, namely SXP 18.3 and SXP 51.0, in Sections \ref{SubSec:Stable} and \ref{SubSec:OrderedUnstable} respectively and connect the two classes to the aforementioned Rayleigh-Taylor accretion regimes.

\subsection{Time-resolved fastness parameter $\omega(t)$}\label{SubSec:TimeDepOmega}

We apply the Kalman filter and nested sampler in Section \ref{Sec:KFAnalysis} to the RXTE PCA data for the 24 objects in Table \ref{Table:SourceProperties}. For each object, we employ the posterior maximum likelihood estimate of $\bm{\Theta}$ returned by the nested sampler as input to the Kalman filter to generate the time series $\omega(t_n)$ ($1 \leq n \leq N$); see Sections 3 and 4 in \cite{OLeary_2024b} and \cite{OLeary_2024a} for details of similar analyses. Each plotted point corresponds to one $\omega(t_n)$ sample. That is, we evaluate $\omega(t_n) = [R_{\rm m}(t_n)/R_{\rm c}(t_n)]^{3/2}$ for every $\Omega(t_n)$, $Q(t_n)$, and $S(t_n)$ returned by the Kalman filter, where $R_{\rm m}(t)$ and $R_{\rm c}(t)$ are defined in terms of the magnetospheric variables by Equations (\ref{Eq:AlfRad}) and (\ref{Eq:CoRot}), respectively. We plot $\omega(t_n)$ versus $t_n$ in the top panels of Figures \ref{Fig:RepSource18.3} and \ref{Fig:RepSource51} for the representative objects SXP 18.3 and SXP 51.0 respectively and in the top panels of Figures \ref{Fig:RepSource4.78}--\ref{Fig:RepSource138} in Appendix \ref{App:A} for the other 22 objects. The top panels are the focus of this section. We also plot the pulse amplitude $A(t_n)$ versus $t_n$ in the middle panels of the same figures, and the associated $A(t_n)$-$\omega(t_n)$ scatter plot in the bottom panels. The middle and bottom panels are the foci of Section \ref{Sec:MagFunnels}.

Visual inspection of the top panels of Figures \ref{Fig:RepSource18.3}, \ref{Fig:RepSource51}, and \ref{Fig:RepSource4.78}--\ref{Fig:RepSource138} reveals that the 24 objects separate cleanly into two classes. In one class, the fastness wanders randomly in the range $\omega(t_n) \gtrsim 0.60$. This class contains 10 objects, represented by SXP 18.3 (Figure \ref{Fig:RepSource18.3}) and also including SXP 4.78, SXP 6.85, SXP 11.5, SXP 11.9, SXP 138, SXP 152, SXP 264, SXP 292, and SXP 293. In the other class, the fastness wanders randomly in the range $\omega(t_n) \lesssim 0.45$. This class contains 14 objects, represented by SXP 51.0 (Figure \ref{Fig:RepSource51}) and also including SXP 8.88, SXP 59.0, SXP 82.4, SXP 95.2, SXP 101, SXP 172, SXP 202A, SXP 214, SXP 323, SXP 523, SXP 565, SXP 756, and SXP 893. None of the 24 objects in Table \ref{Table:SourceProperties} except SXP 293 spend sustained time intervals in the intermediate fastness range $0.45 \lesssim \omega(t_n) \lesssim 0.60$, although some objects make occasional, sporadic excursions into the latter range. We discuss the measured features and physical interpretation of the two classes in Sections \ref{SubSec:Stable} and \ref{SubSec:OrderedUnstable}.

\subsection{Stable regime}\label{SubSec:Stable}

Consider first the top panel of Figure \ref{Fig:RepSource18.3} for SXP 18.3. To guide the theoretical interpretation, we draw grey, dashed, horizontal lines to divide the panel into the ordered unstable [$\omega(t) \lesssim 0.45$], chaotic unstable [$0.45 \lesssim \omega(t) \lesssim 0.6$], and stable [$\omega(t) \gtrsim 0.6$] accretion regimes identified in numerical simulations \citep{Romanova_2008,Kulkarni_2008,Blinova_2016}. In many simulations, accretion still occurs in the weak propeller regime $1 \lesssim \omega(t) \lesssim 1.25$, so we extend the vertical axis to span the range $0 \leq \omega(t) \leq 1.25$ \citep{Spruit_1993,Dangelo_2010,Lii_2014,Papitto_2015}. 

The top panel of Figure \ref{Fig:RepSource18.3} displays three key features. First, we observe that $\omega(t_n)$ stays between the approximate boundaries of the stable accretion regime for $\approx 93\%$ of the total observation time. Second, we observe eight brief excursions, when $\omega(t_n)$ enters the chaotic unstable accretion regime near MJD 51650, MJD 52180, MJD 52320, MJD 53560, MJD 54410, MJD 54740, MJD  54815, and MJD 55575. The eight excursions account for $\approx 3\%$ of the total observation time. None of the excursions are sustained; there is no evidence that SXP 18.3 spends half its time persistently in one accretion regime, then switches, and spends the rest of its time in the other accretion regime, for example. The same holds for the other nine objects in the same class as SXP 18.3, as evidenced by Figures \ref{Fig:RepSource4.78}, \ref{Fig:RepSource11.9}, \ref{Fig:RepSource6.85}, \ref{Fig:RepSource11.5}, \ref{Fig:RepSource152}, \ref{Fig:RepSource264}--\ref{Fig:RepSource293}, and \ref{Fig:RepSource138}. Third, we observe that SXP 18.3 accretes in the weak propeller regime near MJD 53230, between MJD 54035 and MJD 54180, and near MJD 54540 and MJD 54880. Excursions of $\omega(t)$ into the weak propeller regime account for $\approx 3\%$ of the total observation time. For the remaining $\approx 1\%$, the fastness parameter satisfies $1.25 \lesssim \omega(t) \lesssim 1.45$ (not plotted). In short, the top panel of Figure \ref{Fig:RepSource18.3} represents the first time-resolved observational evidence, that SXP 18.3 accretes in the Rayleigh-Taylor stable regime. It also confirms observationally the existence of the weak propeller regime, which has been predicted theoretically by many authors \citep{Ustyugova_2006,Dangelo_2010,Dangelo_2012,Lii_2014,Romanova_2014,Dangelo_2015,Dangelo_2017}. As a corollary, the results imply that the Kalman filter and nested sampler in Section \ref{Sec:KFAnalysis} measure $\omega(t_n)$ in a physically meaningful and self-consistent fashion, absent some unlikely coincidence.

The foregoing qualitative features of $\omega(t_n)$ for SXP 18.3 are shared with the following SMC objects: two  X-ray pulsars classified as spinning up, namely SXP 4.78 and SXP 11.9, visualized in Figures \ref{Fig:RepSource4.78} and \ref{Fig:RepSource11.9}, respectively; six X-ray pulsars classified as being near rotational equilibrium, namely SXP 6.85, SXP 11.5, SXP 152, SXP 264, SXP 292, and SXP 293 visualized in Figures \ref{Fig:RepSource6.85}, \ref{Fig:RepSource11.5}, \ref{Fig:RepSource152}, and \ref{Fig:RepSource264}--\ref{Fig:RepSource293}, respectively; and one X-ray pulsar classified as spinning down, namely SXP 138, visualized in Figure \ref{Fig:RepSource138}. It therefore appears that the sign of $\dot{P}$ does not control entry into the stable accretion regime, nor vice versa. Out of the 10 objects, SXP 152 fluctuates with $\omega(t_n) \gtrsim 0.6$ the longest, staying within the approximate boundaries of the stable regime for $\approx 99\%$ of the observation time, while SXP 293 stays within the stable regime for the shortest time, viz.\ $\approx 70\%$, and spends the rest of the time visiting sporadically the chaotic unstable regime. SXP 11.5 and SXP 11.9 fluctuate stochastically about $\omega(t) \approx 1$, accreting via the weak propeller regime $\omega(t_n) \gtrsim 1$ for $\approx 30\%$ of the time. The root-mean-square of $\omega(t_n)$ ranges from a minimum of 0.035 for SXP 293 to a maximum of 0.18 for SXP 6.85. 

\subsection{Ordered unstable regime}\label{SubSec:OrderedUnstable}

\begin{figure}
\centering{
    \includegraphics[width=0.95\textwidth, keepaspectratio]{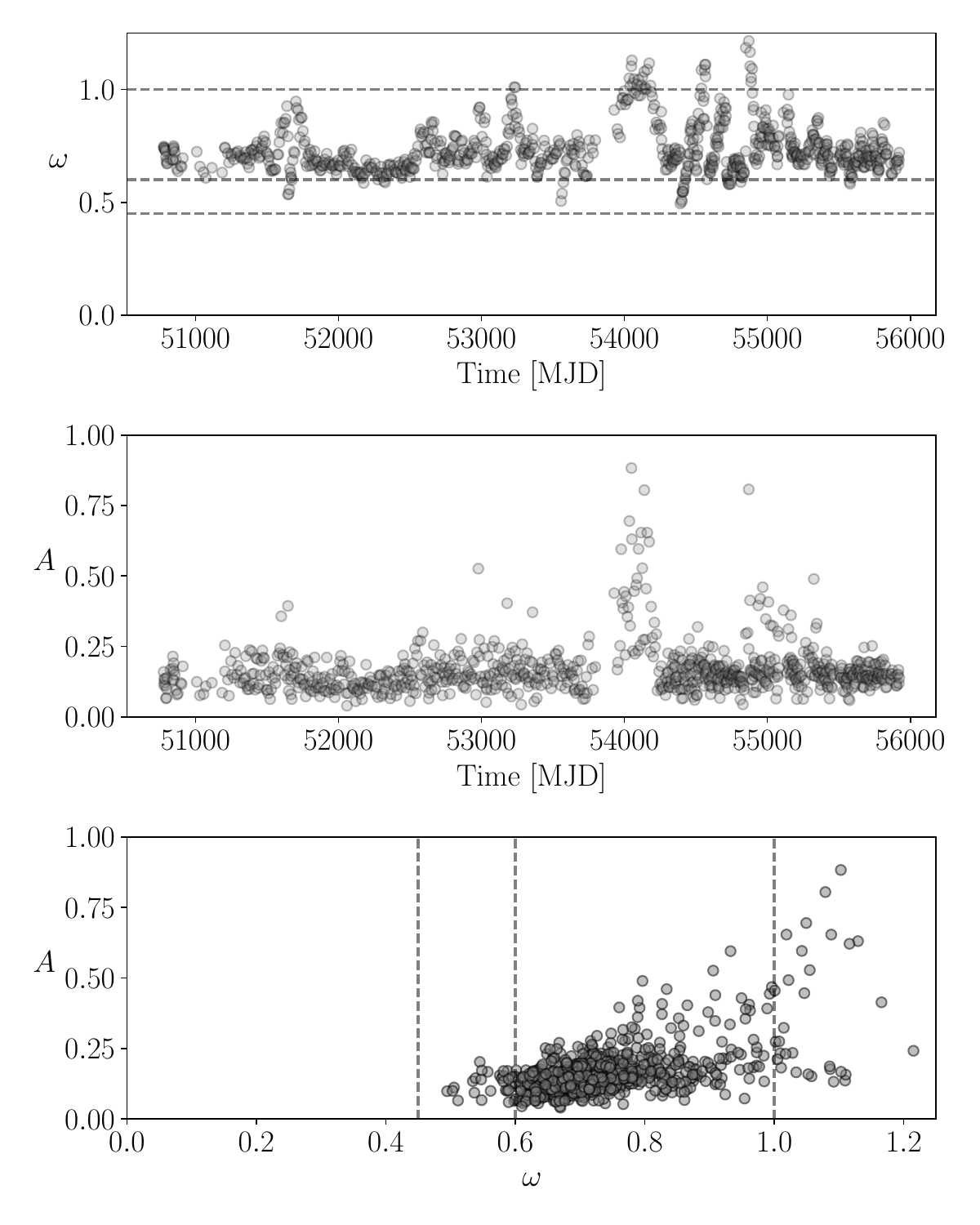}}
    \caption{Magnetocentrifugal accretion history of SXP 18.3, which is representative of the Rayleigh-Taylor stable accretion regime. (Top panel.) Time-resolved fastness parameter $\omega(t_n)$ versus $t_n$ (units: MJD), measured by the Kalman filter and nested sampler in Section \ref{Sec:KFAnalysis}. (Middle panel.) Fractional pulse amplitude $A(t_n)$ versus $t_n$, measured independently of the Kalman filter and nested sampler. (Bottom panel.) Scatter plot of $A(t_n)$ versus $\omega(t_n)$ from the middle and top panels respectively, included to help visualize correlations. The horizontal (top panel) and vertical (bottom panel), dashed, grey lines indicate the approximate boundaries of the ordered unstable [$\omega(t) \lesssim 0.45$], chaotic unstable [$0.45 \lesssim \omega(t) \lesssim 0.6$], and stable [$\omega(t) \gtrsim 0.6$] accretion regimes identified in numerical simulations \citep{Romanova_2008,Kulkarni_2008,Blinova_2016}.}
    \label{Fig:RepSource18.3}
\end{figure}

\begin{figure}
    \includegraphics[width=0.95\textwidth, keepaspectratio]{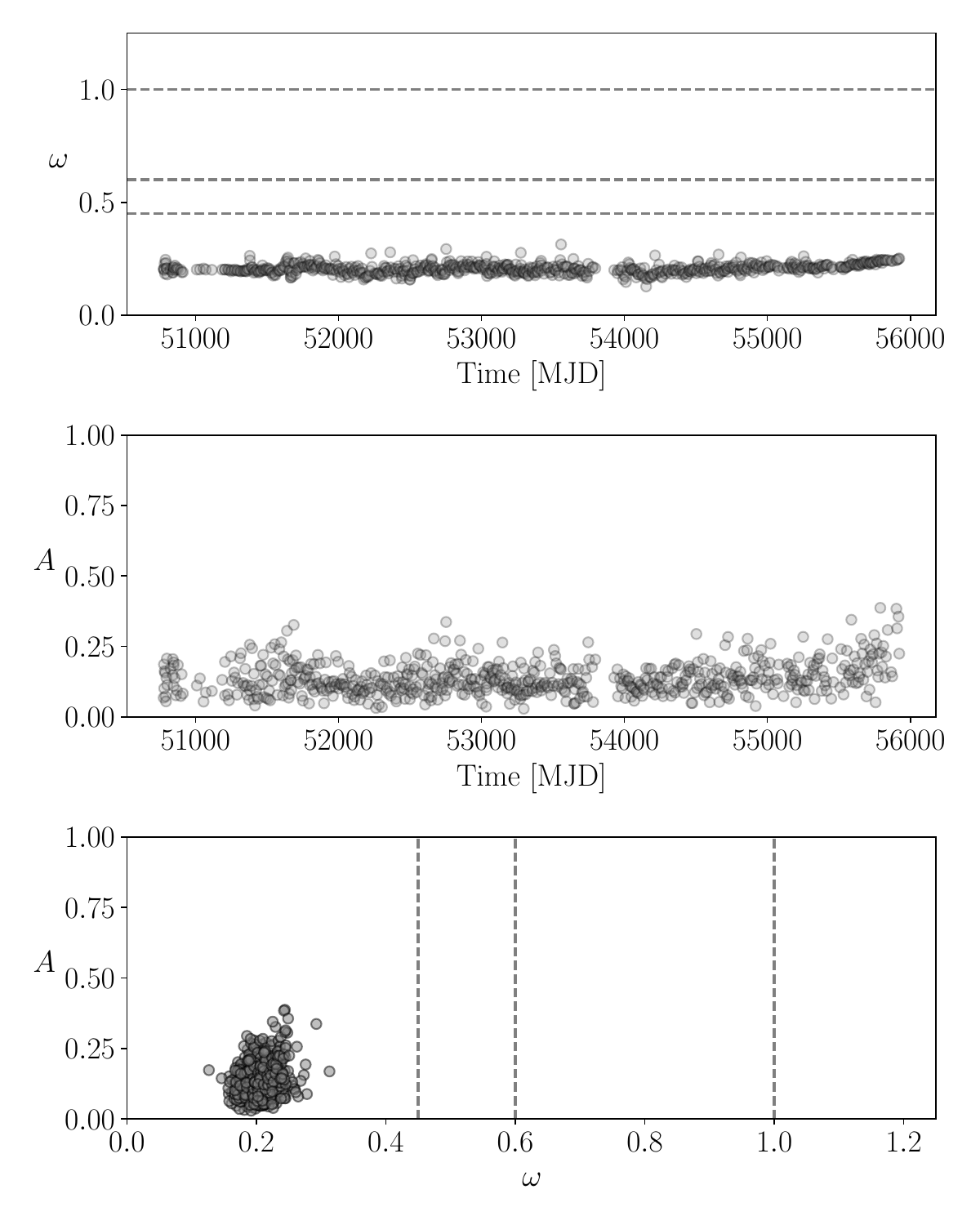}
    \caption{Same as Figure \ref{Fig:RepSource18.3} but for SXP 51.0, which is representative of the Rayleigh-Taylor ordered unstable regime. }
    \label{Fig:RepSource51}
\end{figure}

Now consider the top panel of Figure \ref{Fig:RepSource51}, corresponding to SXP 51.0.
The panel displays two key features. First, we observe that $\omega(t_n)$ remains in the ordered unstable accretion regime for the whole observation time. Other objects in this class do make sporadic excursions out of the ordered unstable regime, as quantified in the next paragraph. However, the excursions are typically rarer and shorter than those discussed in Section \ref{SubSec:Stable}. Second, we measure the fluctuation amplitude to be lower than that measured in the top panel of Figure \ref{Fig:RepSource18.3}. Quantitatively, the root-mean-square of $\omega(t_n)$ equals 0.021 for SXP 51.0, compared to 0.12 for SXP 18.3. The top panel of Figure \ref{Fig:RepSource51} represents the first time-resolved observational evidence, that SXP 51.0 accretes in the  Rayleigh-Taylor ordered unstable regime.

The foregoing qualitative features of $\omega(t_n)$ for SXP 51.0 are shared with the following SMC objects: four X-ray pulsars classified as spinning up, namely SXP 59.0, SXP 172, SXP 323, and SXP 756, visualized in Figures \ref{Fig:RepSource59.0}--\ref{Fig:RepSource756}, respectively; seven X-ray pulsars classified as being near rotational equilibrium, namely SXP 82.4, SXP 101, SXP 202A, SXP 214, SXP 523, SXP 565, and SXP 893, visualized in Figures \ref{Fig:RepSource82.4} \ref{Fig:RepSource101}, \ref{Fig:RepSource202}, \ref{Fig:RepSource214}, and \ref{Fig:RepSource523}--\ref{Fig:RepSource893}, respectively; and two X-ray pulsars classified as spinning down, namely SXP 8.88 and SXP 95.2, visualized in Figures \ref{Fig:RepSource8.80} and \ref{Fig:RepSource95.2}, respectively. Again, the sign of $\dot{P}$ does not appear to control or be controlled by the accretion regime. Out of the 14 objects in the ordered unstable regime, 10 stay within the approximate boundary $ \omega(t_n) \lesssim 0.45$ for the entire observation time. Their root-mean-square fastness fluctuations range from a minimum of 0.0057 for SXP 101 to a maximum of 0.019 for SXP 82.4; that is, they typically fluctuate less than objects in the stable regime. The other four objects, namely SXP 8.88, SXP 59.0, SXP 95.2, and SXP 202A stay in the ordered unstable regime for $\geq 98\%$ of the total observation time, punctuated by brief excursions ($\leq 2 \%$) into the chaotic unstable and stable regimes. Their root-mean-square $\omega(t_n)$ ranges from a minimum of $0.030$ for SXP 95.2 to a maximum of $0.071$ for SXP 59.0. 

As a matter of terminology, it may seem counterintuitive that $\omega(t_n)$ fluctuations are smaller in the ordered unstable regime than in the stable regime, as one tends intuitively to associate instabilities with higher levels of stochasticity. Physically, however, the result is not surprising at all. In the ordered unstable regime, we have $\omega(t) \lesssim 0.45$, and the magnetosphere is compact. The mechanical momentum flux in the accretion flow dominates the Maxwell stress well inside $R_{\rm c}(t)$, and the accreting material forces its way onto the stellar surface via well-defined funnels (see Section \ref{Sec:MagFunnels}), which cannot be perturbed easily by time-varying Maxwell stresses. In contrast, in the stable regime, we have $\omega(t) \gtrsim 0.60$, and the opposite is true: the magnetosphere is larger, the Maxwell stress is more important dynamically inside $R_{\rm c}(t)$, and the funnels bearing accreting material onto the stellar surface can be perturbed more easily. 

The reader may wonder, whether the results about $\omega(t)$ fluctuations are sensitive to the assumptions made about the process and measurement noises in the Kalman filter (see Section \ref{Sec:KFAnalysis}). Preliminary tests are presented in Appendix \ref{App:B}, which suggest provisionally that the sensitivity is low. A full battery of tests is postponed, until the Kalman filter analysis is repeated on more objects with higher $N$, to increase the statistical significance of the exercise.

We remind the reader that \cite{Blinova_2016} resolved the Rayleigh-Taylor stable [$ 0.6 \lesssim \omega(t) \lesssim 1$],  chaotic unstable [$0.45 \lesssim \omega(t) \lesssim 0.6$], and ordered unstable [$\omega(t) \lesssim 0.45$] accretion regimes for rotating, magnetized, compact stellar objects whose rotation and magnetic axes are misaligned by $\Theta \lesssim 5^\circ$. The foregoing boundaries are modified for $\Theta \gtrsim 5^\circ$. For example, preliminary analyses reveal that the boundary between stable and unstable accretion occurs near $\omega(t) \approx 0.54$ for $\Theta = 20^{\circ}$; see Figure 1 of \cite{Blinova_2016}. Similarly, for larger misalignment, e.g.\  $40^{\circ} \lesssim \Theta \lesssim 60^{\circ}$, only stable and chaotic unstable accretion regimes are observed in numerical simulations. This is not surprising: misalignment leads to complicated, nonaxisymmetric funnel flows, as revealed by three-dimensional numerical simulations \citep{Romanova_2003,Romanova_2004}. We refer the reader to Section 8 of \cite{Blinova_2016} for further details.

Recent polarimetric observations of accretion-powered pulsars using the Imaging X-ray Polarimetry Explorer \citep{Soffitta_2021} reveal that some systems, e.g. GRO J1008--57 \citep{Tsygankov_2023} and X Persei \citep{mushtukov_2023}, host nearly orthogonal rotators.  It is challenging to quantify at present how the boundary between stable and unstable Rayleigh-Taylor accretion regimes changes as a function of $\Theta$ without additional -- and computationally expensive -- global, three-dimensional, magnetohydrodynamic numerical simulations of inclined rotators with $5^\circ \lesssim \Theta \lesssim 90^\circ$. As a first step in this direction, however, we present the magnetospheric radius $R_{\rm m}$ and hence fastness $\omega$ as an approximate, phenomenological function of $\Theta$, i.e. $R_{\rm m} = R_{\rm m}(t, \Theta)$ and $\omega = \omega(t, \Theta)$, in Appendix \ref{App:Misaligment} to support future (refined) Kalman filter analysis of accretion-powered pulsars.

%

\begin{table}
\centering
\hspace{-1.5cm}
\begin{tabular}{c c}
\tableline
\tableline
Name (SXP) & $r[A(t), \omega(t)]$ \\
\tableline
    4.78 & 0.39 $\pm$ 0.032\\
    11.9$^*$ &  0.096 $\pm$ 0.050\\
    59.0 &  0.55 $\pm$ 0.028\\
    172$^*$ &  0.087 $\pm$ 0.034\\
    323$^*$ &  0.0014 $\pm$ 0.039\\
    756$^*$ &  0.037 $\pm$ 0.051\\
\tableline
    6.85 &  0.37 $\pm$ 0.038\\
    11.5 &  0.49 $\pm$ 0.036\\
    18.3 &  0.63 $\pm$ 0.027\\
    82.4 &  0.51 $\pm$ 0.029 \\
    101 &  0.31 $\pm$ 0.047\\
    152 &  0.22 $\pm$ 0.041\\
    202A$^*$ &  0.11 $\pm$ 0.042\\
    214 & 0.18 $\pm$ 0.034\\
    264 &  0.16 $\pm$ 0.039\\
    292 &  0.24 $\pm$ 0.038\\
    293 &  0.38 $\pm$ 0.030\\
    523$^*$ &  0.029 $\pm$ 0.043\\
    565 &  0.24 $\pm$ 0.033\\
    893$^*$ & 0.080 $\pm$ 0.039\\
\tableline
    8.88 & 0.71 $\pm$ 0.024\\
    51.0 &  0.25 $\pm$ 0.038\\
    95.2$^*$ &  0.052 $\pm$ 0.034\\
    138 &  0.22 $\pm$ 0.032\\
\tableline
\end{tabular}
\caption{Pearson correlation coefficients relating the independently measured RXTE PCA pulse amplitude and the fastness parameter, $r[A(t) , \omega(t)]$, for the objects in Table \ref{Table:SourceProperties}. The uncertainty corresponds to the Pearson standard error $s_r = [(1-r^2)/(N-2)]^{1/2}$, which satisfies  $0.024 \leq s_r \leq 0.051$. The top, middle, and bottom sections contain X-ray pulsars classified in Table 3 in \cite{Yang_2017} as spinning up ($\epsilon \leq -1.5$), near rotational equilibrium ($-1.5  < \epsilon < 1.5$), and spinning down ($\epsilon \geq 1.5$), respectively. Objects whose names are marked with an asterisk satisfy $|r| \leq 3 s_r$, i.e.\ their correlation coefficients are not statistically different from zero.}\label{Table:TempCorrs}
\end{table}

\section{Magnetospheric accretion funnels}\label{Sec:MagFunnels}

It is important to cross-check the results in Section \ref{Sec:RTRegimes} through tests that are independent of the inference scheme in Section \ref{Sec:KFAnalysis}, which yields $\omega(t_n)$. Fortunately, at least one independent test exists, which can be performed readily with RXTE data stored by the High Energy Astrophysics Science Archive Research Center. Specifically, the test involves the dimensionless pulse amplitude $A(t_n)$, which is measured by folding the RXTE PCA light curves at the known fundamental frequencies of the X-ray sources detected in the RXTE PCA field of view, yielding one $A(t_n)$ sample per $P(t_n)$ sample independently of $\omega(t_n)$.   

Three-dimensional magnetohydrodynamic simulations predict that the three Rayleigh-Taylor accretion regimes discussed in Section \ref{Sec:RTRegimes} are accompanied by distinct accretion flows within the magnetosphere.  In the stable regime, accretion occurs via two funnel streams, which deposit material at antipodal hotspots near the magnetic poles. In the chaotic unstable regime, several transient, stochastic, self-healing accretion tongues penetrate the magnetosphere and deposit material at random locations on the stellar surface. In the ordered unstable regime, the accretion tongues merge, forming one or two equatorial tongues. Spectral analysis of simulated light curves, generated synthetically from the simulation output \citep{Romanova_2003,Romanova_2004,Kulkarni_2008}, reveals that the funnel flows imprint observable signatures on the measured X-ray pulsations. Broadly speaking, $A(t)$ decreases, as the number of funnels and the complexity of their chaotic motion increase \citep{Romanova_2008,Kulkarni_2009}. This makes sense physically; one or two small, well-defined hot spots modulate the X-ray light more strongly, as the star rotates, than a larger number of hot spots which cover a significant and changing fraction of the stellar surface. One therefore predicts that, on balance, $A(t)$ should be higher for objects accreting in the stable regime (Section \ref{SubSec:Stable}) than for objects in the ordered unstable regime (Section \ref{SubSec:OrderedUnstable}), after allowing for the system's intrinsic randomness and other important, unknown, control variables such as the inclination angle between the rotation and magnetic axes \citep{Romanova_2008,Kulkarni_2009,Blinova_2016}. We test this prediction in Sections \ref{SubSec:TimePA} and \ref{SubSec:CrossCorr}.

\subsection{Time-resolved pulse amplitude $A(t)$}\label{SubSec:TimePA}
We start by plotting $A(t_n)$ versus $t_n$ in the middle panels of Figures \ref{Fig:RepSource18.3} and \ref{Fig:RepSource51} for the representative objects SXP 18.3 and SXP 51.0 respectively and in the middle panels of Figures \ref{Fig:RepSource4.78}--\ref{Fig:RepSource138} in Appendix \ref{App:A} for the other 22 objects. The reader is referred to Section 2.4.1 of \cite{Yang_2017} for a step-by-step guide on how to estimate pulse amplitude from RXTE light curves, as well as to \cite{Bildsten_1997} for a summary of related but different X-ray pulsar frequency estimation techniques, e.g.\ via fits to pulse-phase measurements, and to \cite{VanderPlas_2018} for a detailed discussion about practical considerations when processing unevenly sampled data in the Fourier domain. 

Visual inspection of the top and middle panels of Figure \ref{Fig:RepSource18.3} for SXP 18.3 reveals the following features. We observe two brief spikes in $A(t_n)$ near MJD 51650 and MJD 53230, accompanied by temporally correlated $\omega(t_n)$ fluctuations. Near MJD 53560, we also observe a sustained departure ($\gtrsim 1 \, \rm{year}$) of $A(t_n)$ from its time-averaged value,  accompanied by a slow transition of $\omega(t_n)$ from the chaotic unstable regime to the boundary of the weak propeller regime, followed by a decrease in $A(t_n)$ when $\omega(t_n)$ moves from the stable regime to the boundary of the chaotic unstable regime near MJD 54180. Such qualitative features are shared with other objects in the stable regime, namely SXP 4.78, SXP 6.85, and SXP 293. Overall, it is apparent visually that the time series in the top and middle panels of Figure \ref{Fig:RepSource18.3} are correlated temporally to some extent. The degree of correlation is quantified in Section \ref{SubSec:CrossCorr}  with the aid of the bottom panel in Figure \ref{Fig:RepSource18.3}.

It is more challenging to compare visually the fluctuations in the top and middle panels of Figure \ref{Fig:RepSource51} for SXP 51.0, because the fluctuations are weaker than for SXP 18.3. However, it is encouraging that the behaviors in the top and middle panels of Figure \ref{Fig:RepSource51} are consistent. For example, we do not observe strong fluctuations in one panel and not the other. The same behavior is shared with other objects in the ordered unstable regime, namely SXP 95.2, SXP 101, SXP 172, SXP 214, SXP 523, SXP 565, SXP 756, and SXP 893. Some objects in the ordered unstable regime, namely SXP 8.88, SXP 82.4, and SXP 59.0, do exhibit visually discernible spikes in $A(t_n)$ accompanied by temporally correlated spikes in $\omega(t_n)$. For example, between MJD 52500 and 52750 in the middle panel of Figure \ref{Fig:RepSource59.0} for SXP 59.0, we observe a spike in $A(t_n)$, accompanied by a transition of $\omega(t_n)$ from the ordered unstable regime, through the chaotic unstable regime, to the stable regime. A fuller analysis of $A(t_n)$-$\omega(t_n)$ correlations is done in Section \ref{SubSec:CrossCorr}.

\subsection{Cross-correlations between $A(t)$ and $\omega(t)$}\label{SubSec:CrossCorr}

We plot $A(t_n)$ versus $\omega(t_n)$ parametrically (i.e.\ as a scatter plot) in the bottom panels of Figures \ref{Fig:RepSource18.3} and \ref{Fig:RepSource51} for the representative sources SXP 18.3 and SXP 51.0 respectively and in the bottom panels of Figures \ref{Fig:RepSource4.78}--\ref{Fig:RepSource138} in Appendix \ref{App:A} for the other 22 objects. To guide the physical interpretation, we draw grey, dashed, vertical lines to divide the panel into the Rayleigh-Taylor ordered unstable $[\omega(t) \lesssim 0.45]$, chaotic unstable $[0.45 \lesssim \omega(t) \lesssim 0.6]$, and stable $[\omega(t) \gtrsim 0.6]$ accretion regimes identified in numerical simulations \citep{Romanova_2008,Kulkarni_2008,Blinova_2016}. 

Visual inspection of the bottom panel of Figure \ref{Fig:RepSource18.3} confirms the following features, also visible in the top and middle panels of the same figure.  We observe that the fastness parameter satisfies $0.6 \lesssim \omega(t) \lesssim 1$ for the majority of the observation time, punctuated by brief excursions into the Rayleigh-Taylor chaotic accretion regime $0.45 \lesssim \omega(t) \lesssim 0.6$, as well as the weak propeller regime $1 \lesssim \omega(t) \lesssim 1.25$. There is evidence of a moderately significant correlation between $A(t_n)$ and $\omega(t_n)$, visible as a diagonal tilt in the bottom panel of Figure \ref{Fig:RepSource18.3}, and reflected temporally in the top and middle panels of the same figure, as noted in Section \ref{SubSec:TimePA}. Similar behavior is shared with other objects in the stable accretion regime, namely SXP 4.78, SXP 6.85, SXP 11.5, and SXP 293. Visual inspection of the bottom panel of Figure \ref{Fig:RepSource51} for SXP 51.0 likewise reveals a moderately significant $A(t_n)$-$\omega(t_n)$ correlation, visible as a diagonal tilt. Other objects in the ordered unstable accretion regime share similar behavior, namely SXP 8.88, SXP 59.0, and SXP 82.4. 

In some objects it is challenging to discern visually the extent of the $A(t_n)$-$\omega(t_n)$ correlation. In Table \ref{Table:TempCorrs}, therefore, we report the Pearson correlation coefficients $r[A(t_n), \omega(t_n)]$ for the 24 objects analyzed in this paper. The reported uncertainties correspond to the standard errors $s_r$, which satisfy $0.024 \leq s_r \leq 0.051$. Table \ref{Table:TempCorrs} makes several important points. First, we measure $r[A(t_n),\omega(t_n)] > 0$ for all 24 objects. Satisfying the latter inequality in one or two objects could be deemed an accident, but it is significant statistically that there are no exceptions. Second, $r[A(t_n),\omega(t_n)]$ differs significantly from zero in 16 out of 24 objects at the level of (say) three standard deviations, the exceptions being SXP 11.9, SXP 95.2, SXP 172, SXP 202A,  SXP 323, SXP 523, SXP 756, and SXP 893 (identified with an asterisk in Table \ref{Table:TempCorrs}). Third, $r[A(t_n),\omega(t_n)]$ differs significantly from zero in multiple objects in the top, middle, and bottom sections of Table \ref{Table:TempCorrs}. There does not seem to be a dependence on the sign of $\dot{P}$ and hence $\epsilon$. For objects classified as spinning up (top section of Table \ref{Table:TempCorrs}), $r[A(t_n), \omega(t_n)]$ ranges from a minimum of $0.0014 \pm 0.039$ for SXP 323 to a maximum of $0.55 \pm 0.028$ for SXP 59.0. For objects classified as being near rotational equilibrium (middle panel of Table \ref{Table:TempCorrs}), $r[A(t_n), \omega(t_n)]$ ranges from a minimum of $0.029 \pm 0.043$ for SXP 523 to a maximum of $0.63 \pm 0.027$ for SXP 18.3. For objects classified as spinning down (bottom section of Table \ref{Table:TempCorrs}), $r[A(t_n), \omega(t_n)]$ ranges from a minimum of $0.052 \pm 0.034$ for SXP 95.2 to a maximum of $0.71 \pm 0.024$ for SXP 8.88.

\section{Conclusions}\label{Sec:Concl}

Global, three-dimensional numerical simulations of Rayleigh-Taylor instabilities at the disk-magnetosphere boundary of rotating, magnetized, compact objects with $R_{\rm m}(t)/R \lesssim 7$ reveal that accretion occurs in three regimes, namely the ordered unstable regime, the chaotic unstable regime, and the stable regime \citep{Romanova_2014,Romanova_2015,Blinova_2016}. The time-dependent fastness parameter $\omega(t)$ is the main factor governing what accretion regime applies at time $t$ \citep{Blinova_2016}.

In this paper, we apply the signal processing framework developed by \cite{Melatos_2022} to measure the time-resolved fastness parameter $\omega(t)$ of 24 accretion-powered pulsars in the SMC using pulse period $P(t)$ and aperiodic X-ray luminosity $L(t)$ time series measured by the RXTE PCA \citep{Yang_2017}. The framework is applied to six systems classified as spinning up, 14 systems classified as being near rotational equilibrium, and four systems classified as spinning down. The new results in Sections \ref{Sec:RTRegimes} and \ref{Sec:MagFunnels}, summarized and itemized below, connect $\omega(t)$ and the independently measured pulse amplitude time series $A(t)$ with the three Rayleigh-Taylor accretion regimes predicted by three-dimensional numerical simulations \citep{Romanova_2014,Romanova_2015,Blinova_2016}.

The main results of the Kalman filter analysis are itemized as follows.

(i) The 24 objects separate into two groups, classified according to their time-resolved fastness $\omega(t_n)$. The first group, comprising 10 stars, satisfies $0.6 \lesssim \omega(t_n) \lesssim 1$ for $\geq 70\%$ of the total observation time, corresponding to the Rayleigh-Taylor stable accretion regime. Out of the 10 objects, SXP 152 satisfies $\omega(t_n) \gtrsim 0.6$ the longest, staying within the approximate boundaries of the stable regime for $\approx 99\%$ of the observation time. The second group, comprising 14 stars, satisfies $\omega(t_n) \lesssim 0.45$ for $\geq 98\%$ of the total observation time, corresponding to the Rayleigh-Taylor ordered unstable accretion regime. Out of the 14 objects, 10 objects remain in the ordered unstable accretion regime for the whole observation time. 

(ii) None of the 24 objects spend sustained time intervals in the Rayleigh-Taylor chaotic unstable regime $0.45 \lesssim \omega(t_n) \lesssim 0.6$, with the exception of SXP 293, which satisfies $\omega(t_n)\lesssim 0.6$ for $\approx 30\%$ of the total observation time. The remaining 23 objects make occasional, sporadic excursions into the latter regime, with $0.45 \lesssim \omega(t_n) \lesssim 0.6$ for $\leq 3\%$ of the total observation time. 

(iii) The root-mean-square fastness fluctuations are larger typically in the Rayleigh-Taylor stable regime than in the Rayleigh-Taylor ordered unstable regime. The root-mean-square of $\omega(t_n)$ for the 10 objects in the stable regime ranges from a minimum of 0.035 for SXP 293 to a maximum of 0.18 for SXP 6.85. For the 14 objects in the ordered unstable regime, it ranges from a minimum of 0.0057 for SXP 101 to a maximum of 0.071 for SXP 51.0.

(iv) As a bonus, the data confirm the existence of the weak propeller regime, predicted theoretically by many authors \citep{Ustyugova_2006,Dangelo_2010,Dangelo_2012,Lii_2014,Romanova_2014,Dangelo_2015,Dangelo_2017}. Among the 24 objects, 10 make sporadic excursions into the weak propeller regime, $1 \lesssim \omega(t) \lesssim 1.25$, with SXP 11.5 and SXP 11.9 spending the longest time ($\approx 30\%$ of the total observation) doing so. 

(v) The fastness $\omega(t_n)$ correlates temporally with the time-resolved pulse amplitude $A(t_n)$. For example, we observe a sustained departure ($\gtrsim 1 \, \rm{year}$) of $A(t_n)$ from its time-averaged value near MJD 53560 in the middle panel of Figure \ref{Fig:RepSource18.3} for SXP 18.3, accompanied by a transition of $\omega(t_n)$ from the chaotic unstable regime to the boundary of the weak propeller regime in the top panel of Figure \ref{Fig:RepSource18.3}, followed by a decrease in $A(t_n)$ when $\omega(t_n)$ moves from the stable regime to the boundary of the chaotic unstable regime near MJD 54180. Such qualitative features are shared with other objects in Table \ref{Table:SourceProperties}, e.g.\ SXP 4.78, SXP 6.85, and SXP 293.

(vi) We measure a positive cross-correlation between $A(t_n)$ and $\omega(t_n)$. That is, the Pearson correlation coefficients satisfy $r[A(t_n), \omega(t_n)]>0$ for all 24 objects in Table \ref{Table:SourceProperties}. The significance of the correlation exceeds three standard deviations for 16 out of the 24 objects, whose Pearson correlation coefficients range from a minimum of 0.16 $\pm$ 0.039 for SXP 264  to a maximum of 0.71 $\pm$ 0.024 for SXP 8.88. The Pearson correlation coefficients listed in Table \ref{Table:TempCorrs} are consistent with the different kinds of magnetospheric funnel flows predicted by magnetohydrodynamic simulations in the three Rayleigh-Taylor accretion regimes, as predicted by spectral analysis of simulated light curves from the simulation output \citep{Romanova_2008,Kulkarni_2009, Romanova_2015}. Specifically, the simulations predict, that $A(t)$ decreases, as the number of funnels and the complexity of their motion increase, consistent with the results in Section \ref{Sec:MagFunnels}. This consistency check carries special weight, because $A(t)$ is measured directly from RXTE PCA data and is therefore independent of the Kalman filter analysis which measures $\omega(t)$. 

The agreement between observations and theory established in Sections \ref{Sec:RTRegimes} and \ref{Sec:MagFunnels} is encouraging. However, it is vital to emphasize that the simulations referenced throughout this paper, which predict the existence of the three Rayleigh-Taylor accretion regimes and their $\omega(t)$ boundaries, focus mostly on compact magnetospheres with $R_{\rm m}(t) / R \lesssim 7$ and small misalignment angles $\Theta \leq 5^\circ$ \citep{Romanova_2014,Romanova_2015,Blinova_2016}. In contrast, the SMC X-ray pulsars analyzed in this paper have $R_{\rm m}(t) / R \gtrsim 10^2$. It is certainly plausible that the simulation results may extend to less compact magnetospheres, but one must be cautious, because it gets harder to resolve the relevant length and time scales and simulate realistic values of key transport coefficients [e.g.\ \cite{Blinova_2016} assumed a small viscosity parameter $\alpha = 0.02$ in their simulations], as the system size increases. Nevertheless, the result in item (ii) above --- that the 24 objects analyzed here do not spend sustained time intervals in the chaotic unstable regime --- is consistent with preliminary simulations with $R_{\rm m}(t) / R > 10$, which conclude the same thing \citep{Romanova_2014,Romanova_2015}. There are excellent prospects in the long term to leverage the Kalman filter framework in this paper to deepen the contact between theory and observations, as more data are analyzed, and new generations of simulations become available. 

In terms of future work in the near term, we mention briefly one illustrative possibility out of many. It is straightforward to apply the Kalman filter to more complicated accretion scenarios, such as torque reversals \citep{Bildsten_1997}. One can do this in several ways, e.g.\ by analyzing the secular intervals between reversals separately and comparing the static parameters ($\bar{Q}$, $\bar{S}$, $\bar{\eta}$, and so on) in one interval and the next, or by running the Kalman filter to straddle two secular intervals and infer the time-resolved history of $\omega(t_n)$ during the reversal itself. The results may shed light on the reversal observed in 4U 1626$-$67, where it is claimed that the system enters the Rayleigh-Taylor chaotic unstable accretion regime near MJD 54500 \citep{turkouglu_2017}. 

This research was supported by the Australian Research Council Centre of Excellence for Gravitational Wave Discovery, grant number CE170100004. NJO’N is the recipient of a Melbourne Research Scholarship. DMC acknowledges funding through the National Science Foundation Astronomy and astrophysics research grant 2109004.  DMC,
SB, and STGL acknowledge funding through the National Aeronautics and Space Administration Astrophysics Data Analysis Program grant NNX14-AF77G.

\appendix

\section{Sensitivity analysis}\label{App:B}
The unscented Kalman filter \citep{Julier_1997,Wan_2000,Wan_2001}, described in Appendix A of \cite{OLeary_2024b}, ingests as input the $P(t_n)$ and $L(t_n)$ uncertainties, quantified by the measurement noise covariance $\bm{\Sigma}_n$ at time $t_n$. In this appendix, we test provisionally the sensitivity of the Kalman filter to changes in $\bm{\Sigma}_n$ by modifying the $L(t_n)$ uncertainty and comparing the output with the results in the main body of the text. The numerical experiment follows the same procedure outlined briefly in Section \ref{Sec:KFAnalysis} above, and described in detail in Section 4.1 of \cite{OLeary_2024b}. A complete sensitivity analysis is postponed, until the signal processing framework in this paper is applied to more objects with higher $N$, to increase the statistical significance of the exercise. 

The measurement uncertainties in $\bm{\Sigma}_{n}$ are approximated in this paper as follows. For each $P(t_n)$ sample, \cite{Yang_2017} calculated the uncertainty independently using Equation (14) in \cite{Horne_1986}, which is based on the standard deviation of the frequency assuming Gaussian noise \citep{Kovacs_1981}. The aperiodic X-ray luminosity $L(t_n)$ is not measured directly by the RXTE PCA, so the uncertainty per $L(t_n)$ sample is unknown independently a priori. For each $L(t_n)$ sample, therefore, we approximate the uncertainty by the variance of the $L(t_n)$ time series $\sigma_{L}^2$ for each of the 24 objects analyzed in Sections \ref{Sec:RTRegimes} and \ref{Sec:MagFunnels}. That is, we assign a unique error bar to each $P(t_n)$ sample per object, calculated independently according to Equation (14) in \cite{Horne_1986}, but we assign the same $L(t_n)$ uncertainty for all $1\leq n \leq N$ in any particular object, calculated from the variance of $L(t_1),\hdots,L(t_N)$ and denoted by $\sigma^{2}_{L}$.

\begin{figure}[h]
    \centering
\includegraphics[width=1\textwidth, keepaspectratio]{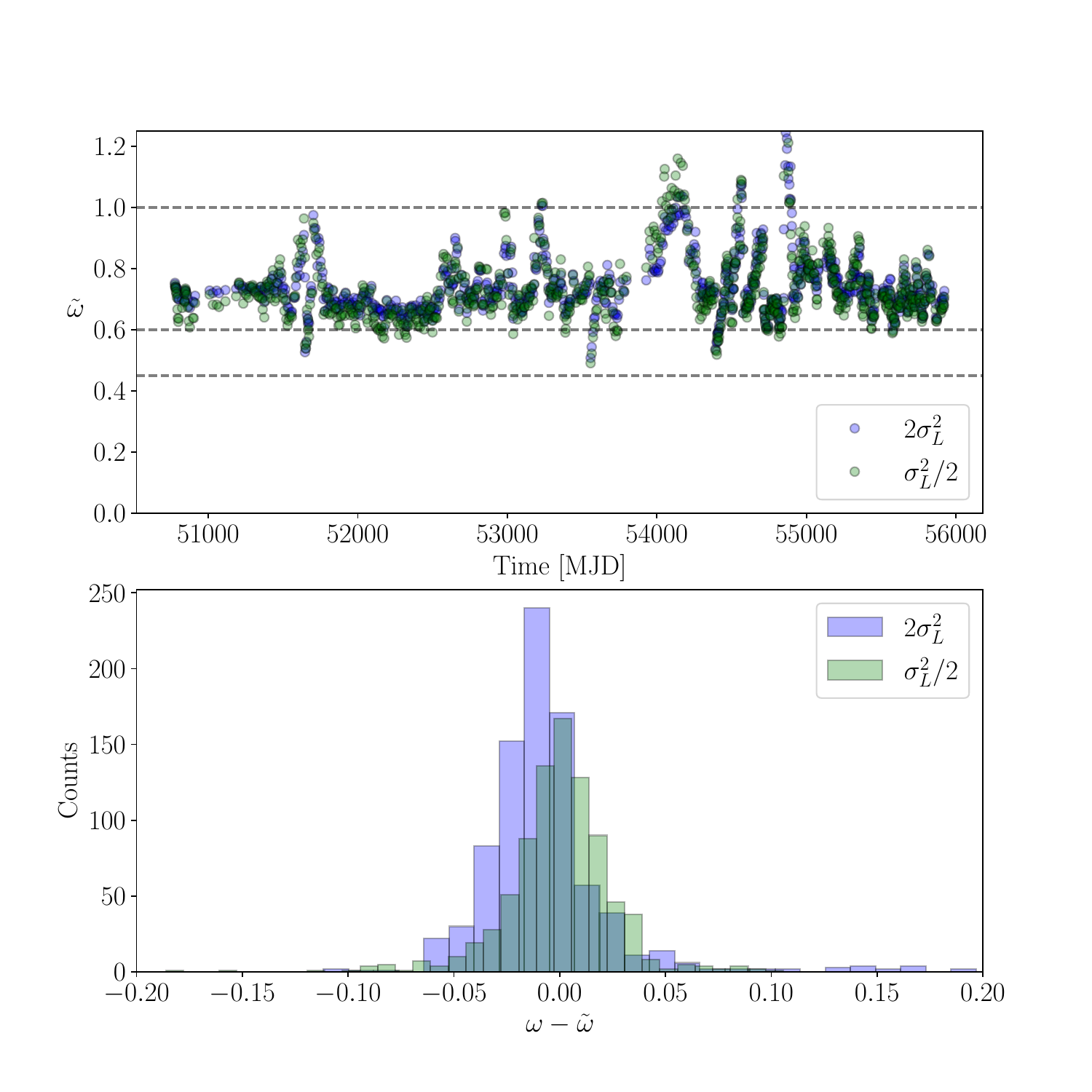}
    \caption{Sensitivity analysis of the Kalman filter output when accepting different $\bm{\Sigma}_n$ inputs. (Top panel.) Time-resolved fastness parameter $\tilde{\omega}(t_n)$ versus $t_n$ (units: MJD) for SXP 18.3, inferred by the Kalman filter using $2\sigma_{L}^2$ (blue points) and $\sigma_{L}^2/2$ (green points) to approximate the variance of the $L(t_n)$ uncertainty. The two fastness histories resemble one another qualitatively. The horizontal, dashed, grey lines indicate the approximate boundaries of the ordered unstable [$\omega(t) \lesssim 0.45$], chaotic unstable [$0.45 \lesssim \omega(t) \lesssim 0.6$], and stable [$\omega(t) \gtrsim 0.6$] accretion regimes identified in numerical simulations \citep{Romanova_2008,Kulkarni_2008,Blinova_2016}. (Bottom panel.) Histogram of fastness residuals $\omega(t_n) - \tilde{\omega}(t_n)$, inferred by the Kalman filter using $2\sigma_{L}^2$ (blue histogram) and $\sigma_{L}^2/2$ (green histogram) to approximate the variance of the $L(t_n)$ uncertainty. Here $\omega(t_n)$ denotes the $n$-th fastness sample from the top panel of the original analysis (with uncertainty variance $\sigma_L^2$) in Figure \ref{Fig:RepSource18.3}.}
    \label{Fig:FastnessCompare}
\end{figure}

How much do the results in Sections \ref{Sec:RTRegimes} and \ref{Sec:MagFunnels} change, if the uncertainty in $L(t_n)$ is approximated differently? Here we conduct a simple, provisional test by halving and doubling the variance above. That is, in the top panel of Figure \ref{Fig:FastnessCompare}, we plot the time-resolved fastness $\tilde{\omega}(t_n)$ as a function of $t_n$ for the representative object SXP 18.3, inferred using $2\sigma^2_{L}$ (blue points) and $\sigma_{L}^2/2$ (green points) to approximate the aperiodic X-ray luminosity measurement noise. Upon inspecting visually the top panels of Figures \ref{Fig:RepSource18.3} and \ref{Fig:FastnessCompare}, we observe that the three fastness histories share the same qualitative features and do not deviate significantly amongst themselves. For example, we observe a spike between MJD 53975 and MJD 54175 in the top panel of Figure \ref{Fig:RepSource18.3}, which also appears in the two fastness histories in the top panel of Figure \ref{Fig:FastnessCompare}. 

To get a preliminary sense of the quantitative difference between the fastness history $\omega(t_n)$ presented in Figure \ref{Fig:RepSource18.3} for SXP 18.3 
and the fastness histories $\tilde{\omega}(t_n)$ in the top panel of Figure \ref{Fig:FastnessCompare},
we plot the $\omega(t_n) - \tilde{\omega}(t_n)$ residuals as two histograms in the bottom panel of Figure \ref{Fig:FastnessCompare}. It is encouraging that both distributions are narrow and centered near zero, with $\approx 92\%$ (blue histogram) and $\approx 95\%$ (green histogram) of the distribution satisfying $|\omega(t_n) - \tilde{\omega}(t_n)| \lesssim 0.05$. 

The same numerical experiment is performed on the other 23 objects in Figures \ref{Fig:RepSource51}--\ref{Fig:RepSource138} (histograms not plotted for brevity). The conclusions regarding $\omega(t_n) - \tilde{\omega}(t_n)$ match those in the previous paragraph; the fastness histories are similar, when the variance of the $L(t_n)$ uncertainty equals $\sigma_L^2/2$, $\sigma_L^2$, and $2 \sigma_L^2$. Importantly, even when the Kalman filter accepts different $L(t_n)$ uncertainties as an input, the 24 objects separate into the same two distinct groups found in the main body of the text, with 10 objects accreting in the stable regime, and 14 objects accreting in the ordered unstable regime.

\section{Orbital modulation of $P(\lowercase{t})$}\label{App:PPVariations}
The pulse period time series $P(t)$ analyzed in Sections \ref{Sec:RTRegimes} and \ref{Sec:MagFunnels} are not corrected for orbital motion. Hence, the reader may wonder how sensitively the results about $\omega(t)$ depend on orbital motion. Broadly speaking, to account fully for orbital motion requires a complete reprocessing of the 36316 X-ray timing measurements collected by the RXTE PCA and analyzed by \cite{Yang_2017}, correcting the X-ray photon time-of-arrivals using (for example) the parabolic, cubic, or sinusoidal models reported in Table 3 of \cite{Singh_2002}, practical examples of which are given in \cite{LaPalombara_2016} for the SMC high-mass X-ray binary SXP 2.37 (also known as SMC X--2) and in \cite{Singh_2024} for the Galactic accretion-powered millisecond X-ray pulsar IGR J17591--2342. The output of the foregoing correction is a different set of time-ordered pulse period measurements $P_{\rm mod}(t)$ for each of the 24 objects analyzed in Sections \ref{Sec:RTRegimes} and \ref{Sec:MagFunnels}, whose associated Kalman filter analysis would yield different time-resolved estimates of $\Omega(t)$, $Q(t)$, $S(t)$, and hence $\omega(t)$.

Within the Kalman filter framework of Section \ref{Sec:KFAnalysis}, the relatively coarse sampling $t_{n+1}-t_n \gg P_{\rm b}$ means that the Doppler modulations are approximately uncorrelated at $t_{n+1}$ and $t_n$ and can be regarded as a form of measurement noise, which is absorbed into $N_P(t)$ in Equation (\ref{Eq:PulsePeriodNoise}). The Doppler measurement noise due to binary motion is comparable to or smaller than other sources of measurement noise. This implies that Kalman analyses of $P(t)$ and $P_{\rm mod}(t)$ yield similar results in terms of resolving stable, chaotic unstable, and ordered unstable Rayleigh-Taylor accretion regimes. As just one example, consider the timing and spectral analysis of the X-ray transient SXP 2.37 during its 2015 outburst in \cite{LaPalombara_2016}. The latter authors corrected the X-ray timing data of SXP 2.37 for orbital motion and found $|\bar{P}_{\rm mod} - \bar{P}| \sim 1 \times 10^{-4} \, \rm{s}$. In general, the expected difference for HMXBs is as large as $|\bar{P}_{\rm mod} - \bar{P}| \sim 1\times 10^{-3} \, \rm{s}$ \citep{Townsend_2011,Vasilopoulos_2017}. Such differences, i.e.\ $10^{-4} \lesssim |\bar{P}_{\rm mod} - \bar{P}|/(1 \, \rm{s}) \lesssim 10^{-3}$, are comparable to, and in some cases smaller than, the RXTE pulse period measurement uncertainties $\sigma_{P} \lesssim 10^{-3} \, \rm{s}$ for many of the 24 SMC X-ray pulsars analyzed in the present paper (see Table \ref{Table:SourceProperties}) and can be absorbed in $N_P(t)$. The unscented Kalman filter employed in the present paper is adept at handling uncertainties in the measurement as well as dynamical processes, encoded in the (time-dependent) noise covariances, details of which can be found in Appendix A of \cite{OLeary_2024b}.

\section{Kalman filter output}\label{App:A}
\begin{figure}[h]
\centering
\includegraphics[width=0.45\textwidth, keepaspectratio]{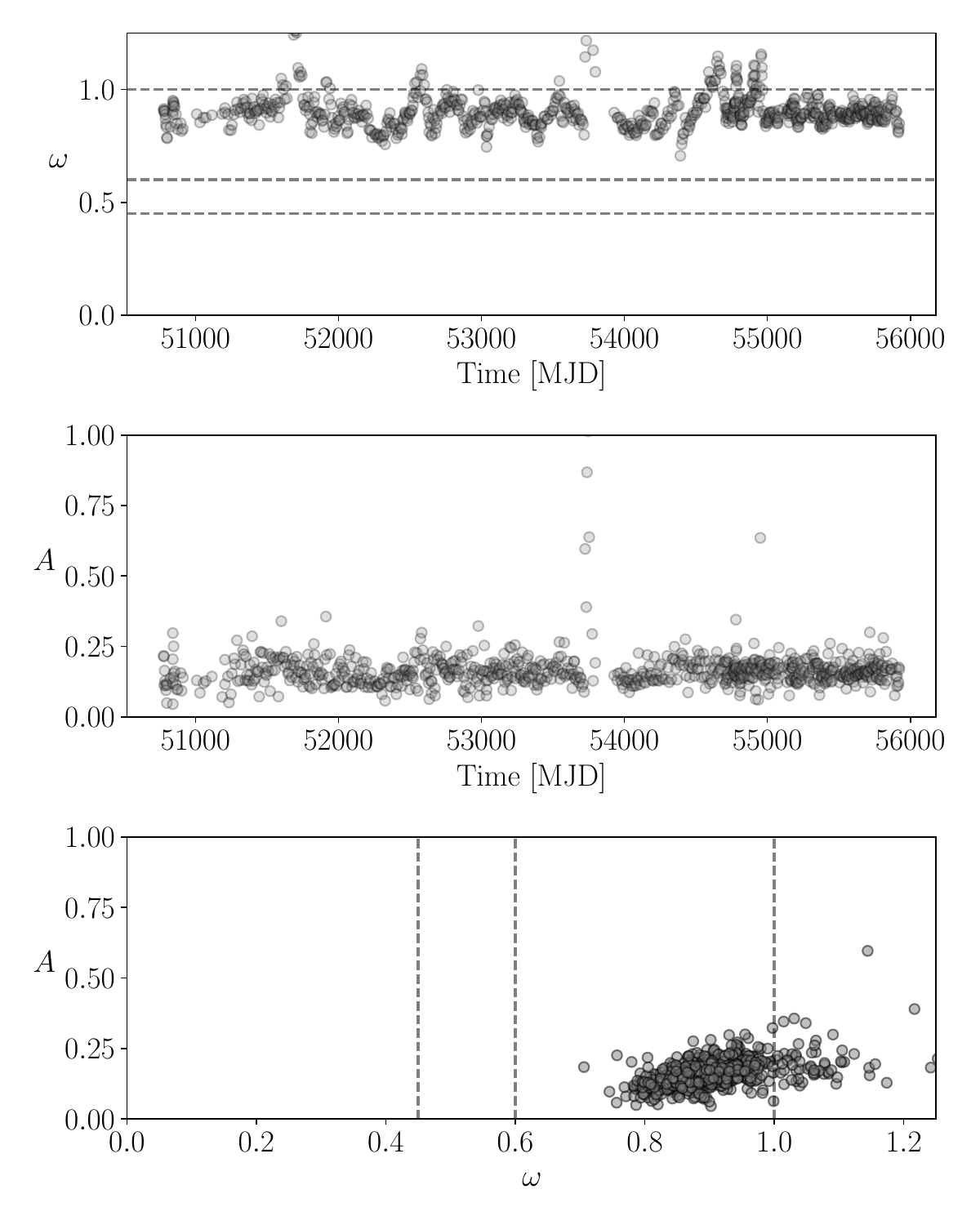}
    \caption{Same as Figure \ref{Fig:RepSource18.3} but for SXP 4.78.}
    \label{Fig:RepSource4.78}
\end{figure}

\begin{figure}[h]
\centering
\includegraphics[width=0.45\textwidth, keepaspectratio]{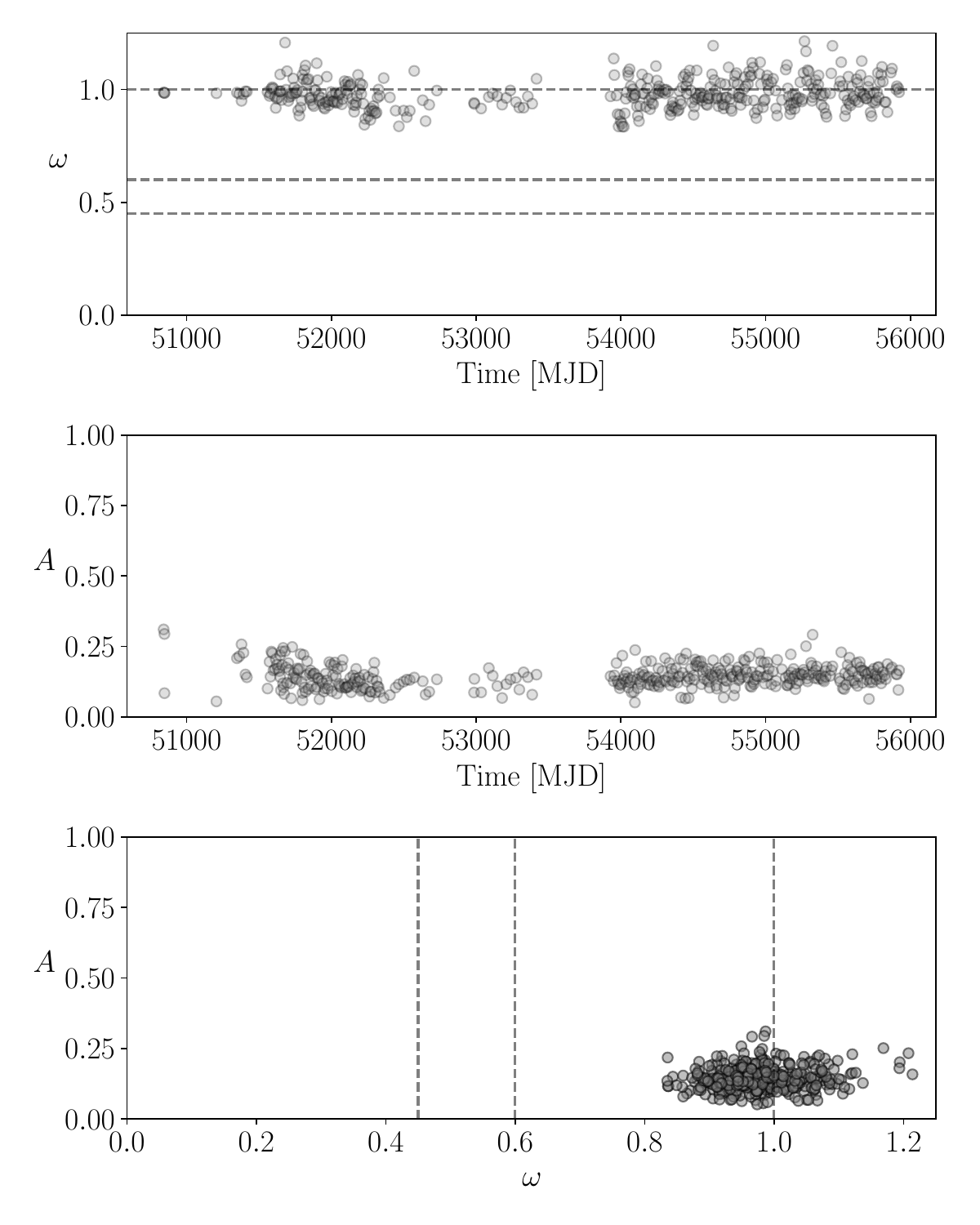}
    \caption{Same as Figure \ref{Fig:RepSource18.3} but for SXP 11.9.}
    \label{Fig:RepSource11.9}
\end{figure}

\begin{figure}[h]
\centering
\includegraphics[width=0.45\textwidth, keepaspectratio]{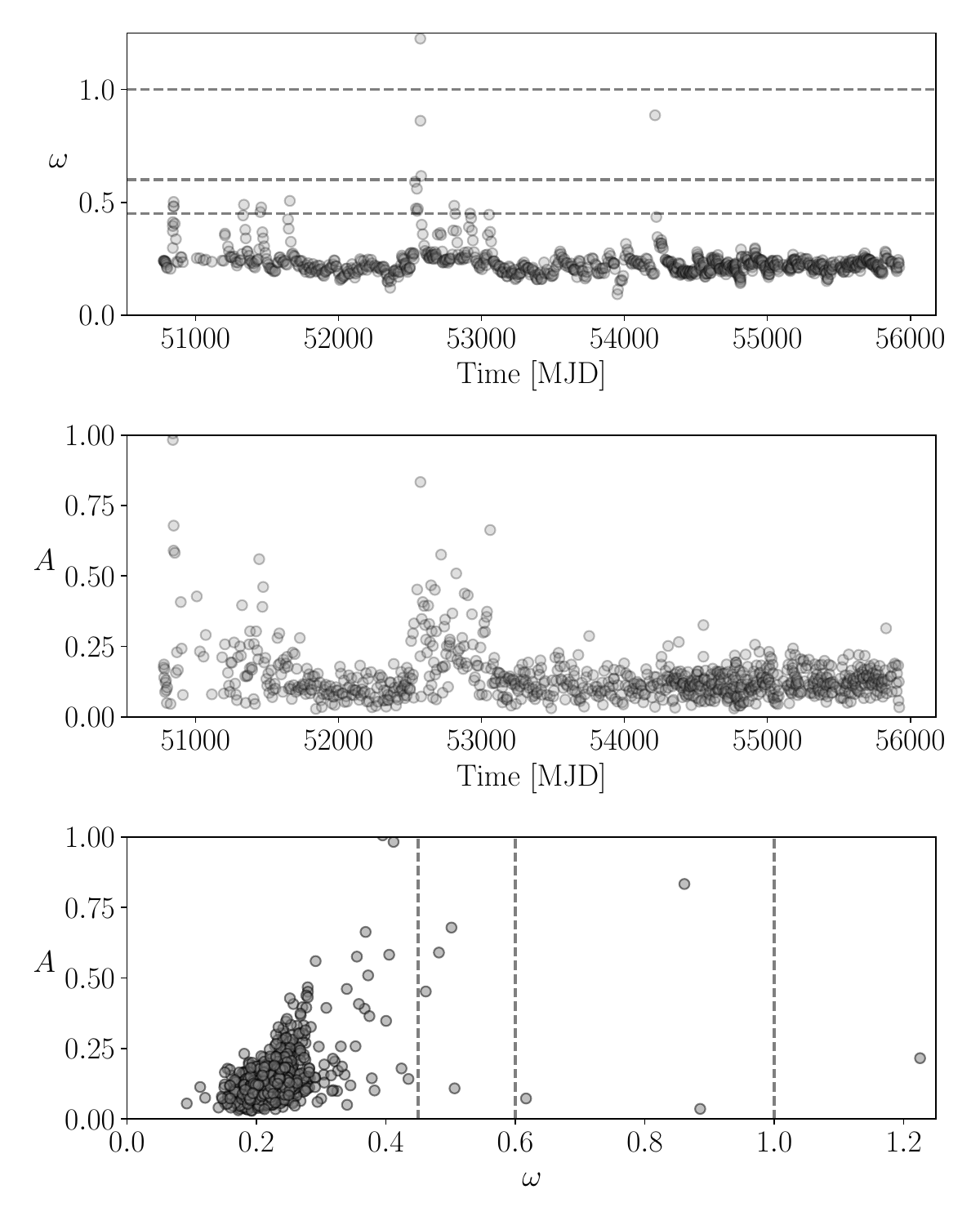}
    \caption{Same as Figure \ref{Fig:RepSource18.3} but for SXP 59.0.}
    \label{Fig:RepSource59.0}
\end{figure}

\begin{figure}[h]
\centering
\includegraphics[width=0.45\textwidth, keepaspectratio]{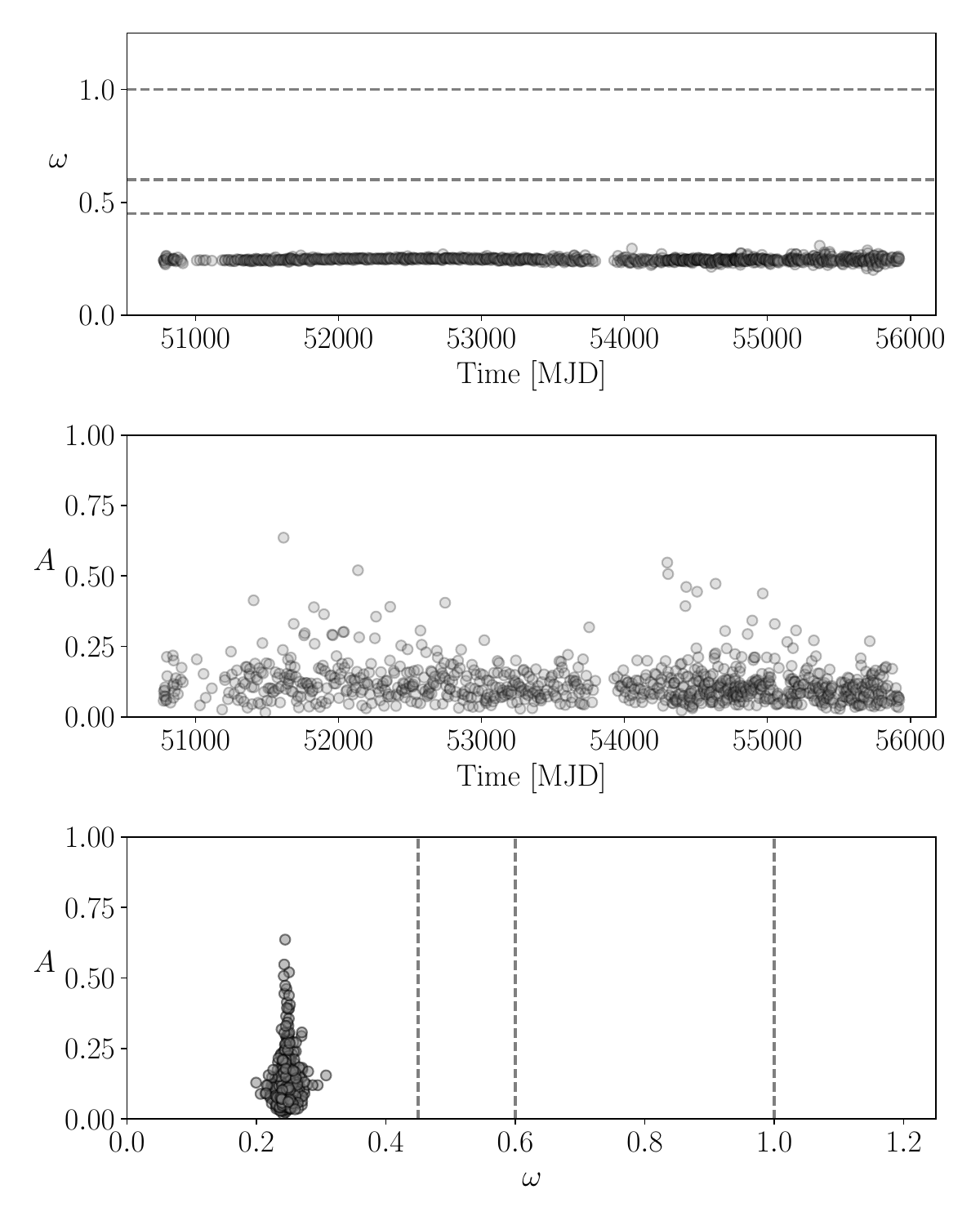}
    \caption{Same as Figure \ref{Fig:RepSource18.3} but for SXP 172.}
    \label{Fig:RepSource172}
\end{figure}

\begin{figure}[h]
\centering
\includegraphics[width=0.45\textwidth, keepaspectratio]{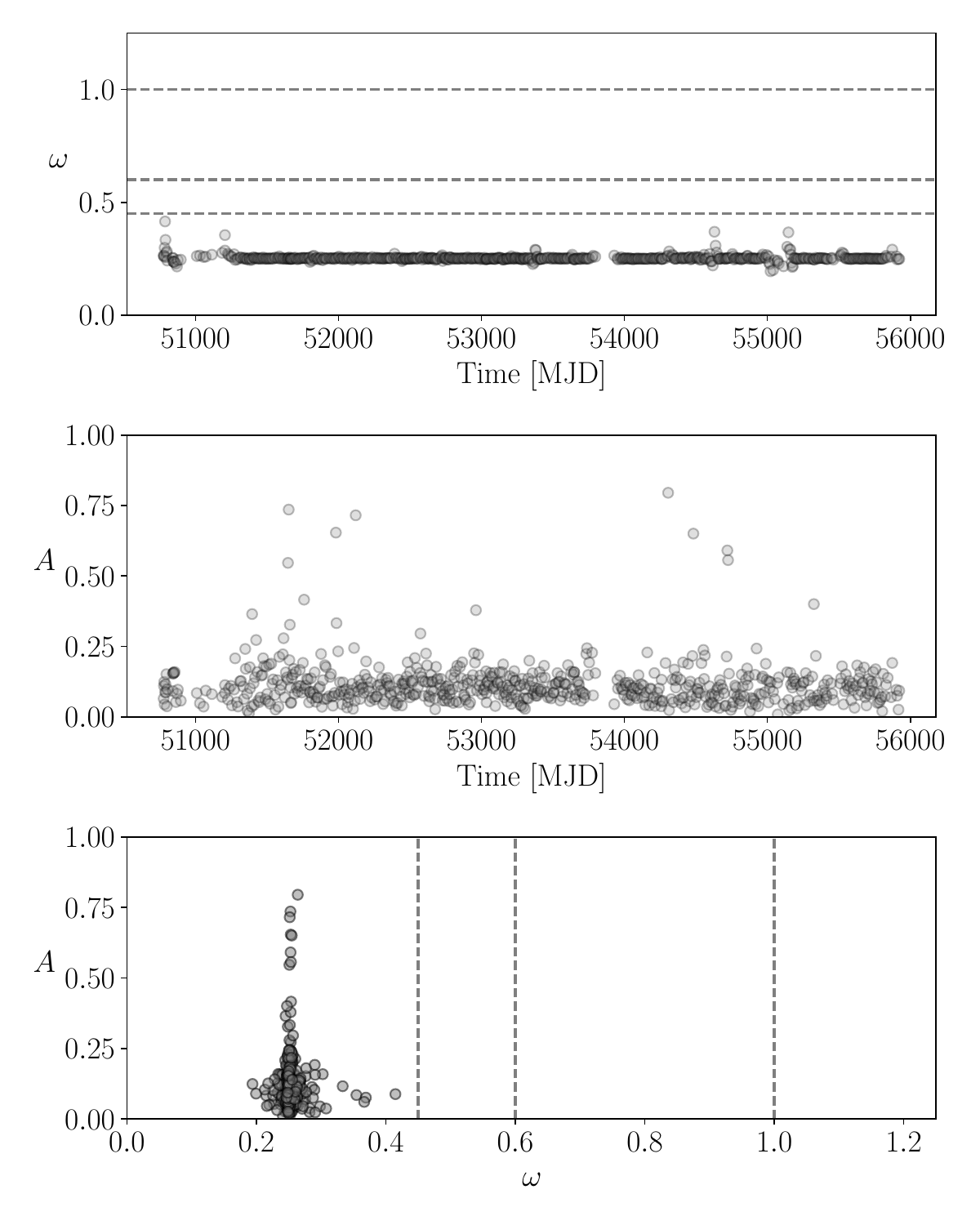}
    \caption{Same as Figure \ref{Fig:RepSource18.3} but for SXP 323.}
    \label{Fig:RepSource323}
\end{figure}

\begin{figure}[h]
\centering
\includegraphics[width=0.45\textwidth, keepaspectratio]{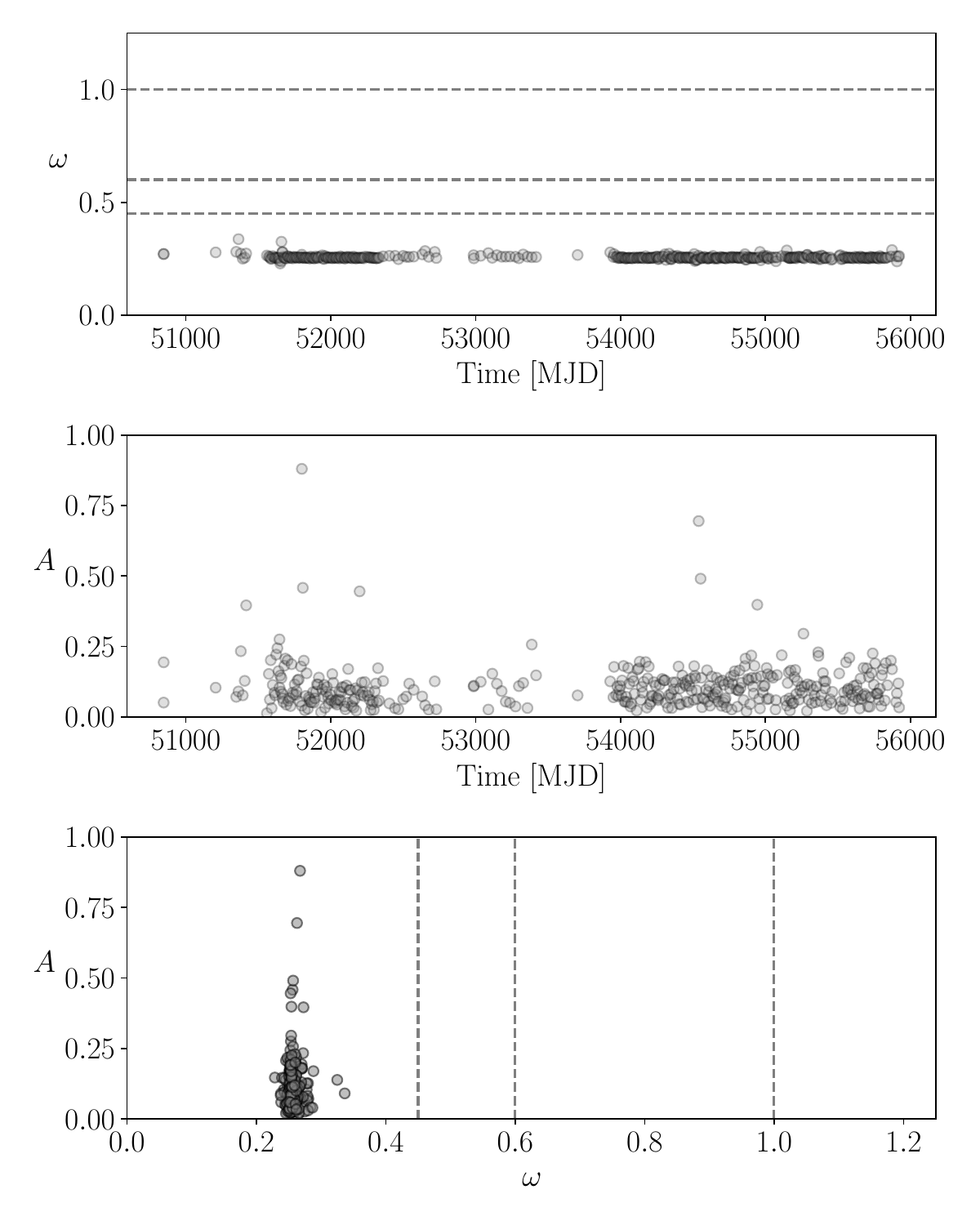}
    \caption{Same as Figure \ref{Fig:RepSource18.3} but for SXP 756.}
    \label{Fig:RepSource756}
\end{figure}

\begin{figure}[h]
\centering \includegraphics[width=0.45\textwidth, keepaspectratio]{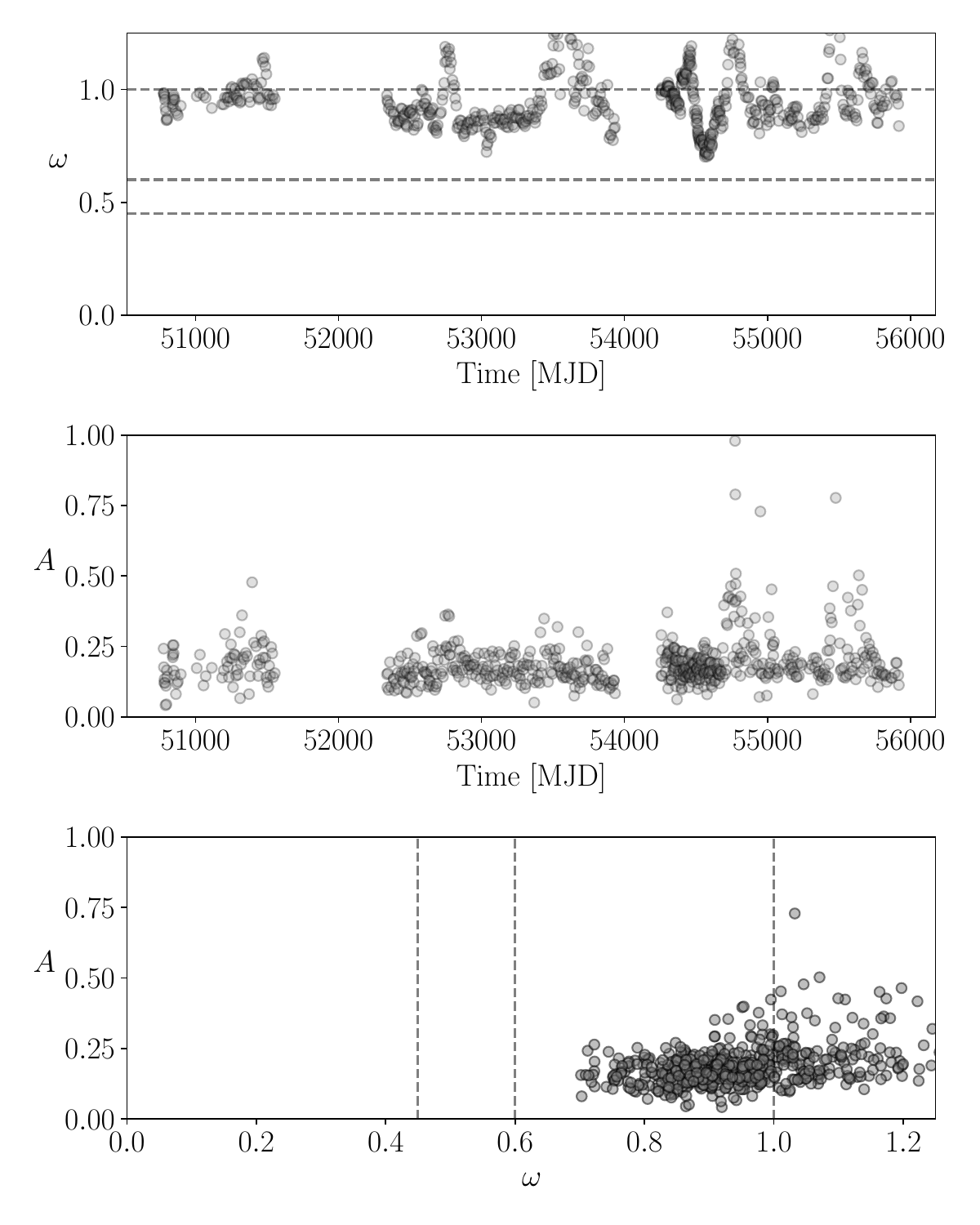}
    \caption{Same as Figure \ref{Fig:RepSource18.3} but for SXP 6.85.}
    \label{Fig:RepSource6.85}
\end{figure}

\begin{figure}[h]
\centering
\includegraphics[width=0.45\textwidth, keepaspectratio]{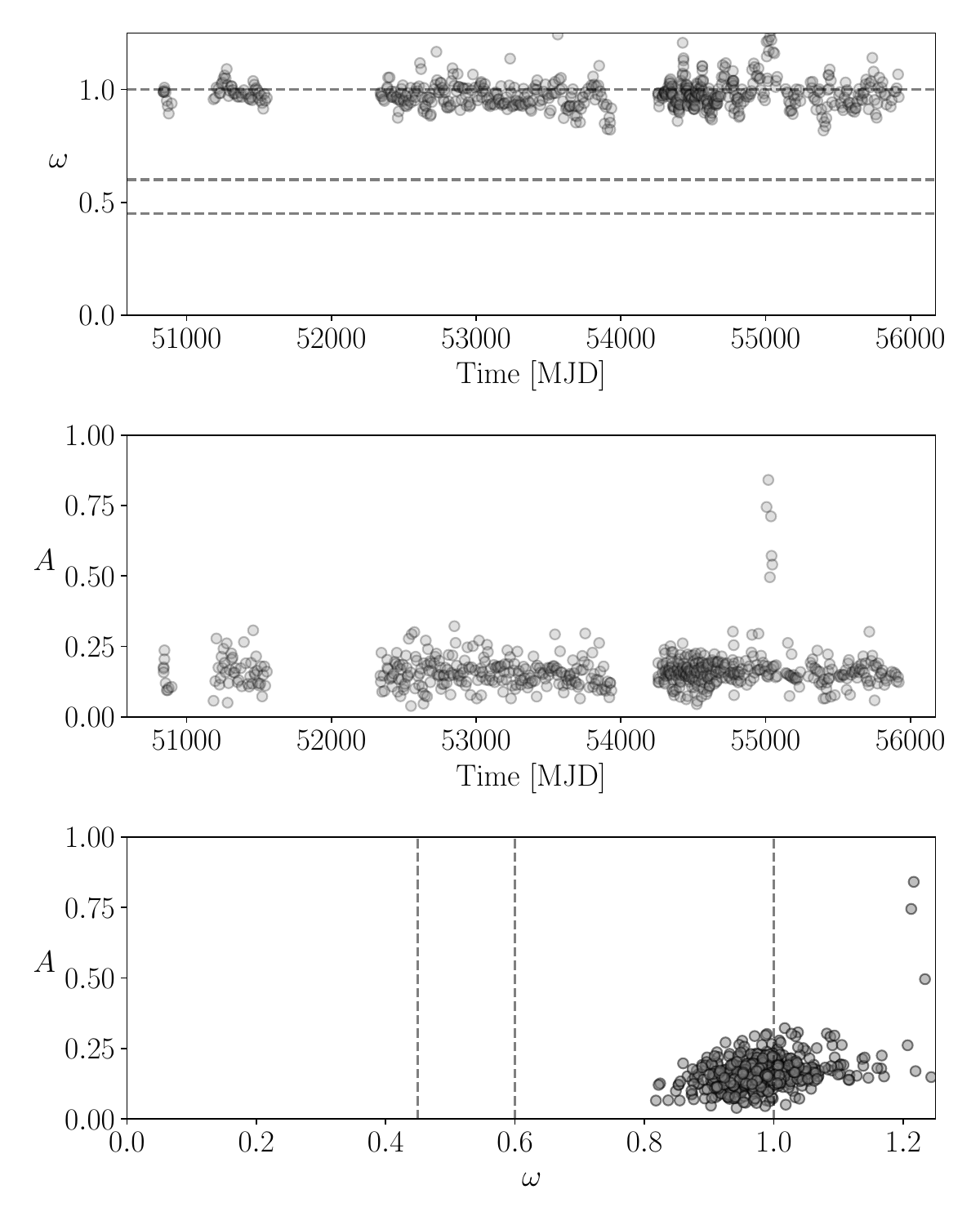}
    \caption{Same as Figure \ref{Fig:RepSource18.3} but for SXP 11.5.}
    \label{Fig:RepSource11.5}
\end{figure}

\begin{figure}[h]
\centering
\includegraphics[width=0.45\textwidth, keepaspectratio]{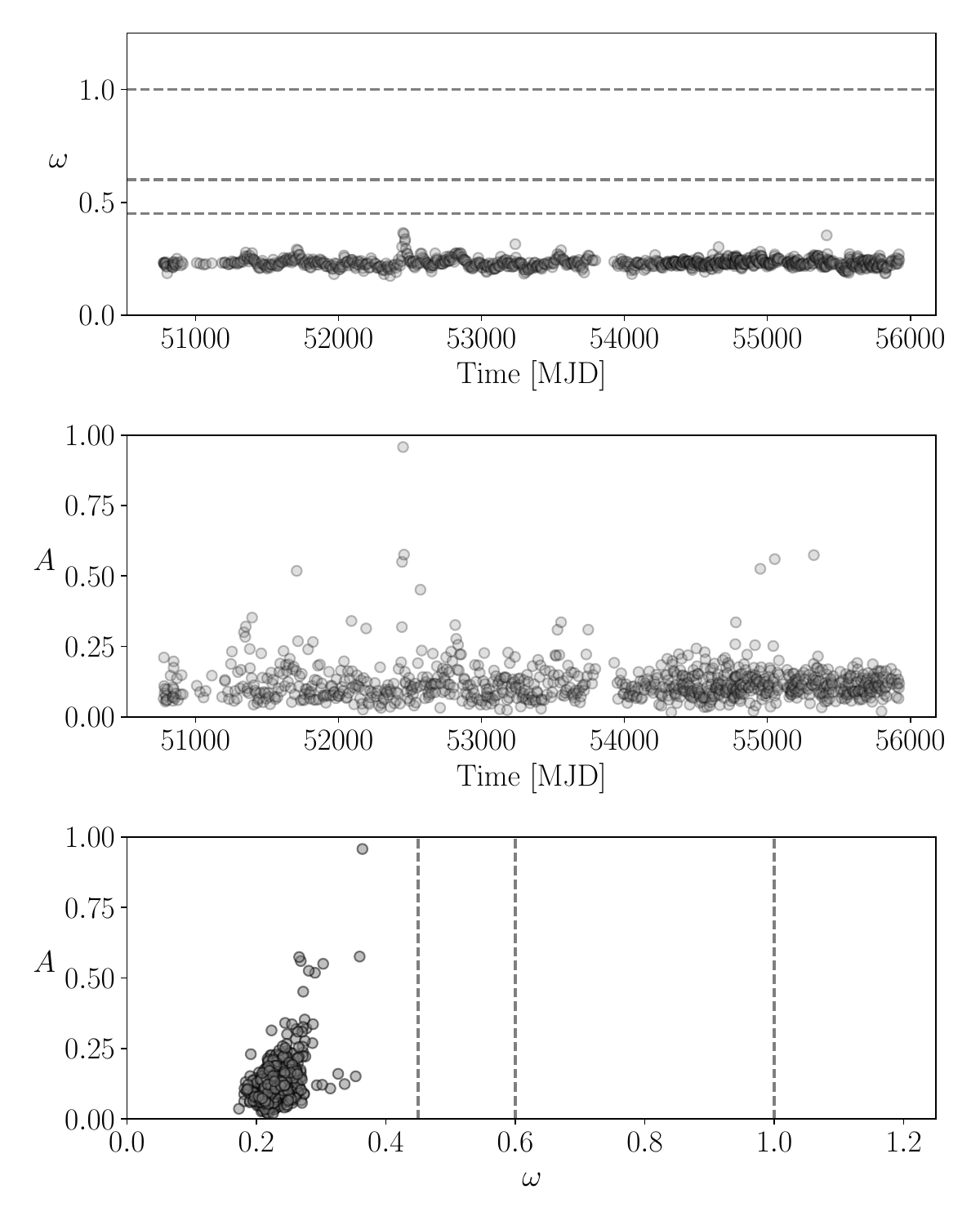}
    \caption{Same as Figure \ref{Fig:RepSource18.3} but for SXP 82.4.}
    \label{Fig:RepSource82.4}
\end{figure}

\begin{figure}[h]
\centering
\includegraphics[width=0.45\textwidth, keepaspectratio]{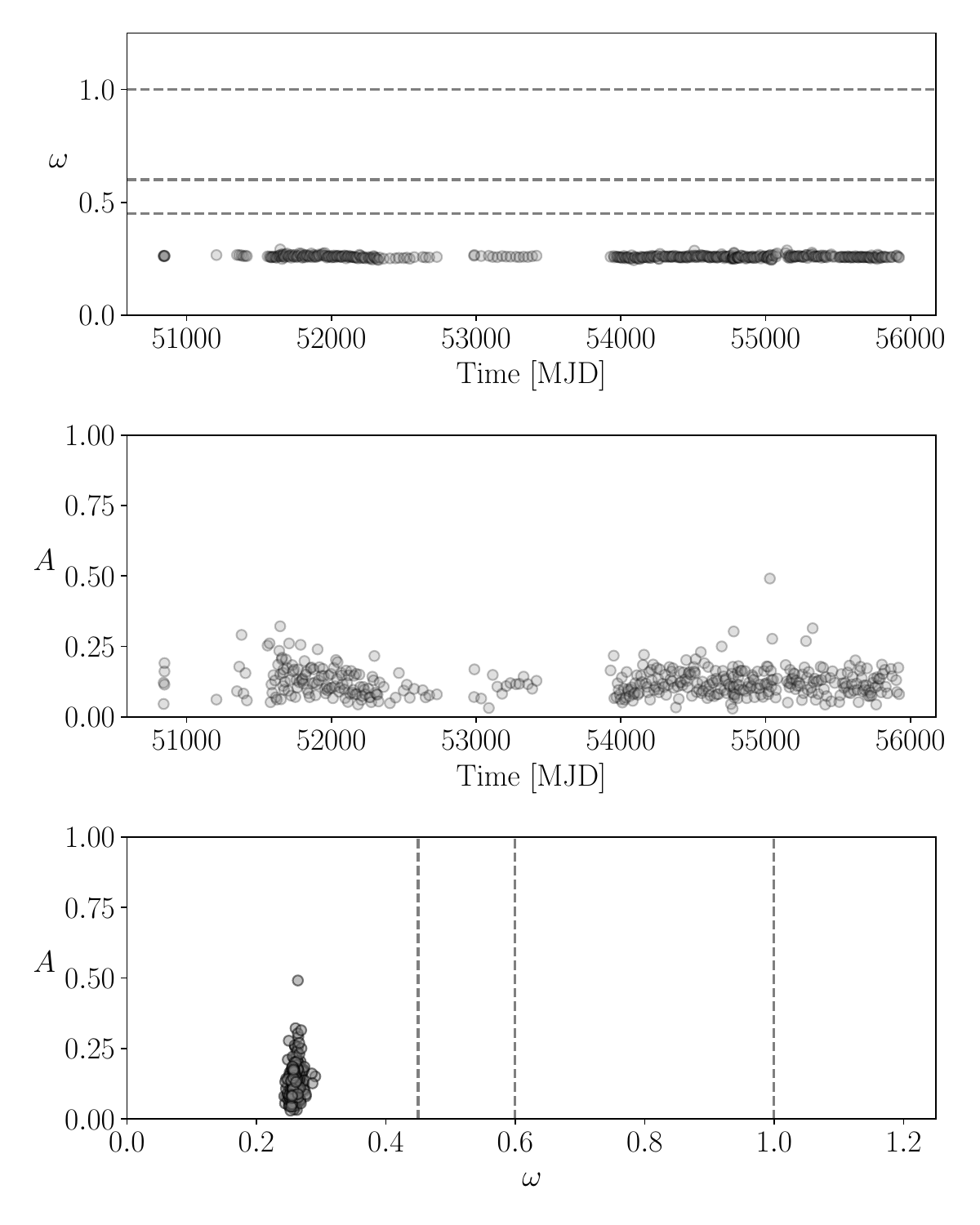}
    \caption{Same as Figure \ref{Fig:RepSource18.3} but for SXP 101.}
    \label{Fig:RepSource101}
\end{figure}

\begin{figure}[h]
\centering
\includegraphics[width=0.45\textwidth, keepaspectratio]{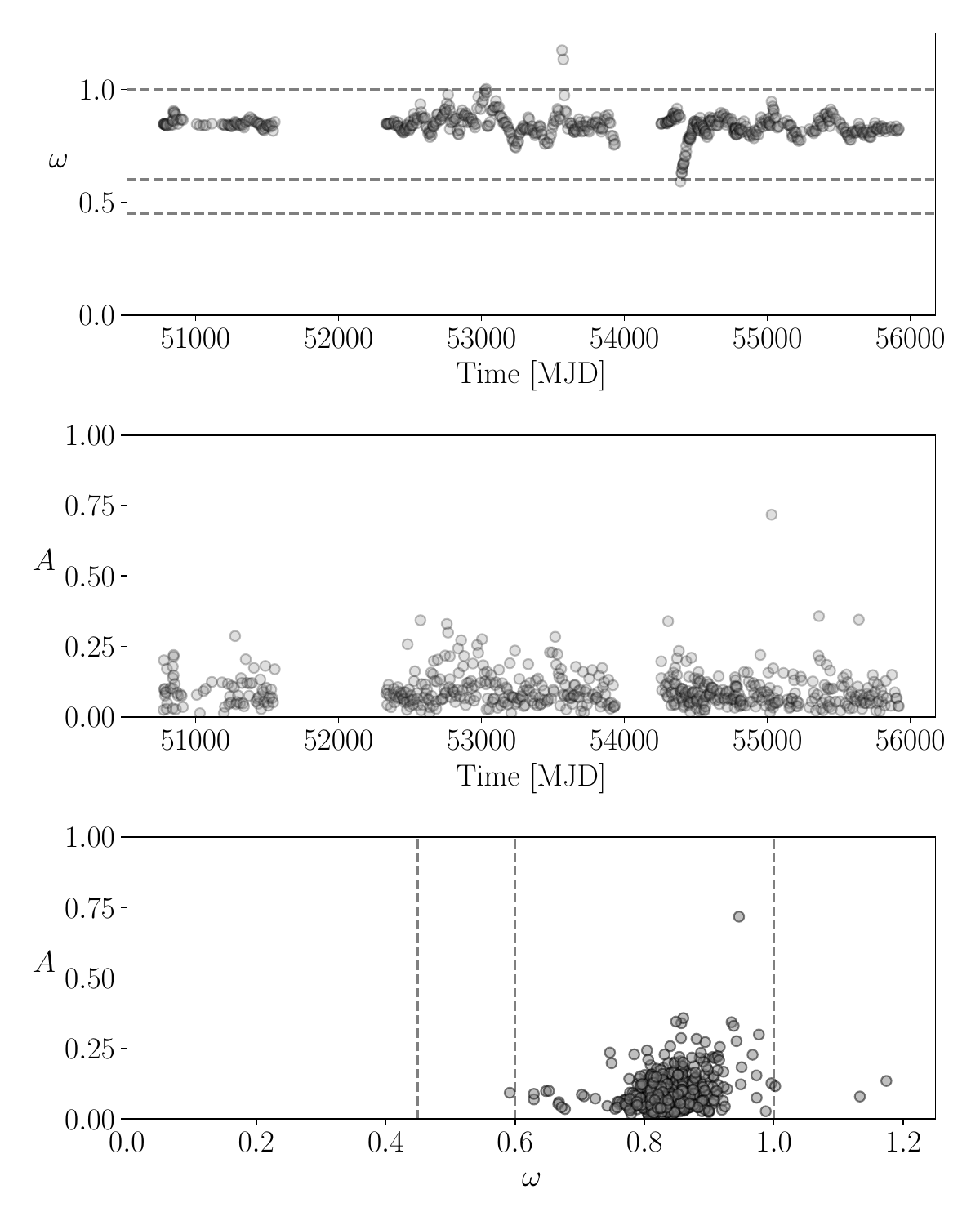}
    \caption{Same as Figure \ref{Fig:RepSource18.3} but for SXP 152.}
    \label{Fig:RepSource152}
\end{figure}

\begin{figure}[h]
\centering
\includegraphics[width=0.45\textwidth, keepaspectratio]{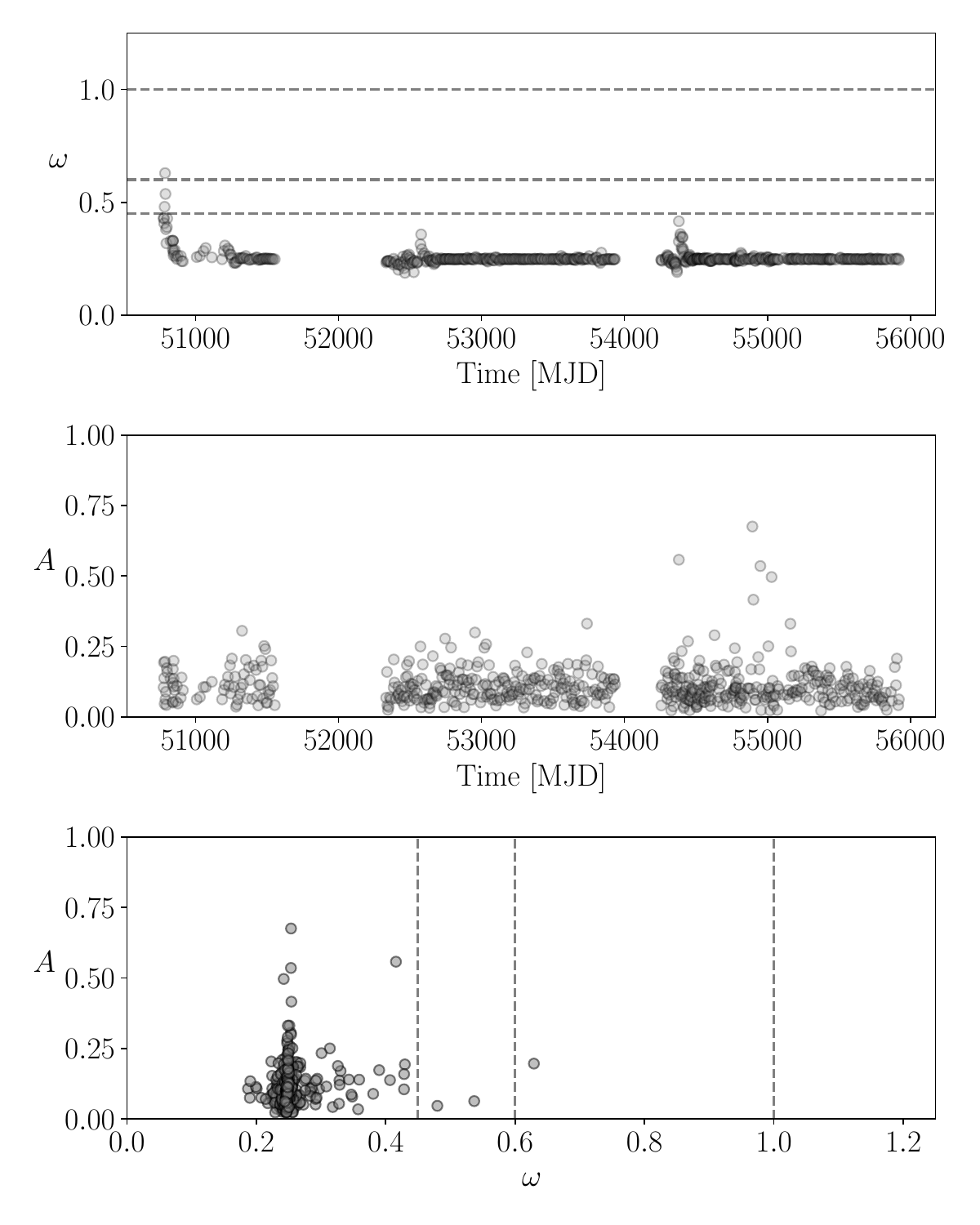}
    \caption{Same as Figure \ref{Fig:RepSource18.3} but for SXP 202A.}
    \label{Fig:RepSource202}
\end{figure}

\begin{figure}[h]
\centering
\includegraphics[width=0.45\textwidth, keepaspectratio]{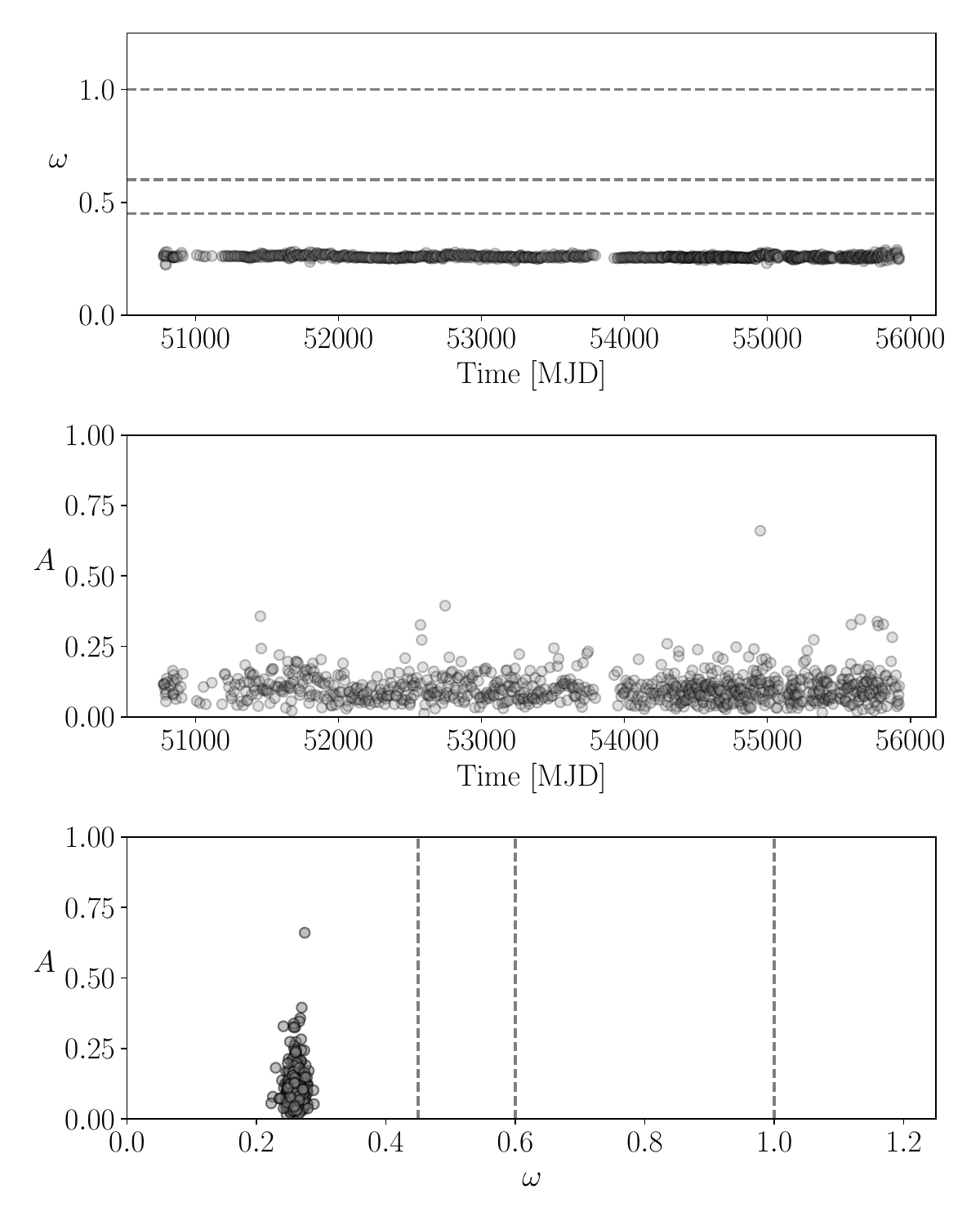}
    \caption{Same as Figure \ref{Fig:RepSource18.3} but for SXP 214.}
    \label{Fig:RepSource214}
\end{figure}

\begin{figure}[h]
\centering
\includegraphics[width=0.45\textwidth, keepaspectratio]{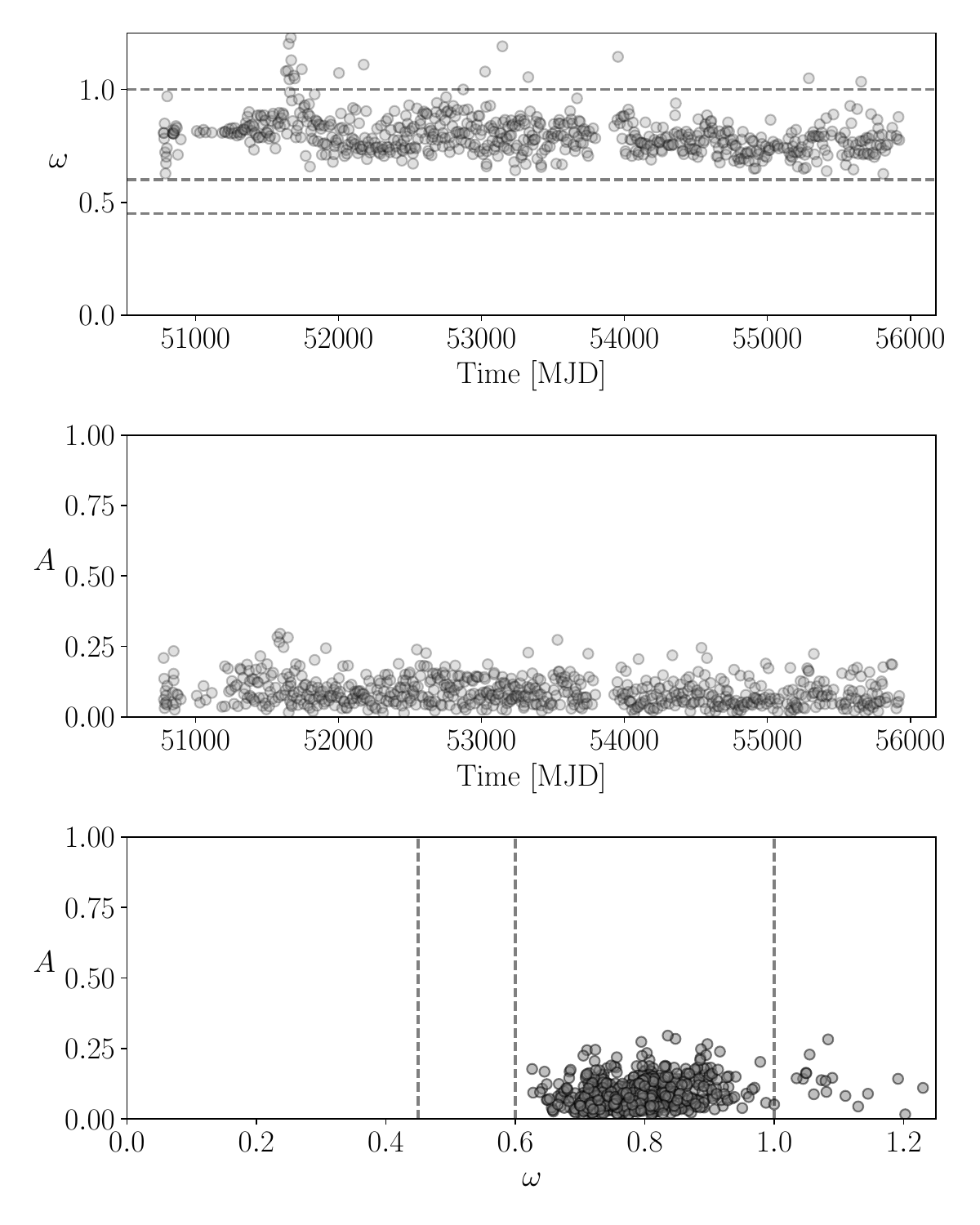}
    \caption{Same as Figure \ref{Fig:RepSource18.3} but for SXP 264.}
    \label{Fig:RepSource264}
\end{figure}

\begin{figure}[h]
   \centering \includegraphics[width=0.45\textwidth, keepaspectratio]{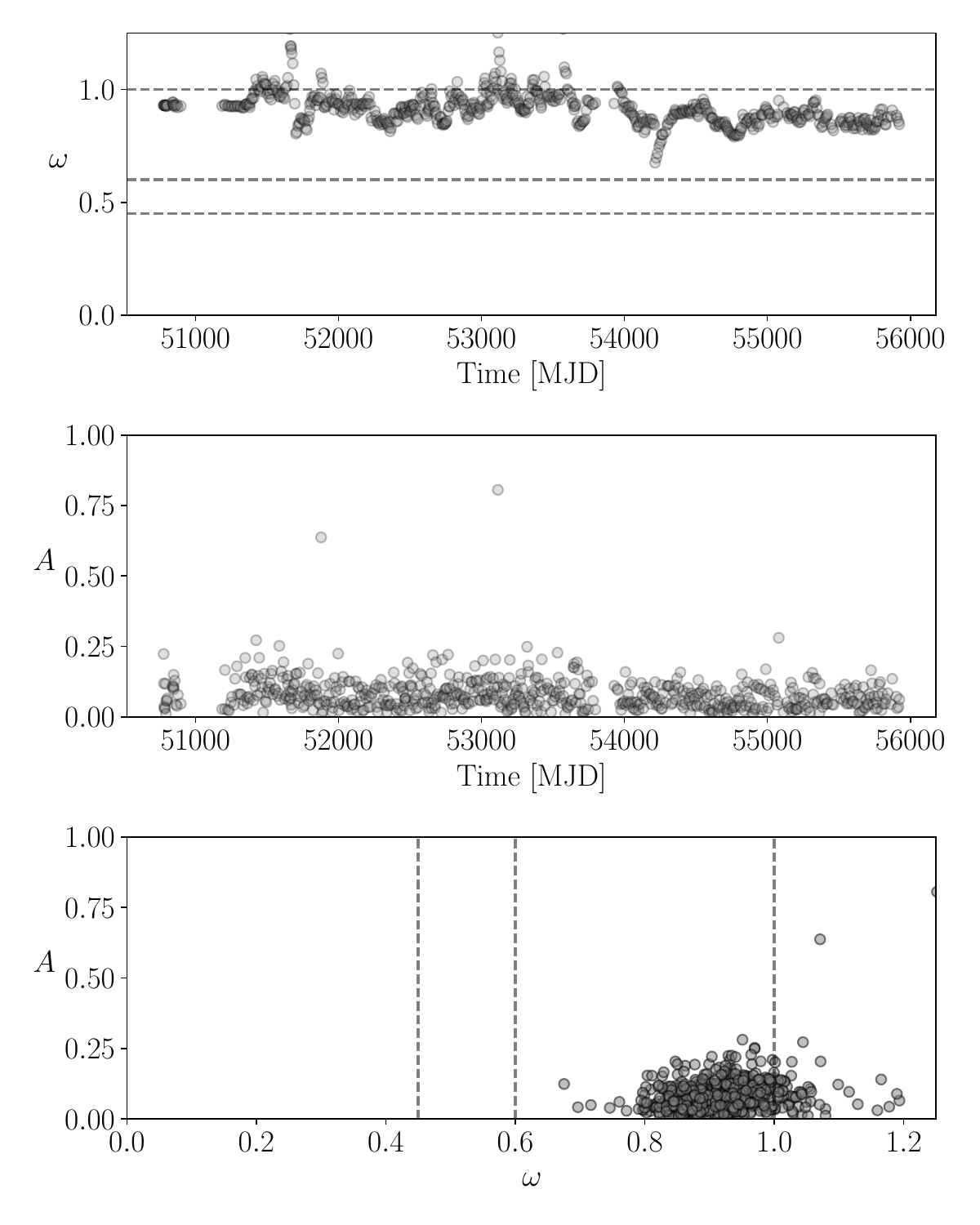}
    \caption{Same as Figure \ref{Fig:RepSource18.3} but for SXP 292.}
    \label{Fig:RepSource292}
\end{figure}

\begin{figure}[h]
\centering
\includegraphics[width=0.45\textwidth, keepaspectratio]{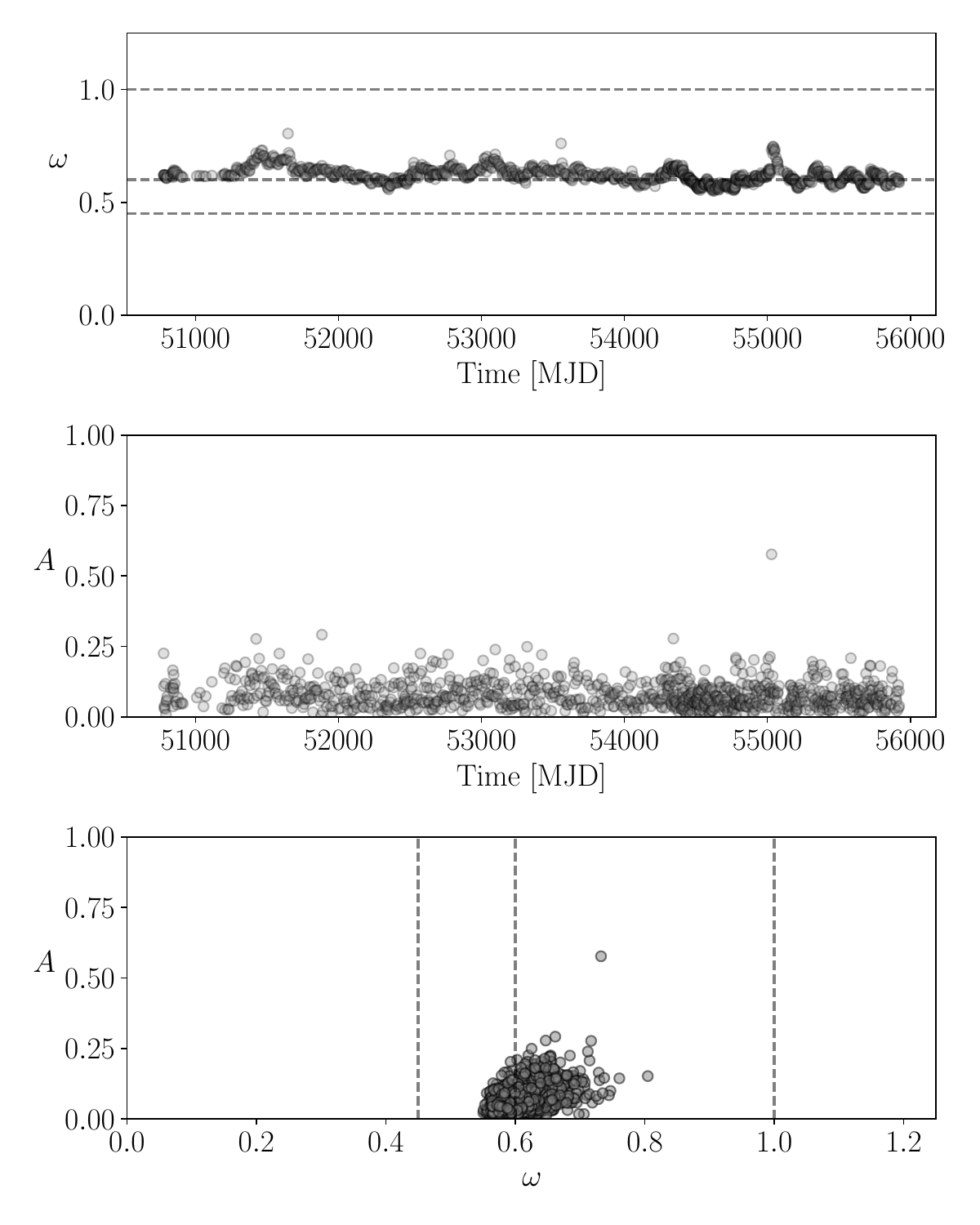}
    \caption{Same as Figure \ref{Fig:RepSource18.3} but for SXP 293.}
    \label{Fig:RepSource293}
\end{figure}

\begin{figure}[h]
    \centering
    \includegraphics[width=0.45\textwidth, keepaspectratio]{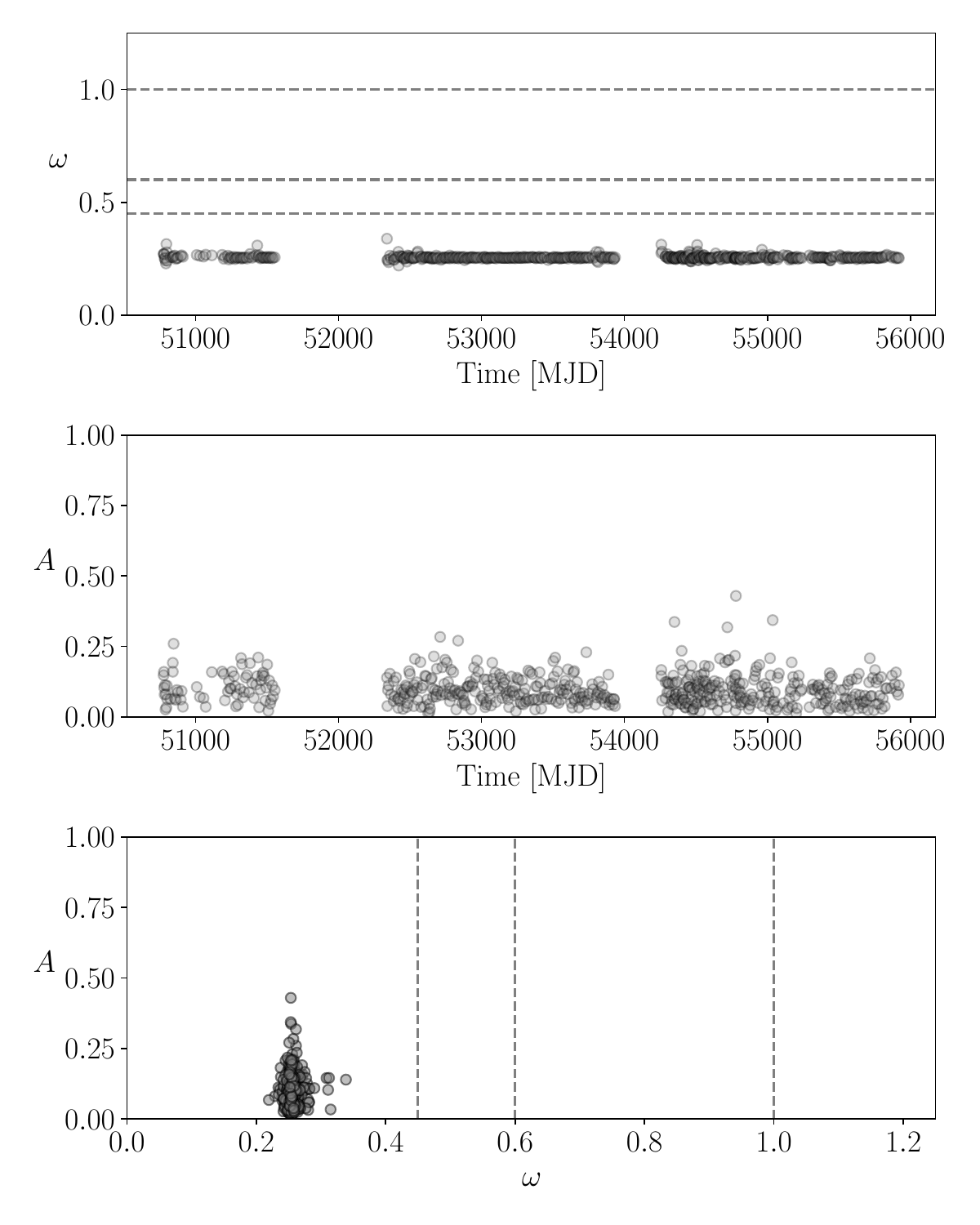}
    \caption{Same as Figure \ref{Fig:RepSource18.3} but for SXP 523.}
    \label{Fig:RepSource523}
\end{figure}

\begin{figure}[h]
    \centering
    \includegraphics[width=0.45\textwidth, keepaspectratio]{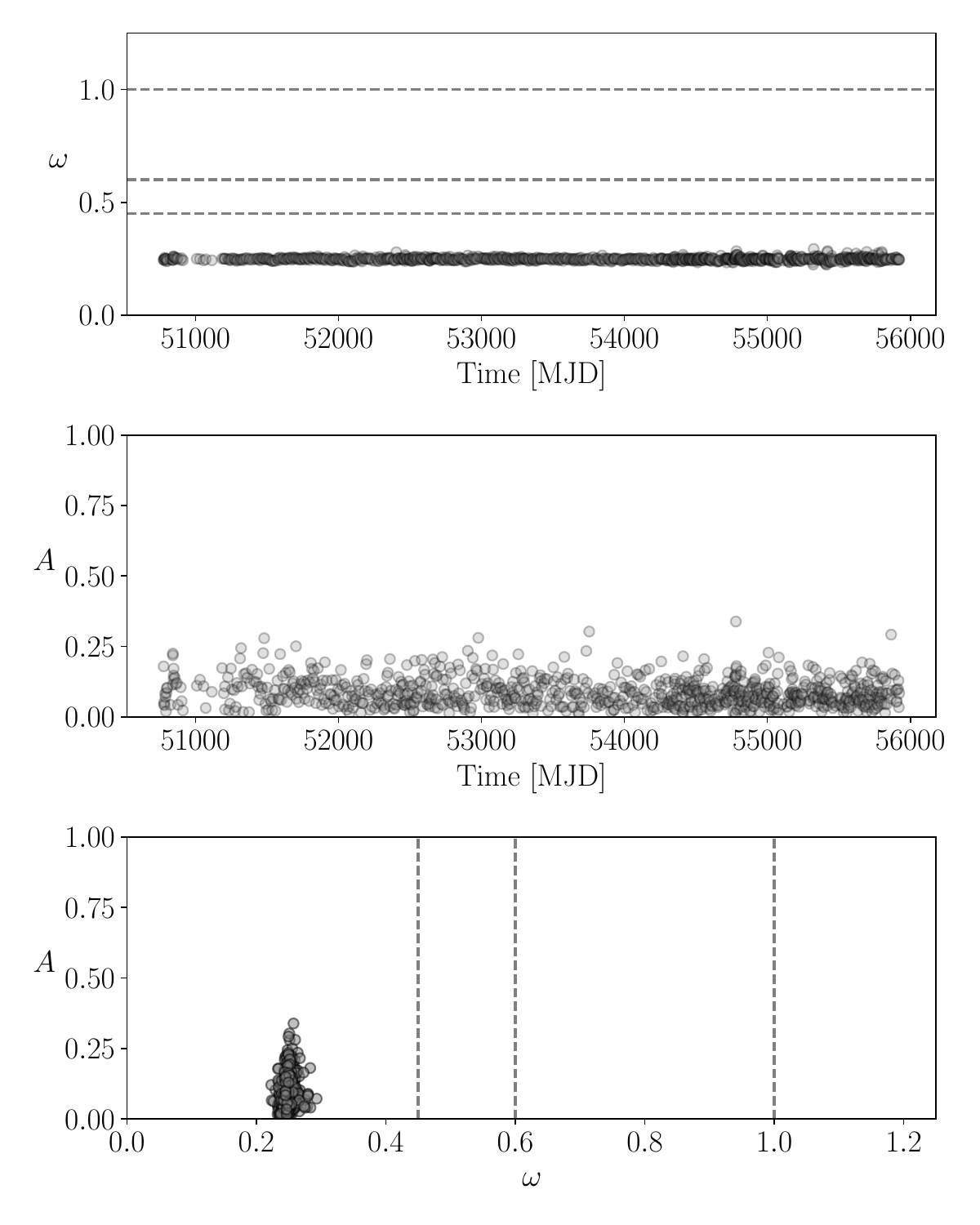}
    \caption{Same as Figure \ref{Fig:RepSource18.3} but for SXP 565.}
    \label{Fig:RepSource565}
\end{figure}

\begin{figure}[h]
    \centering
    \includegraphics[width=0.45\textwidth, keepaspectratio]{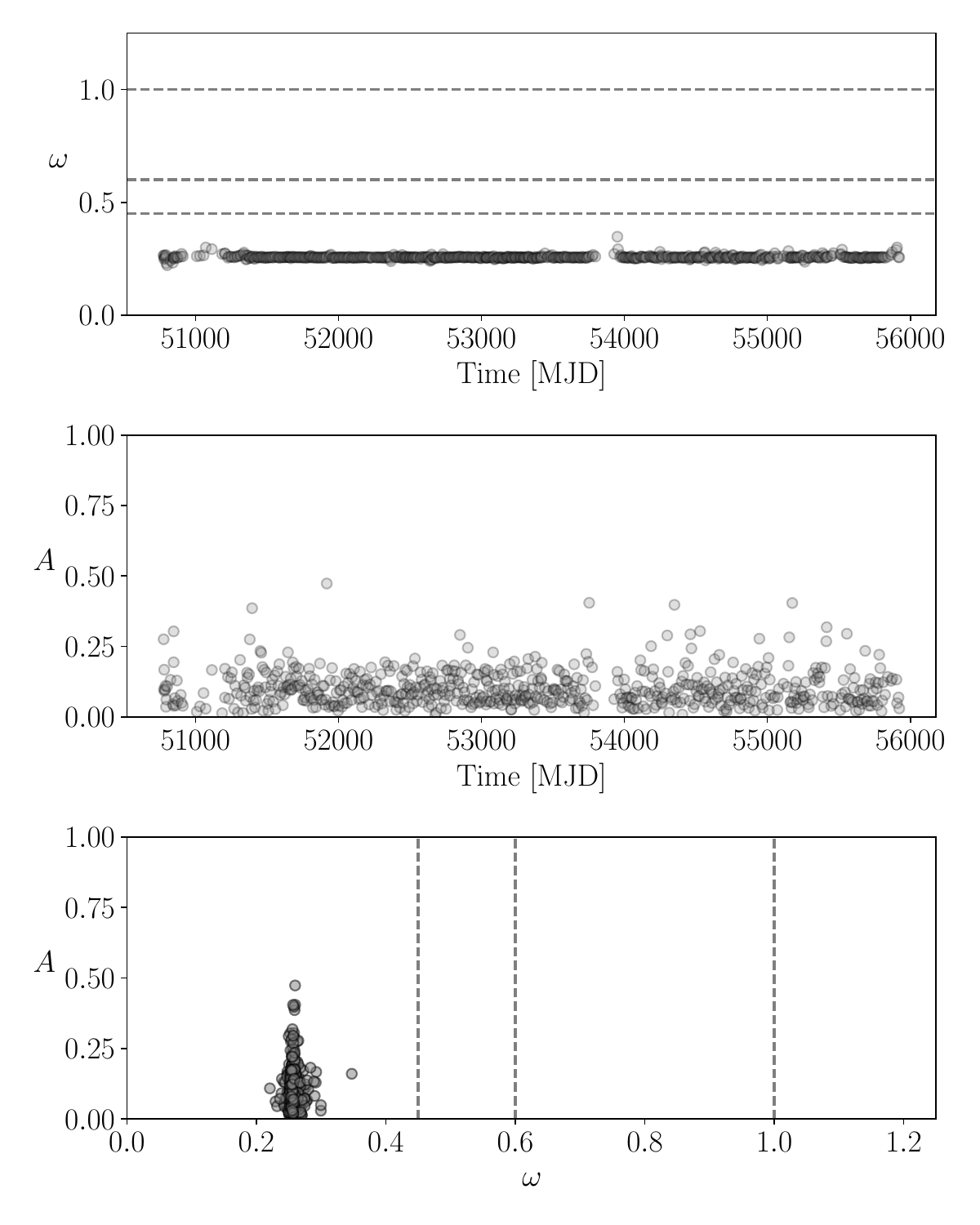}
    \caption{Same as Figure \ref{Fig:RepSource18.3} but for SXP 893.}
    \label{Fig:RepSource893}
\end{figure}

\begin{figure}[h]
    \centering
    \includegraphics[width=0.45\textwidth, keepaspectratio]{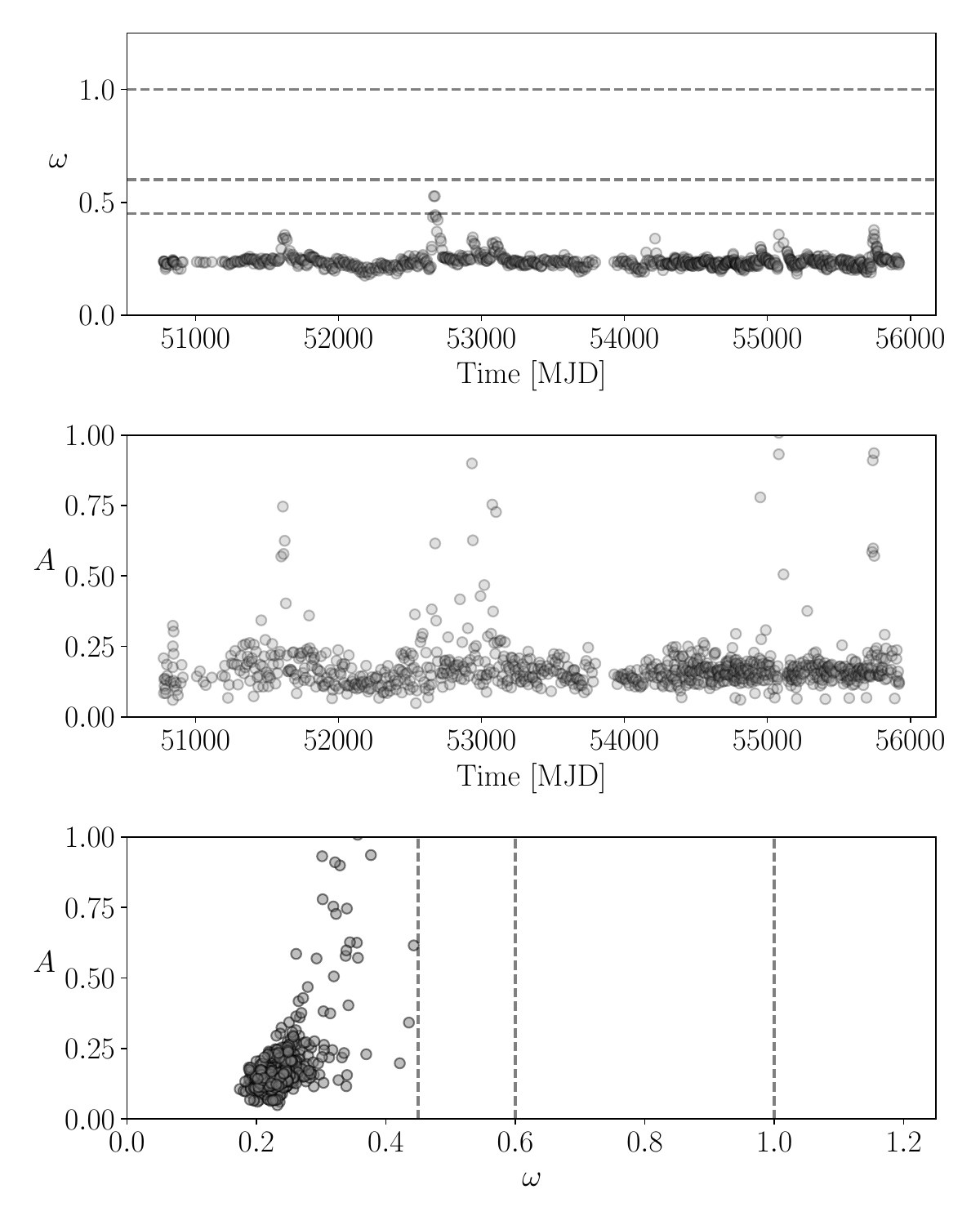}
    \caption{Same as Figure \ref{Fig:RepSource18.3} but for SXP 8.88.}
    \label{Fig:RepSource8.80}
\end{figure}

\begin{figure}[h]
    \centering
    \includegraphics[width=0.45\textwidth, keepaspectratio]{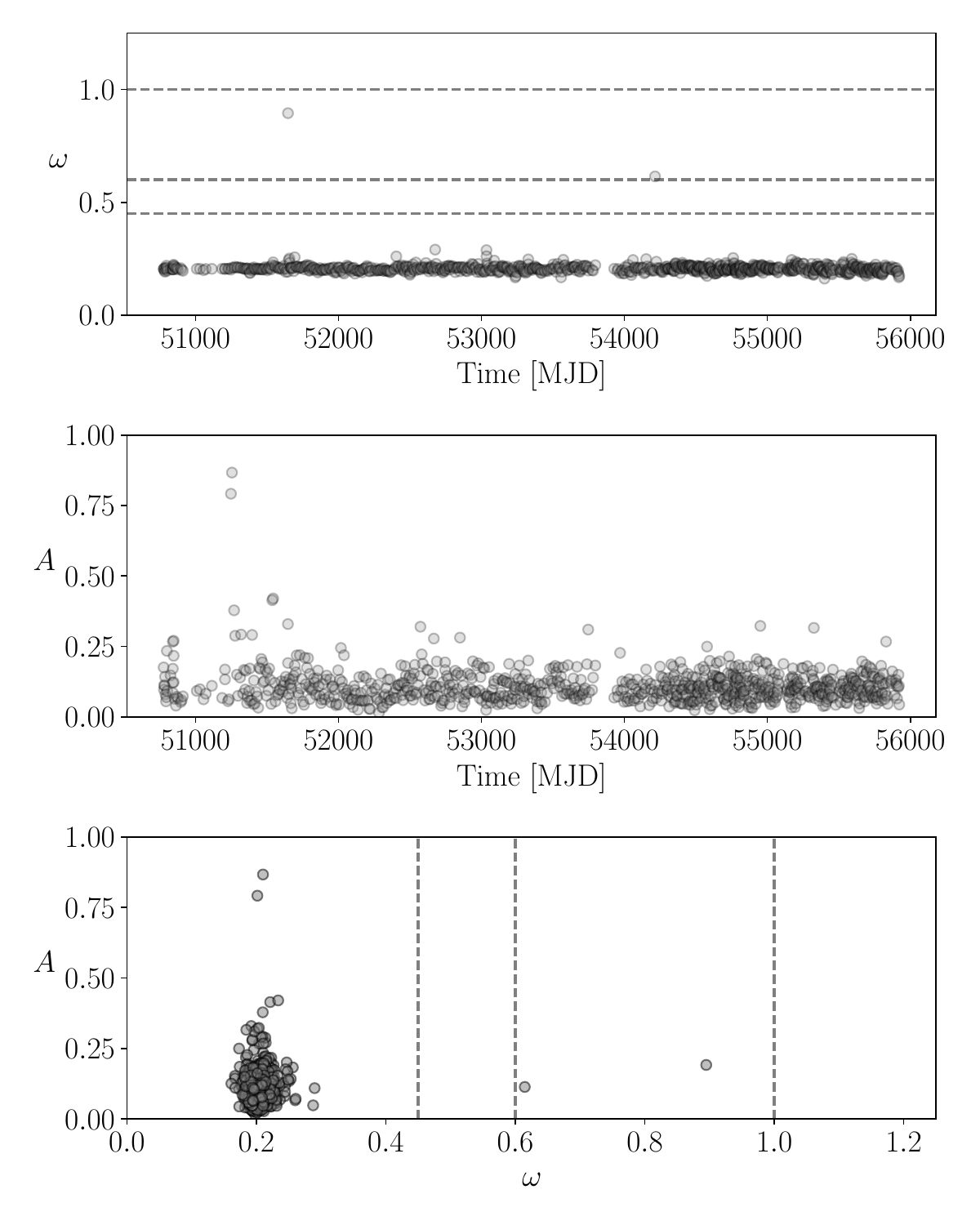}
    \caption{Same as Figure \ref{Fig:RepSource18.3} but for SXP 95.2.}
    \label{Fig:RepSource95.2}
\end{figure}

\begin{figure}[h]
    \centering
    \includegraphics[width=0.45\textwidth, keepaspectratio]{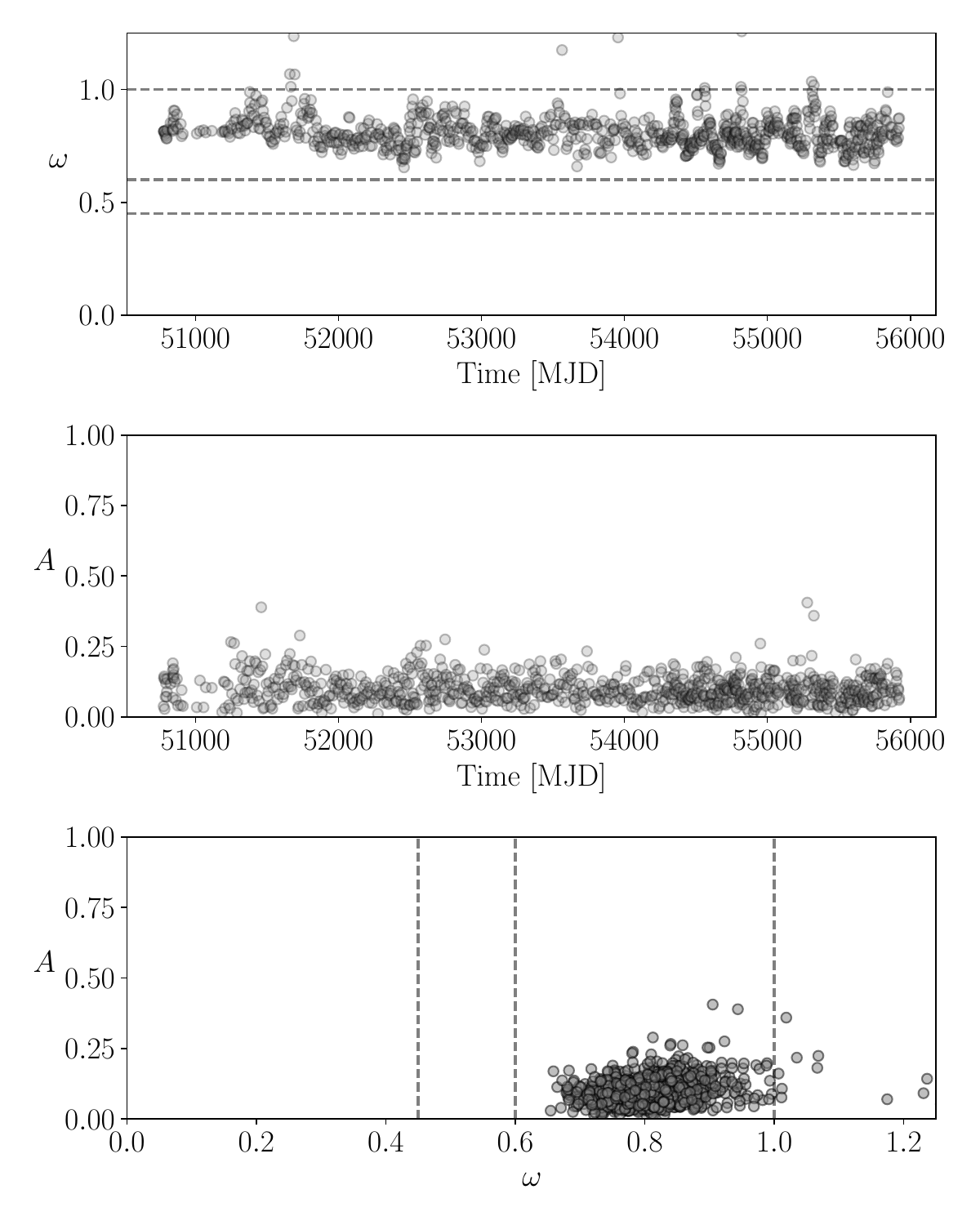}
    \caption{Same as Figure \ref{Fig:RepSource18.3} but for SXP 138.}
    \label{Fig:RepSource138}
\end{figure}

In this appendix, we present for the sake of reproducibility the Kalman filter output for 22 out of the 24 SMC X-ray pulsars studied in this paper in Figures \ref{Fig:RepSource4.78}--\ref{Fig:RepSource138}. The same information is presented for the other two objects, SXP 18.3 and SXP 51.0, which represent the stable and ordered unstable accretion regimes respectively, in Figures \ref{Fig:RepSource18.3} and \ref{Fig:RepSource51} in the main body of the paper. The formats of Figure \ref{Fig:RepSource18.3}, \ref{Fig:RepSource51}, and \ref{Fig:RepSource4.78}--\ref{Fig:RepSource138} are identical. The top panel plots the time-resolved fastness $\omega(t_n)$ as a function of $t_n$, as measured by the Kalman filter. The middle panel plots the pulse amplitude $A(t_n)$ as a function of time $t_n$. Importantly, $A(t_n)$ is measured independently using RXTE PCA data, without reference to the Kalman filter. The bottom panel plots $A(t_n)$ parametrically as a function of $\omega(t_n)$ (i.e.\ scatter plot), to help the reader visualize the $A(t_n)$-$\omega(t_n)$ cross-correlations, whose Pearson coefficients are reported in Table \ref{Table:TempCorrs}. The results are grouped according to the sign of $\dot{P}$, as in Tables \ref{Table:SourceProperties} and \ref{Table:TempCorrs}, i.e.\ spinning up ($\epsilon \leq -1.5$; Appendix \ref{SubSec:SpinUp}), near rotational equilibrium ($-1.5 < \epsilon < 1.5$; Appendix \ref{SubSec:NearRotEq}), and spinning down ($\epsilon \geq 1.5$; Appendix \ref{SubSec:SpinDown}). The grouping helps to make the point, that the accretion regime (stable or ordered unstable) does not depend on the sign of $\dot{P}$.
\subsection{Spinning up}\label{SubSec:SpinUp}
\clearpage
\subsection{Near rotational equilibrium}\label{SubSec:NearRotEq}
\clearpage
\subsection{Spinning down}\label{SubSec:SpinDown}

\section{Magnetic obliquity}\label{App:Misaligment}

Within the canonical magnetocentrifugal paradigm \citep{Ghosh_1977,Ghosh_1978,Ghosh_1979}, it is implicitly assumed that the rotation and magnetic axes are aligned, i.e. $\Theta = 0^\circ$. Recent polarimetric observations of accretion-powered pulsars reveal that some systems harbor nearly orthogonal rotators \citep{mushtukov_2023,Tsygankov_2023}, so that the Rayleigh-Taylor boundaries resolved in \cite{Blinova_2016} do not necessarily apply. Although additional -- and computationally expensive -- global, three dimensional, magnetohydrodynamic numerical simulations of inclined rotators with $5^\circ \leq \Theta \leq 90^\circ$ fall outside the scope of the present paper, we present here for completeness the fastness $\omega = \omega(t, \Theta)$ as an approximate, phenomenological function of $\Theta$. 

Within the canonical picture of magnetocentrifugal accretion, the $\Theta$-dependent disk-magnetosphere boundary occurs at the magnetospheric radius $R_{\rm m} = R_{\rm m}(t, \Theta)$, where the Maxwell stress balances the disk ram pressure, i.e. 
\begin{equation}\label{Eq:MaxwellMagnetic}
S = B^2/(8 \pi) \approx \rho v^2/2,
\end{equation}
where $\rho = Q/(4 \pi R_{\rm m}^2 v)$ and $v = (2 GM/R_{\rm m })^{1/2}$ are the mass
density and infall speed, respectively, in free fall \citep{Menou_1999,Frank_2002}. Assuming a dipolar magnetic field $\bm{B}$, the components of $\bm{B}$ are given in cylindrical coordinates by Equations (1)--(3) of \cite{Jetzer_1998} and imply $S \propto  B^2 = \mu^2 R_{\rm m}^{-6} f(\Theta)$ with $f(\Theta) = 1 + 3 \sin^2 \Theta$. Substituting $\rho, v$, as well as the foregoing expression for $S\propto B^2$ into Equation (\ref{Eq:MaxwellMagnetic}) yields 
\begin{equation}\label{Eq:RmTheta}
R_{\rm{m}}(t,\Theta) = (2 \pi^{2/5})^{-1} (GM)^{1/5} Q(t)^{2/5} [S(t) f(\Theta)]^{-2/5}
\end{equation}
and hence the fastness
\begin{equation}\label{Eq:OmTheta}
\omega(t,\Theta) = [R_{\rm m}(t,\Theta)/R_{\rm c}(t)]^{3/2}.
\end{equation}
The reader may combine Equations (\ref{Eq:RmTheta}) and (\ref{Eq:OmTheta}) with preferred values of $\Theta \neq 0$ to generate revised fastness time series for the 24 objects in this paper, mindful of the idealizations noted above.

\clearpage

\clearpage
\footnotesize
\bibliography{main.bib}{}

\begin{thebibliography}{}
\expandafter\ifx\csname natexlab\endcsname\relax\def\natexlab#1{#1}\fi
\providecommand{\url}[1]{\href{#1}{#1}}
\providecommand{\dodoi}[1]{doi:~\href{http://doi.org/#1}{\nolinkurl{#1}}}
\providecommand{\doeprint}[1]{\href{http://ascl.net/#1}{\nolinkurl{http://ascl.net/#1}}}
\providecommand{\doarXiv}[1]{\href{https://arxiv.org/abs/#1}{\nolinkurl{https://arxiv.org/abs/#1}}}

\bibitem[{J. Aly \& J. Kuijpers(1990)Aly \& Kuijpers}]{Aly_1990}
Aly, J., \& Kuijpers, J. 1990, \bibinfo{title}{Flaring interactions between accretion disk and neutron star magnetosphere,} Astronomy and Astrophysics (ISSN 0004-6361), vol. 227, no. 2, Jan. 1990, p. 473-482., 227, 473

\bibitem[{T. Amari {et~al.}(1996)Amari, Luciani, Aly, \& Tagger}]{Amari_1996}
Amari, T., Luciani, J., Aly, J., \& Tagger, M. 1996, \bibinfo{title}{Very fast opening of a three-dimensional twisted magnetic flux tube,} The Astrophysical Journal, 466, L39

\bibitem[{U. Anzer \& G. Börner(1980)Anzer \& Börner}]{Anzer_1980}
Anzer, U., \& Börner, G. 1980, \bibinfo{title}{Accretion by neutron stars-Accretion disk and rotating magnetic field,} Astronomy and Astrophysics, 83, 133

\bibitem[{J. Arons \& S.~M. Lea(1976)Arons \& Lea}]{Arons_1976}
Arons, J., \& Lea, S.~M. 1976, \bibinfo{title}{Accretion onto magnetized neutron stars-Structure and interchange instability of a model magnetosphere,} Astrophysical Journal, vol. 207, Aug. 1, 1976, pt. 1, p. 914-936. Research supported by the Science Research Council, 207, 914

\bibitem[{G. Ashton {et~al.}(2019)Ashton, Hübner, Lasky, Talbot, Ackley, Biscoveanu, Chu, Divarkala, Easter, \& Goncharov}]{Ashton_2019}
Ashton, G., Hübner, M., Lasky, P.~D., {et~al.} 2019, \bibinfo{title}{Bilby: Bayesian inference library,} Astrophysics Source Code Library, ascl: 1901.011

\bibitem[{G. Ashton {et~al.}(2022)Ashton, Bernstein, Buchner, Chen, Cs{\'a}nyi, Fowlie, Feroz, Griffiths, Handley, Habeck, {et~al.}}]{Ashton_2022}
Ashton, G., Bernstein, N., Buchner, J., {et~al.} 2022, \bibinfo{title}{Nested sampling for physical scientists,} Nature Reviews Methods Primers, 2, 1

\bibitem[{M. Bachetti {et~al.}(2010)Bachetti, Romanova, Kulkarni, Burderi, \& di~Salvo}]{Bachetti_2010}
Bachetti, M., Romanova, M.~M., Kulkarni, A., Burderi, L., \& di~Salvo, T. 2010, \bibinfo{title}{QPO emission from moving hot spots on the surface of neutron stars: a model,} Monthly Notices of the Royal Astronomical Society, 403, 1193

\bibitem[{S.~A. Balbus \& J.~F. Hawley(1991)Balbus \& Hawley}]{Balbus_1991}
Balbus, S.~A., \& Hawley, J.~F. 1991, \bibinfo{title}{A powerful local shear instability in weakly magnetized disks. I-Linear analysis. II-Nonlinear evolution,} The Astrophysical Journal, 376, 214

\bibitem[{A. Baykal(1997)Baykal}]{Baykal_1997}
Baykal, A. 1997, \bibinfo{title}{The torque and X-ray flux changes of OAO 1657-415.,} Astronomy and Astrophysics, 319, 515

\bibitem[{A. Baykal \& H. Oegelman(1993)Baykal \& Oegelman}]{Baykal_1993}
Baykal, A., \& Oegelman, H. 1993, \bibinfo{title}{An empirical torque noise and spin-up model for accretion-powered X-ray pulsars,} Astronomy and Astrophysics, 267, 119

\bibitem[{L. Bildsten {et~al.}(1997)Bildsten, Chakrabarty, Chiu, Finger, Koh, Nelson, Prince, Rubin, Scott, Stollberg, {et~al.}}]{Bildsten_1997}
Bildsten, L., Chakrabarty, D., Chiu, J., {et~al.} 1997, \bibinfo{title}{Observations of accreting pulsars,} The Astrophysical Journal Supplement Series, 113, 367

\bibitem[{A. Blinova {et~al.}(2016)Blinova, Romanova, \& Lovelace}]{Blinova_2016}
Blinova, A., Romanova, M., \& Lovelace, R. 2016, \bibinfo{title}{Boundary between stable and unstable regimes of accretion. Ordered and chaotic unstable regimes,} Monthly Notices of the Royal Astronomical Society, 459, 2354

\bibitem[{J. Bouvier {et~al.}(2006)Bouvier, Alencar, Harries, Johns-Krull, \& Romanova}]{Bouvier_2006}
Bouvier, J., Alencar, S., Harries, T., Johns-Krull, C., \& Romanova, M. 2006, \bibinfo{title}{Magnetospheric accretion in classical T Tauri stars,} arXiv preprint astro-ph/0603498

\bibitem[{K. Burdonov {et~al.}(2022)Burdonov, Yao, Sladkov, Bonito, Chen, Ciardi, Korzhimanov, Soloviev, Starodubtsev, Zemskov, {et~al.}}]{Burdonov_2022}
Burdonov, K., Yao, W., Sladkov, A., {et~al.} 2022, \bibinfo{title}{Laboratory modelling of equatorial ‘tongue’accretion channels in young stellar objects caused by the Rayleigh-Taylor instability,} Astronomy \& Astrophysics, 657, A112

\bibitem[{D.~J. Burnard {et~al.}(1983)Burnard, Arons, \& Lea}]{Burnard_1983}
Burnard, D.~J., Arons, J., \& Lea, S. 1983, \bibinfo{title}{Accretion onto magnetized neutron stars-X-ray pulsars with intermediate rotation rates,} The Astrophysical Journal, 266, 175

\bibitem[{S. Challa {et~al.}(2011)Challa, Morelande, Mušicki, \& Evans}]{Challa_2011}
Challa, S., Morelande, M.~R., Mušicki, D., \& Evans, R.~J. 2011, Fundamentals of object tracking (Cambridge University Press)

\bibitem[{M. Coe {et~al.}(2015)Coe, Bartlett, Bird, Haberl, Kennea, McBride, Townsend, \& Udalski}]{Coe_2015}
Coe, M., Bartlett, E., Bird, A., {et~al.} 2015, \bibinfo{title}{SXP 5.05= IGR J00569-7226: using X-rays to explore the structure of a Be star's circumstellar disc,} Monthly Notices of the Royal Astronomical Society, 447, 2387

\bibitem[{R. Corbet {et~al.}(2003)Corbet, Markwardt, Coe, Edge, Laycock, \& Marshall}]{Corbet_2003}
Corbet, R., Markwardt, C., Coe, M., {et~al.} 2003, \bibinfo{title}{A New Transient X-ray Pulsar in the SMC (XTE J0055-727),} The Astronomer's Telegram, 214, 1

\bibitem[{C. D'Angelo {et~al.}(2015)D'Angelo, Fridriksson, Messenger, \& Patruno}]{Dangelo_2015}
D'Angelo, C., Fridriksson, J., Messenger, C., \& Patruno, A. 2015, \bibinfo{title}{The radiative efficiency of a radiatively inefficient accretion flow,} Monthly Notices of the Royal Astronomical Society, 449, 2803

\bibitem[{C.~R. D'Angelo \& H.~C. Spruit(2010)D'Angelo \& Spruit}]{Dangelo_2010}
D'Angelo, C.~R., \& Spruit, H.~C. 2010, \bibinfo{title}{Episodic accretion on to strongly magnetic stars,} Monthly Notices of the Royal Astronomical Society, 406, 1208

\bibitem[{M. de~Kool \& U. Anzer(1993)de~Kool \& Anzer}]{deKool_1993}
de~Kool, M., \& Anzer, U. 1993, \bibinfo{title}{A simple analysis of period noise in binary X-ray pulsars,} Monthly Notices of the Royal Astronomical Society, 262, 726

\bibitem[{S. Dyda {et~al.}(2013)Dyda, Lovelace, Ustyugova, Lii, Romanova, \& Koldoba}]{Dyda_2013}
Dyda, S., Lovelace, R., Ustyugova, G., {et~al.} 2013, \bibinfo{title}{Advection of matter and B-fields in alpha-discs,} Monthly Notices of the Royal Astronomical Society, 432, 127

\bibitem[{C. D’Angelo(2017)D’Angelo}]{Dangelo_2017}
D’Angelo, C. 2017, \bibinfo{title}{Spin equilibrium in strongly magnetized accreting stars,} Monthly Notices of the Royal Astronomical Society, 470, 3316

\bibitem[{C.~R. D’Angelo \& H.~C. Spruit(2012)D’Angelo \& Spruit}]{Dangelo_2012}
D’Angelo, C.~R., \& Spruit, H.~C. 2012, \bibinfo{title}{Accretion discs trapped near corotation,} Monthly Notices of the Royal Astronomical Society, 420, 416

\bibitem[{F. Foucart \& D. Lai(2011)Foucart \& Lai}]{Foucart_2011}
Foucart, F., \& Lai, D. 2011, \bibinfo{title}{Evolution of spin direction of accreting magnetic protostars and spin--orbit misalignment in exoplanetary systems--II. Warped discs,} Monthly Notices of the Royal Astronomical Society, 412, 2799

\bibitem[{J. Frank {et~al.}(2002)Frank, King, \& Raine}]{Frank_2002}
Frank, J., King, A., \& Raine, D. 2002, Accretion power in astrophysics (Cambridge university press)

\bibitem[{C.~W. Gardiner {et~al.}(1985)Gardiner {et~al.}}]{Gardiner_1985}
Gardiner, C.~W., {et~al.} 1985, Handbook of stochastic methods, Vol.~3 (springer Berlin)

\bibitem[{P. Ghosh \& F. Lamb(1978)Ghosh \& Lamb}]{Ghosh_1978}
Ghosh, P., \& Lamb, F. 1978, \bibinfo{title}{Disk accretion by magnetic neutron stars,} Astrophysical Journal, Part 2-Letters to the Editor, vol. 223, July 15, 1978, p. L83-L87., 223, L83

\bibitem[{P. Ghosh \& F. Lamb(1979)Ghosh \& Lamb}]{Ghosh_1979}
Ghosh, P., \& Lamb, F. 1979, \bibinfo{title}{Accretion by rotating magnetic neutron stars. III-Accretion torques and period changes in pulsating X-ray sources,} The Astrophysical Journal, 234, 296

\bibitem[{P. Ghosh {et~al.}(1977)Ghosh, Lamb, \& Pethick}]{Ghosh_1977}
Ghosh, P., Lamb, F., \& Pethick, C. 1977, \bibinfo{title}{Accretion by rotating magnetic neutron stars. I-Flow of matter inside the magnetosphere and its implications for spin-up and spin-down of the star,} The Astrophysical Journal, 217, 578

\bibitem[{D. Gruber {et~al.}(1996)Gruber, Blanco, Heindl, Pelling, Rothschild, \& Hink}]{Gruber_1996}
Gruber, D., Blanco, P., Heindl, W., {et~al.} 1996, \bibinfo{title}{The high energy X-ray timing experiment on XTE.,} Astronomy and Astrophysics Supplement, v. 120, p. 641-644, 120, 641

\bibitem[{L. Hartmann {et~al.}(2016)Hartmann, Herczeg, \& Calvet}]{Hartmann_2016}
Hartmann, L., Herczeg, G., \& Calvet, N. 2016, \bibinfo{title}{Accretion onto pre-main-sequence stars,} Annual Review of Astronomy and Astrophysics, 54, 135

\bibitem[{J.~H. Horne \& S.~L. Baliunas(1986)Horne \& Baliunas}]{Horne_1986}
Horne, J.~H., \& Baliunas, S.~L. 1986, \bibinfo{title}{A prescription for period analysis of unevenly sampled time series,} Astrophysical Journal, Part 1 (ISSN 0004-637X), vol. 302, March 15, 1986, p. 757-763. Research, supported by the National Geographic Society and, the Smithsonian Institution., 302, 757

\bibitem[{K. Jahoda {et~al.}(2006)Jahoda, Markwardt, Radeva, Rots, Stark, Swank, Strohmayer, \& Zhang}]{Jahoda_2006}
Jahoda, K., Markwardt, C.~B., Radeva, Y., {et~al.} 2006, \bibinfo{title}{Calibration of the Rossi X-ray timing explorer proportional counter array,} The Astrophysical Journal Supplement Series, 163, 401

\bibitem[{P. Jetzer {et~al.}(1998)Jetzer, Str{\"a}ssle, \& Straumann}]{Jetzer_1998}
Jetzer, P., Str{\"a}ssle, M., \& Straumann, N. 1998, \bibinfo{title}{On the variation of pulsar periods in close binary systems,} New Astronomy, 3, 619

\bibitem[{S.~J. Julier \& J.~K. Uhlmann(1997)Julier \& Uhlmann}]{Julier_1997}
Julier, S.~J., \& Uhlmann, J.~K. 1997, in Signal processing, sensor fusion, and target recognition VI, Vol. 3068, Spie, 182--193

\bibitem[{A. Karaferias {et~al.}(2023)Karaferias, Vasilopoulos, Petropoulou, Jenke, Wilson-Hodge, \& Malacaria}]{karaferias_2023}
Karaferias, A., Vasilopoulos, G., Petropoulou, M., {et~al.} 2023, \bibinfo{title}{A Bayesian approach for torque modelling of BeXRB pulsars with application to super-Eddington accretors,} Monthly Notices of the Royal Astronomical Society, 520, 281

\bibitem[{J. Kennea {et~al.}(2018)Kennea, Coe, Evans, Waters, \& Jasko}]{Kennea_2018}
Kennea, J., Coe, M., Evans, P., Waters, J., \& Jasko, R. 2018, \bibinfo{title}{The first year of S-CUBED: the swift small magellanic cloud survey,} The Astrophysical Journal, 868, 47

\bibitem[{T.~L. Killestein {et~al.}(2023)Killestein, Mould, Steeghs, Casares, Galloway, \& Whelan}]{Killestein_2023}
Killestein, T.~L., Mould, M., Steeghs, D., {et~al.} 2023, \bibinfo{title}{Precision Ephemerides for Gravitational-wave Searches--IV. Corrected and refined ephemeris for Scorpius X-1,} Monthly Notices of the Royal Astronomical Society, 520, 5317

\bibitem[{H. Klus {et~al.}(2014)Klus, Ho, Coe, Corbet, \& Townsend}]{Klus_2014}
Klus, H., Ho, W.~C., Coe, M.~J., Corbet, R.~H., \& Townsend, L.~J. 2014, \bibinfo{title}{Spin period change and the magnetic fields of neutron stars in Be X-ray binaries in the Small Magellanic Cloud,} Monthly Notices of the Royal Astronomical Society, 437, 3863

\bibitem[{A. K\"onigl(1991)K\"onigl}]{Koenigl_1991}
K\"onigl, A. 1991, \bibinfo{title}{Disk accretion onto magnetic T Tauri stars,} The Astrophysical Journal, 370, L39

\bibitem[{G. Kovacs(1981)Kovacs}]{Kovacs_1981}
Kovacs, G. 1981, \bibinfo{title}{The frequency analysis of the low-amplitude Delta Scuti star HD 73763,} Acta Astronomica, vol. 31, no. 1, 1981, p. 75-83., 31, 75

\bibitem[{A. Kulkarni \& M. Romanova(2009)Kulkarni \& Romanova}]{Kulkarni_2009}
Kulkarni, A., \& Romanova, M. 2009, \bibinfo{title}{Possible quasi-periodic oscillations from unstable accretion: 3D magnetohydrodynamic simulations,} Monthly Notices of the Royal Astronomical Society, 398, 701

\bibitem[{A.~K. Kulkarni \& M.~M. Romanova(2008)Kulkarni \& Romanova}]{Kulkarni_2008}
Kulkarni, A.~K., \& Romanova, M.~M. 2008, \bibinfo{title}{Accretion to magnetized stars through the Rayleigh--Taylor instability: global 3D simulations,} Monthly Notices of the Royal Astronomical Society, 386, 673

\bibitem[{N. La~Palombara {et~al.}(2016)La~Palombara, Sidoli, Pintore, Esposito, Mereghetti, \& Tiengo}]{LaPalombara_2016}
La~Palombara, N., Sidoli, L., Pintore, F., {et~al.} 2016, \bibinfo{title}{Spectral analysis of SMC X-2 during its 2015 outburst,} Monthly Notices of the Royal Astronomical Society: Letters, 458, L74

\bibitem[{D. Lai(1999)Lai}]{Lai_1999}
Lai, D. 1999, \bibinfo{title}{Magnetically driven warping, precession, and resonances in accretion disks,} The Astrophysical Journal, 524, 1030

\bibitem[{D. Lai {et~al.}(2014)Lai, Bozzo, Kretschmar, Audard, Falanga, \& Ferrigno}]{Lai_2014}
Lai, D., Bozzo, E., Kretschmar, P., {et~al.} 2014, \bibinfo{title}{EPJ Web of Conferences, Vol. 64, Physics at the Magnetospheric Boundary,} EDP Sciences Geneva, Switzerland

\bibitem[{F. Lamb(1989)Lamb}]{Lamb_1989}
Lamb, F. 1989, in Timing Neutron Stars (Springer), 649--722

\bibitem[{D. Lazzati \& L. Stella(1997)Lazzati \& Stella}]{Lazzati_1997}
Lazzati, D., \& Stella, L. 1997, \bibinfo{title}{On the relationship between the periodic and aperiodic variability of accreting X-ray pulsars,} The Astrophysical Journal, 476, 267

\bibitem[{A.~M. Levine {et~al.}(1996)Levine, Bradt, Cui, Jernigan, Morgan, Remillard, Shirey, \& Smith}]{Levine_1996}
Levine, A.~M., Bradt, H., Cui, W., {et~al.} 1996, \bibinfo{title}{First results from the all-sky monitor on the rossi x-ray timing explorer,} The Astrophysical Journal, 469, L33

\bibitem[{P.~S. Lii {et~al.}(2014)Lii, Romanova, Ustyugova, Koldoba, \& Lovelace}]{Lii_2014}
Lii, P.~S., Romanova, M.~M., Ustyugova, G.~V., Koldoba, A.~V., \& Lovelace, R.~V. 2014, \bibinfo{title}{Propeller-driven outflows from an MRI disc,} Monthly Notices of the Royal Astronomical Society, 441, 86

\bibitem[{M.~S. Longair(2010)Longair}]{Longair_2010}
Longair, M.~S. 2010, High energy astrophysics (Cambridge university press)

\bibitem[{A. Lyne \& F. Graham-Smith(2012)Lyne \& Graham-Smith}]{Lyne_2012}
Lyne, A., \& Graham-Smith, F. 2012, Pulsar astronomy No.~48 (Cambridge University Press)

\bibitem[{A. Melatos {et~al.}(2023)Melatos, O’Neill, Meyers, \& O’Leary}]{Melatos_2022}
Melatos, A., O’Neill, N., Meyers, P., \& O’Leary, J. 2023, \bibinfo{title}{Tracking hidden magnetospheric fluctuations in accretion-powered pulsars with a Kalman filter,} The Astrophysical Journal, 944, 64

\bibitem[{K. Menou {et~al.}(1999)Menou, Esin, Narayan, Garcia, Lasota, \& McClintock}]{Menou_1999}
Menou, K., Esin, A.~A., Narayan, R., {et~al.} 1999, \bibinfo{title}{Black hole and neutron star transients in quiescence,} The Astrophysical Journal, 520, 276

\bibitem[{P.~M. Meyers {et~al.}(2021)Meyers, O’Neill, Melatos, \& Evans}]{Meyers_2021}
Meyers, P.~M., O’Neill, N.~J., Melatos, A., \& Evans, R.~J. 2021, \bibinfo{title}{Rapid parameter estimation of a two-component neutron star model with spin wandering using a Kalman filter,} Monthly Notices of the Royal Astronomical Society, 506, 3349

\bibitem[{A. Mushtukov {et~al.}(2023)Mushtukov, Tsygankov, Poutanen, Doroshenko, Salganik, Costa, Marco, Heyl, Monaca, Lutovinov, {et~al.}}]{mushtukov_2023}
Mushtukov, A., Tsygankov, S., Poutanen, J., {et~al.} 2023, \bibinfo{title}{X-ray polarimetry of X-ray pulsar X Persei: another orthogonal rotator?} Monthly Notices of the Royal Astronomical Society, 524, 2004

\bibitem[{J. O’Leary {et~al.}(2024{\natexlab{a}})O’Leary, Melatos, Kimpson, O’Neill, Meyers, Christodoulou, Bhattacharya, \& Laycock}]{OLeary_2024b}
O’Leary, J., Melatos, A., Kimpson, T., {et~al.} 2024{\natexlab{a}}, \bibinfo{title}{Measuring the magnetic dipole moment and magnetospheric fluctuations of accretion-powered pulsars in the Small Magellanic Cloud with an unscented Kalman filter,} The Astrophysical Journal, 971, 126

\bibitem[{J. O’Leary {et~al.}(2024{\natexlab{b}})O’Leary, Melatos, O’Neill, Meyers, Christodoulou, Bhattacharya, \& Laycock}]{OLeary_2024a}
O’Leary, J., Melatos, A., O’Neill, N.~J., {et~al.} 2024{\natexlab{b}}, \bibinfo{title}{Measuring the magnetic dipole moment and magnetospheric fluctuations of SXP 18.3 with a Kalman filter,} The Astrophysical Journal, 965, 102

\bibitem[{N.~J. O’Neill {et~al.}(2024)O’Neill, Meyers, \& Melatos}]{ONeill_2024}
O’Neill, N.~J., Meyers, P.~M., \& Melatos, A. 2024, \bibinfo{title}{Analysing radio pulsar timing noise with a Kalman filter: a demonstration involving PSR J1359- 6038,} Monthly Notices of the Royal Astronomical Society, stae770

\bibitem[{B. Paczynski(1985)Paczynski}]{Paczynski_1985}
Paczynski, B. 1985, in Cataclysmic Variables and Low-Mass X-Ray Binaries: Proceedings of the 7th North American Workshop held in Campbridge, Massachusetts, USA, January 12--15, 1983, Springer, 1--14

\bibitem[{A. Papitto \& D. Torres(2015)Papitto \& Torres}]{Papitto_2015}
Papitto, A., \& Torres, D. 2015, \bibinfo{title}{A propeller model for the sub-luminous state of the transitional millisecond pulsar PSR J1023+ 0038,} The Astrophysical Journal, 807, 33

\bibitem[{K. Parfrey {et~al.}(2012)Parfrey, Beloborodov, \& Hui}]{Parfrey_2012}
Parfrey, K., Beloborodov, A.~M., \& Hui, L. 2012, \bibinfo{title}{Twisting, reconnecting magnetospheres and magnetar spindown,} The Astrophysical Journal Letters, 754, L12

\bibitem[{M. Romanova {et~al.}(2021)Romanova, Koldoba, Ustyugova, Blinova, Lai, \& Lovelace}]{Romanova_2021}
Romanova, M., Koldoba, A., Ustyugova, G., {et~al.} 2021, \bibinfo{title}{3D MHD simulations of accretion on to stars with tilted magnetic and rotational axes,} Monthly Notices of the Royal Astronomical Society, 506, 372

\bibitem[{M. Romanova {et~al.}(2014)Romanova, Lovelace, Bachetti, Blinova, Koldoba, Kurosawa, Lii, \& Ustyugova}]{Romanova_2014}
Romanova, M., Lovelace, R., Bachetti, M., {et~al.} 2014, in EPJ Web of Conferences, Vol.~64, EDP Sciences, 05001

\bibitem[{M. Romanova {et~al.}(2012)Romanova, Ustyugova, Koldoba, \& Lovelace}]{Romanova_2012}
Romanova, M., Ustyugova, G., Koldoba, A., \& Lovelace, R. 2012, \bibinfo{title}{MRI-driven accretion on to magnetized stars: global 3D MHD simulations of magnetospheric and boundary layer regimes,} Monthly Notices of the Royal Astronomical Society, 421, 63

\bibitem[{M.~M. Romanova \& A.~K. Kulkarni(2009)Romanova \& Kulkarni}]{Romanova_2009}
Romanova, M.~M., \& Kulkarni, A.~K. 2009, \bibinfo{title}{Discovery of drifting high-frequency quasi-periodic oscillations in global simulations of magnetic boundary layers,} Monthly Notices of the Royal Astronomical Society, 398, 1105

\bibitem[{M.~M. Romanova {et~al.}(2008)Romanova, Kulkarni, \& Lovelace}]{Romanova_2008}
Romanova, M.~M., Kulkarni, A.~K., \& Lovelace, R.~V. 2008, \bibinfo{title}{Unstable disk accretion onto magnetized stars: First global three-dimensional magnetohydrodynamic simulations,} The Astrophysical Journal, 673, L171

\bibitem[{M.~M. Romanova \& S.~P. Owocki(2015)Romanova \& Owocki}]{Romanova_2015}
Romanova, M.~M., \& Owocki, S.~P. 2015, \bibinfo{title}{Accretion, outflows, and winds of magnetized stars,} Space Science Reviews, 191, 339

\bibitem[{M.~M. Romanova {et~al.}(2004)Romanova, Ustyugova, Koldoba, \& Lovelace}]{Romanova_2004}
Romanova, M.~M., Ustyugova, G.~V., Koldoba, A.~V., \& Lovelace, R.~V. 2004, \bibinfo{title}{Three-dimensional simulations of disk accretion to an inclined dipole. II. Hot spots and variability,} The Astrophysical Journal, 610, 920

\bibitem[{M.~M. Romanova {et~al.}(2003)Romanova, Ustyugova, Koldoba, Wick, \& Lovelace}]{Romanova_2003}
Romanova, M.~M., Ustyugova, G.~V., Koldoba, A.~V., Wick, J.~V., \& Lovelace, R.~V. 2003, \bibinfo{title}{Three-dimensional simulations of disk accretion to an inclined dipole. I. Magnetospheric flows at different $\Theta$,} The Astrophysical Journal, 595, 1009

\bibitem[{R. Rothschild {et~al.}(1998)Rothschild, Blanco, Gruber, Heindl, MacDonald, Marsden, Pelling, Wayne, \& Hink}]{Rothschild_1998}
Rothschild, R., Blanco, P., Gruber, D., {et~al.} 1998, \bibinfo{title}{In-flight performance of the high energy X-ray timing experiment on the Rossi X-ray Timing Explorer,} The Astrophysical Journal, 496, 538

\bibitem[{A. Singh {et~al.}(2024)Singh, Sanna, Bhattacharyya, Chakraborty, Jangle, Katoch, Antia, \& Bijewar}]{Singh_2024}
Singh, A., Sanna, A., Bhattacharyya, S., {et~al.} 2024, \bibinfo{title}{AstroSat timing and spectral analysis of the accretion-powered millisecond X-ray pulsar IGR J17591--2342,} Monthly Notices of the Royal Astronomical Society, stae2640

\bibitem[{N. Singh {et~al.}(2002)Singh, Naik, Paul, Agrawal, Rao, \& Singh}]{Singh_2002}
Singh, N., Naik, S., Paul, B., {et~al.} 2002, \bibinfo{title}{New measurements of orbital period change in Cygnus X-3,} Astronomy \& Astrophysics, 392, 161

\bibitem[{J. Skilling(2004)Skilling}]{Skilling_2004}
Skilling, J. 2004, \bibinfo{title}{Nested sampling,}

\bibitem[{J. Skilling(2006)Skilling}]{Skilling_2006}
Skilling, J. 2006, \bibinfo{title}{Nested sampling for general Bayesian computation,} Bayesian analysis, 1, 833

\bibitem[{J. Smak(1984)Smak}]{Smak_1984}
Smak, J. 1984, \bibinfo{title}{Accretion in cataclysmic binaries. IV-Accretion disks in dwarf novae,} Acta Astronomica, 34, 161

\bibitem[{P. Soffitta {et~al.}(2021)Soffitta, Baldini, Bellazzini, Costa, Latronico, Muleri, Del~Monte, Fabiani, Minuti, Pinchera, {et~al.}}]{Soffitta_2021}
Soffitta, P., Baldini, L., Bellazzini, R., {et~al.} 2021, \bibinfo{title}{The instrument of the imaging x-ray polarimetry explorer,} The Astronomical Journal, 162, 208

\bibitem[{J.~S. Speagle(2020)Speagle}]{Speagle_2020}
Speagle, J.~S. 2020, \bibinfo{title}{dynesty: a dynamic nested sampling package for estimating Bayesian posteriors and evidences,} Monthly Notices of the Royal Astronomical Society, 493, 3132

\bibitem[{H.~C. Spruit \& R.~E. Taam(1993)Spruit \& Taam}]{Spruit_1993}
Spruit, H.~C., \& Taam, R.~E. 1993, \bibinfo{title}{An instability associated with a magnetosphere-disk interaction,} Astrophysical Journal, Part 1 (ISSN 0004-637X), vol. 402, no. 2, p. 593-604., 402, 593

\bibitem[{J.~M. Stone \& T. Gardiner(2007{\natexlab{a}})Stone \& Gardiner}]{Stone_2007}
Stone, J.~M., \& Gardiner, T. 2007{\natexlab{a}}, \bibinfo{title}{The magnetic Rayleigh-Taylor instability in three dimensions,} The Astrophysical Journal, 671, 1726

\bibitem[{J.~M. Stone \& T. Gardiner(2007{\natexlab{b}})Stone \& Gardiner}]{Stone2007a}
Stone, J.~M., \& Gardiner, T. 2007{\natexlab{b}}, \bibinfo{title}{Nonlinear evolution of the magnetohydrodynamic Rayleigh-Taylor instability,} Physics of Fluids, 19

\bibitem[{T. Takagi {et~al.}(2016)Takagi, Mihara, Sugizaki, Makishima, \& Morii}]{Takagi_2016}
Takagi, T., Mihara, T., Sugizaki, M., Makishima, K., \& Morii, M. 2016, \bibinfo{title}{Application of the Ghosh \& Lamb relation to the spin-up/down behavior in the X-ray binary pulsar 4U 1626- 67,} Publications of the Astronomical Society of Japan, 68, S13

\bibitem[{L. Townsend {et~al.}(2011)Townsend, Coe, Corbet, \& Hill}]{Townsend_2011}
Townsend, L., Coe, M., Corbet, R., \& Hill, A. 2011, \bibinfo{title}{On the orbital parameters of Be/X-ray binaries in the Small Magellanic Cloud,} Monthly Notices of the Royal Astronomical Society, 416, 1556

\bibitem[{S.~S. Tsygankov {et~al.}(2023)Tsygankov, Doroshenko, Mushtukov, Poutanen, Di~Marco, Heyl, La~Monaca, Forsblom, Malacaria, Marshall, {et~al.}}]{Tsygankov_2023}
Tsygankov, S.~S., Doroshenko, V., Mushtukov, A.~A., {et~al.} 2023, \bibinfo{title}{X-ray pulsar GRO J1008- 57 as an orthogonal rotator,} Astronomy \& Astrophysics, 675, A48

\bibitem[{M.~M. T{\"u}rko{\u{g}}lu {et~al.}(2017)T{\"u}rko{\u{g}}lu, {\"O}zs{\"u}kan, Erkut, \& Ek{\c{s}}i}]{turkouglu_2017}
T{\"u}rko{\u{g}}lu, M.~M., {\"O}zs{\"u}kan, G., Erkut, M.~H., \& Ek{\c{s}}i, K.~Y. 2017, \bibinfo{title}{Constraints on the disc--magnetosphere interaction in accreting pulsar 4U 1626- 67,} Monthly Notices of the Royal Astronomical Society, 471, 422

\bibitem[{G. Ustyugova {et~al.}(2006)Ustyugova, Koldoba, Romanova, \& Lovelace}]{Ustyugova_2006}
Ustyugova, G., Koldoba, A., Romanova, M.~M., \& Lovelace, R. 2006, \bibinfo{title}{“Propeller” regime of disk accretion to rapidly rotating stars,} The Astrophysical Journal, 646, 304

\bibitem[{J.~T. VanderPlas(2018)VanderPlas}]{VanderPlas_2018}
VanderPlas, J.~T. 2018, \bibinfo{title}{Understanding the lomb--scargle periodogram,} The Astrophysical Journal Supplement Series, 236, 16

\bibitem[{G. Vasilopoulos {et~al.}(2017)Vasilopoulos, Zezas, Antoniou, \& Haberl}]{Vasilopoulos_2017}
Vasilopoulos, G., Zezas, A., Antoniou, V., \& Haberl, F. 2017, \bibinfo{title}{SXP 15.6: X-ray spectral and temporal properties of a newly discovered pulsar in the Small Magellanic Cloud,} Monthly Notices of the Royal Astronomical Society, 470, 4354

\bibitem[{E.~A. Wan \& R. Van Der~Merwe(2000)Wan \& Van Der~Merwe}]{Wan_2000}
Wan, E.~A., \& Van Der~Merwe, R. 2000, in Proceedings of the IEEE 2000 Adaptive Systems for Signal Processing, Communications, and Control Symposium (Cat. No. 00EX373), Ieee, 153--158

\bibitem[{E.~A. Wan \& R. Van Der~Merwe(2001)Wan \& Van Der~Merwe}]{Wan_2001}
Wan, E.~A., \& Van Der~Merwe, R. 2001, \bibinfo{title}{The unscented Kalman filter,} Kalman filtering and neural networks, 221

\bibitem[{N. White {et~al.}(1983)White, Swank, \& Holt}]{White_1983}
White, N., Swank, J., \& Holt, S. 1983, \bibinfo{title}{Accretion powered X-ray pulsars,} The Astrophysical Journal, 270, 711

\bibitem[{J. Yang {et~al.}(2017)Yang, Laycock, Christodoulou, Fingerman, Coe, \& Drake}]{Yang_2017}
Yang, J., Laycock, S., Christodoulou, D., {et~al.} 2017, \bibinfo{title}{A comprehensive library of X-ray pulsars in the Small Magellanic Cloud: Time evolution of their luminosities and spin periods,} The Astrophysical Journal, 839, 119

\bibitem[{F. Yatabe {et~al.}(2018)Yatabe, Makishima, Mihara, Nakajima, Sugizaki, Kitamoto, Yoshida, \& Takagi}]{Yatabe_2018}
Yatabe, F., Makishima, K., Mihara, T., {et~al.} 2018, \bibinfo{title}{An application of the Ghosh \& Lamb model to the accretion-powered X-ray pulsar X Persei,} Publications of the Astronomical Society of Japan, 70, 89

\bibitem[{P. Zarchan(2005)Zarchan}]{Zarchan_2005}
Zarchan, P. 2005, Progress in astronautics and aeronautics: fundamentals of Kalman filtering: a practical approach, Vol. 208 (Aiaa)

\end{thebibliography}
\bibliographystyle{aasjournal}
\end{document}